\documentclass[reprint,amsmath,amssymb,aps,superscriptaddress,longbibliography,prb]{revtex4-2}

\usepackage{mathtools}
\usepackage{graphicx}
\usepackage{dcolumn}
\usepackage{bm}
\usepackage{hyperref}
\usepackage{tikz}
\usepackage[italicdiff]{physics}
\usepackage{textcomp}

\usetikzlibrary{intersections, calc}
\usetikzlibrary{decorations.pathmorphing}
\usetikzlibrary{decorations.markings}
\usetikzlibrary{arrows.meta}

\hypersetup{
	colorlinks=true,
	citecolor=red,
	linkcolor=blue,
	urlcolor=magenta,
}

\newcommand{\leftmixing}{\rceil}
\newcommand{\rightmixing}{\lceil}

\begin{document}

\title{Weak-coupling tensor cross interpolation impurity solver for nonequilibrium dynamical mean-field theory}
\author{Shuta Matsuura}
\affiliation{Department of Physics, University of Tokyo, Hongo, Tokyo 113-0033, Japan}
\author{Hiroshi Shinaoka}
\affiliation{Department of Physics, Saitama University, Saitama 338-8570, Japan}
\author{Philipp Werner}
\affiliation{Department of Physics, University of Fribourg, 1700 Fribourg, Switzerland}
\author{Naoto Tsuji}
\affiliation{Department of Physics, University of Tokyo, Hongo, Tokyo 113-0033, Japan}
\affiliation{RIKEN Center for Emergent Matter Science (CEMS), Wako, Saitama 351-0198, Japan}
\affiliation{Trans-scale Quantum Science Institute, University of Tokyo, Hongo, Tokyo 113-0033, Japan}

\date{\today}

\begin{abstract}
Simulating nonequilibrium quantum many-body systems remains a major challenge due to the exponential growth of the computational complexity with real time.
Here we implement a nonequilibrium impurity solver based on the weak-coupling expansion and the tensor cross interpolation, and apply it to nonequilibrium dynamical mean-field theory (DMFT).
The method approximates the integrands of the high-dimensional integrals arising in the weak-coupling expansion in a tensor-train form, enabling efficient evaluations without stochastic sampling and thereby mitigating the sign problem affecting continuous-time quantum Monte Carlo (CT-QMC) methods.
Benchmark calculations for an exactly solvable nonequilibrium impurity model agree well with the exact results and reveal a low-rank structure of the integrands. 
When applied to interaction-quench problems in the half-filled Hubbard model, the method reproduces fast thermalization at a critical interaction strength with accuracy comparable to CT-QMC.
Away from half filling, where the sign problem becomes even more severe, the present approach remains well controlled, revealing a crossover instead of a sharply defined fast thermalization point in the 3/4-filled case.
The solver can also be applied to steady-state DMFT problems, yielding accurate spectral functions in the metallic regime without analytic continuation.
\end{abstract}

\maketitle

\section{Introduction}
\label{sec:intro}

Recent advances in ultrafast pump-probe techniques have made it possible to observe and manipulate nonequilibrium dynamics in strongly correlated electron systems on their intrinsic time scales \cite{giannetti2016ultrafast,basov2017towards,delatorre2021nonthermal}. 
In parallel, ultracold-atom systems provide a complementary route to study nonequilibrium many-body physics, offering clean realizations of Hubbard-type models and well-controlled driving protocols \cite{bloch2008manybody,jordens2008mott,schneider2008metallic}. 
In such experiments, the system does not always relax to an equilibrium state within a short time, but can exhibit long-lived nonthermal states which are 
distinct from equilibrium ones.
Prominent examples include light-induced transient superconducting-like states in stripe-ordered cuprates \cite{fausti2011light} and ultrafast switching to a robust low-resistance state in a layered dichalcogenide \cite{stojchevska2014ultrafast}. 

To understand such phenomena from a microscopic perspective, one needs theoretical frameworks that can accurately treat the effect of strong correlations in real time. 
One of the powerful approaches for this purpose is nonequilibrium dynamical mean-field theory (DMFT) \cite{schmidt2002nonequilibrium,freericks2006nonequilibrium,aoki2014nonequilibrium}. 
In this framework, a lattice model is mapped onto an effective impurity model coupled to a self-consistently determined bath, enabling an exact calculation of the dynamics of lattice models in the limit of infinite dimensions. 
The power of this framework has been demonstrated, for example, in studies of thermalization after interaction quenches \cite{eckstein2008nonthermal,eckstein2009thermalization,eckstein2010interaction}, dielectric breakdown of Mott insulators \cite{eckstein2010dielectric}, and nonthermal critical behavior observed in the dynamics of a symmetry-broken phase  \cite{werner2012nonthermal,tsuji2013nonequilibrium,tsuji2013nonthermal}.

Despite these successes, accessing the long-time regime in a numerically controlled manner within nonequilibrium DMFT remains challenging. 
A major obstacle is the lack of accurate impurity solvers that can efficiently simulate the real-time dynamics of impurity models over long times.
Among the numerically exact methods, a widely used approach is the continuous-time quantum Monte Carlo (CT-QMC) method \cite{rubtsov2005continuous,werner2006continuous,gull2008continuous,muhlbacher2008real,werner2009diagrammatic,werner2010weak,gull2011continuous}, where
the high-dimensional integrals arising in perturbative expansions are evaluated by stochastic sampling.
Although CT-QMC has been very successful in solving equilibrium impurity problems, its application to nonequilibrium systems is severely limited by the dynamical sign problem.
This has restricted existing nonequilibrium DMFT studies based on CT-QMC solvers to short times and the particle-hole-symmetric case, where the sign problem is less severe \cite{eckstein2010interaction,eckstein2009thermalization,tsuji2011,eckstein2011,canovi2014}.

This limitation motivates the search for alternatives to CT-QMC that can evaluate high-dimensional integrals arising in impurity problems without relying on the Monte Carlo sampling. 
A promising candidate is the tensor-train method based on the tensor cross interpolation (TCI) algorithm, which was originally developed in the field of applied mathematics \cite{oseledets2010tt,oseledets2011tensor,savostyanov2011fast,savostyanov2014quasioptimality,dolgov2020parallel,nunez2022learning,nunez2025learning}. 
In this approach, the integrand is regarded as a high-dimensional tensor after discretization, and is approximated by a tensor-train form (analogous to a matrix product state for one-dimensional quantum systems) by the TCI algorithm. 
A key feature of the TCI algorithm is that even when the original tensor can only be evaluated on demand but cannot be fully stored, 
it is still possible to construct a tensor-train approximation \cite{oseledets2010tt,savostyanov2014quasioptimality,dolgov2020parallel}, which is in stark contrast to singular value decomposition-based (SVD-based) methods.
Once a low-rank tensor-train representation is obtained, the original high-dimensional integral is reduced to a sequence of one-dimensional integrals followed by tensor contractions, which can be performed efficiently.
Because the method does not rely on stochastic sampling, it is expected to remain useful even in situations where Monte Carlo approaches suffer from severe sign cancellations, provided that a low-rank tensor-train structure exists.

Based on these advantages, TCI impurity solvers have recently been explored for both equilibrium and nonequilibrium problems \cite{erpenbeck2023tensor,matsuura2025tensor,ishida2025lowrank,nunez2022learning,jeannin2025comprehensive,eckstein2024solving,kim2025strong,geng2025third,geng2025photoinduced,geng2026high}. 
In the nonequilibrium context, TCI has been combined with the strong-coupling expansion, and applications to nonequilibrium DMFT have been demonstrated \cite{eckstein2024solving,kim2025strong,geng2025third,geng2025photoinduced,geng2026high}. 
In contrast, previous studies of the weak-coupling TCI approach  \cite{nunez2022learning,jeannin2025comprehensive} have been limited to impurity models, and focused on temporally local observables such as the charge and current at the impurity site. They therefore did not compute the two-time Green's functions required for nonequilibrium DMFT, and the application of the weak-coupling TCI approach to nonequilibrium DMFT has not yet been demonstrated.

In this paper, we address this problem by developing a nonequilibrium impurity solver that combines the weak-coupling expansion with the TCI algorithm to compute the two-time Green's functions required for nonequilibrium DMFT.
We first benchmark the method against an exactly solvable nonequilibrium impurity model, and show that the integrands appearing in the weak-coupling expansion indeed exhibit low-rank structures that can be efficiently captured by TCI. 
This makes it possible to evaluate the perturbation series up to twentieth order with moderate computational cost. 

We then apply the solver to nonequilibrium DMFT to study interaction quenches in the Hubbard model.
At half filling, the method reaches a level of accuracy comparable to that of CT-QMC, and reproduces the previous results on the relaxation behavior, including rapid thermalization near the dynamical transition point \cite{eckstein2009thermalization}. 
More importantly, the method remains numerically well controlled even away from half filling.
In this regime, the sign problem becomes even more severe for CT-QMC, significantly limiting its applicability to real-time nonequilibrium DMFT calculations. 
We show that within the present TCI approach the 3/4-filled case can be treated with nearly the same level of numerical effort as the particle-hole-symmetric case. We find that the sharp fast-thermalization behavior observed at half filling is absent within the accessible interaction range.
We also apply the solver to a nonequilibrium steady-state formalism in the metallic regime that gives direct access to real-frequency spectra without analytic continuation.

The rest of this paper is organized as follows: 
In Sec.~\ref{sec:noneq DMFT}, we introduce the model and review the nonequilibrium 
Green's function and DMFT formalism.
In Sec.~\ref{sec:tensor-train methods}, we derive the weak-coupling expansion for Green's functions and explain how to evaluate the expansion to high order using the TCI algorithm.
In Sec.~\ref{sec:exactly solvable impurity model}, we benchmark our solver against an exactly solvable impurity model, and show that even the system without particle-hole symmetry can be studied accurately.
In Sec.~\ref{sec:Hubbard quench}, we incorporate our solver into a nonequilibrium DMFT loop, apply it to interaction quenches in the Hubbard model, and discuss the thermalization behavior.
Section~\ref{sec:Hubbard steady state} demonstrates the steady-state DMFT calculation of the Hubbard model.
Finally, in Sec.~\ref{sec:discussions}, we summarize the results and discuss future perspectives.

\section{Nonequilibrium DMFT} \label{sec:noneq DMFT}
In this section, we introduce the lattice model considered in this work and review the formulation of nonequilibrium DMFT \cite{schmidt2002nonequilibrium,freericks2006nonequilibrium,aoki2014nonequilibrium}. In particular, we summarize the formulation for transient dynamics starting from an initial thermal equilibrium state and the steady-state formulation of Ref.~\cite{kunzel2024numerically}, which directly targets the stationary limit.

\subsection{Model}
\label{sec:model}
In this paper, we consider a lattice model described by the following time-dependent Hamiltonian,
\begin{align}
  &H(t) = \sum_{\ev*{i,j},\sigma} \frac{v_{\sigma}}{\sqrt{Z}} c_{i\sigma}^\dagger c_{j\sigma}
  + U(t) \sum_{i} \qty(n_{i\uparrow} - \frac{1}{2}) \qty(n_{i\downarrow} - \frac{1}{2}) \notag \\
  &\hspace{5.5cm} - \delta \mu \sum_{i,\sigma} n_{i\sigma},
  \label{eq:lattice model}
\end{align}
where $c_{i\sigma}^\dagger$ ($c_{i\sigma}$) is the creation (annihilation) operator of an electron with spin $\sigma$ at site $i$, $n_{i\sigma} = c_{i\sigma}^\dagger c_{i\sigma}$ the number operator,
$v_{\sigma}$ the spin-dependent hopping amplitude, $U(t)$ the time-dependent on-site interaction, and $\delta \mu$ the chemical-potential shift from the half-filled point [$\mu = U(0)/2$].
We assume that the system is defined on the Bethe lattice with connectivity $Z$ and the hopping is only between nearest neighbors $\ev*{i,j}$.
The hopping amplitude is scaled by $1/\sqrt{Z}$ to ensure a finite kinetic energy in the limit of $Z \to \infty$.
This model, with $v_\uparrow\neq v_\downarrow$, is known as the mass-imbalanced Hubbard model, which reduces to the standard Hubbard model \cite{hubbard1963electron} when $v_{\uparrow} = v_{\downarrow}$, and to the Falicov--Kimball model \cite{falicov1969simple,freericks2003exact} when $v_{\uparrow} \neq 0$ and $v_{\downarrow} = 0$ (or $v_{\uparrow}=0, v_{\downarrow} \neq 0$).
For simplicity, we focus only on the paramagnetic phase, and do not consider the possibility of symmetry breaking.

\subsection{Nonequilibrium Green's functions}
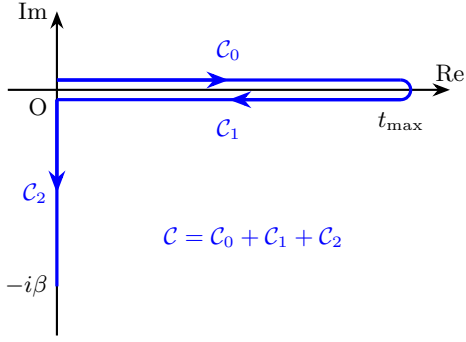
\begin{figure}[t]
	\centering
  \begin{tikzpicture}[>=Stealth, thick, scale=1.3]
    \coordinate (t0) at (0,0);
    \coordinate (tmax) at (5,0);
    \coordinate (t0rev) at (0,-0.8);
    \coordinate (t0beta) at (0,-3);
  
    \draw[->,thick] (-0.5,0) -- (4.0,0) node[above] {Re};
    \draw[<-,thick] (0,0.8) node[left]{Im} -- (0,-2.5);

    \draw[very thick, blue] (0,0.1) -- (3.5,0.1) node[midway, above=4pt] {$\mathcal{C}_{0}$};
    \draw[->,very thick, blue] (0,0.1) -- (1.75,0.1);
    \node[below left] at (t0) {O};
    \node[below=5pt] at (3.5, 0.0) {\(t_{\text{max}}\)};
  
    \draw[very thick, blue] (3.5,0.1) arc[start angle=90, end angle=-90, radius=0.1];
  
    \draw[very thick, blue] (3.5,-0.1) -- (0.0,-0.1) node[midway, below=4pt] {$\mathcal{C}_{1}$};
    \draw[->,very thick, blue] (3.5,-0.1) -- (1.75,-0.1);
  
    \draw[very thick, blue] (0.0,-0.1) -- (0.0,-2.0) node[midway, left] {$\mathcal{C}_{2}$};
    \draw[->,very thick, blue] (0.0,-0.1) -- (0.0,-1.05);
    \draw (0.0,-2.0) node[left]{$-i\beta$};

    \draw[blue] (2.0, -1.5) node{$\mathcal{C} = \mathcal{C}_{0} + \mathcal{C}_{1} + \mathcal{C}_{2}$};
  \end{tikzpicture}
	\caption{\textsf{L}-shaped contour $\mathcal{C}$ that consists of three branches: the forward real-time branch $\mathcal{C}_{0}$, the backward real-time branch $\mathcal{C}_{1}$, and the imaginary-time branch $\mathcal{C}_{2}$.}
	\label{fig:kb_contour}
\end{figure}

We first introduce the nonequilibrium Green's function and summarize the notations and conventions used in the rest of the paper.
The nonequilibrium dynamics starting from a thermal equilibrium state is formulated on the Kadanoff--Baym contour $\mathcal{C}$, which is an \textsf{L}-shaped contour in the complex time plane as shown in Fig.~\ref{fig:kb_contour} \cite{kadanoff1962quantum,danielewicz1984quantum1,danielewicz1984quantum2,wagner1991expansions}.
The contour $\mathcal{C}$ consists of three branches: the forward real-time branch $\mathcal{C}_{0}$ from zero to $t_{\mathrm{max}}$, the backward real-time branch $\mathcal{C}_{1}$ from $t_{\mathrm{max}}$ to zero, and the imaginary-time branch $\mathcal{C}_{2}$ from zero to $-i\beta$.
Here, $t_{\mathrm{max}}$ is the maximum real time up to which we simulate the dynamics, and $\beta$ is the inverse temperature of the initial equilibrium state.

To specify the position of an operator on the contour, we use a complex variable $z$.
If $z\in\mathcal C_2$, it can be identified with the imaginary time $-i\tau$, and we simply write $z=-i\tau$.
For $z\in\mathcal C_0\cup\mathcal C_1$, the real time $t$ alone does not uniquely specify the contour point; one must also indicate the branch index $s$, with $s=0$ for $\mathcal C_0$ and $s=1$ for $\mathcal C_1$. We therefore write $z=t^{(s)}$.

The nonequilibrium Green's function is defined as a two-point correlation function on the contour $\mathcal{C}$,
\begin{align}
  G_{\sigma}(z,z') = -i \frac{\tr [ T_{\mathcal{C}} c_{\sigma}(z) c_{\sigma}^{\dag}(z') e^{i \mathcal{S}}]}{\tr [ T_{\mathcal{C}} e^{i \mathcal{S}}]},
\end{align}
where $\mathcal{S}$ is the action of the system and $T_{\mathcal{C}}$ is the contour-ordering operator.
Since both $z$ and $z'$ may lie on any of the three branches, $G_\sigma(z,z')$ consists of $3\times3=9$ components:
\begin{subequations}
  \begin{align}
    &G_{\sigma}^{ss'}(t,t') = G_{\sigma}(t^{(s)}, t'^{(s')}) & &\quad (s,s' = 0,1), \\
    &G_{\sigma}^{s2}(t,\tau) = G_{\sigma}(t^{(s)}, -i \tau) & &\quad (s = 0,1), \\
    &G_{\sigma}^{2s}(\tau,t) = G_{\sigma}(-i \tau, t^{(s)}) & &\quad (s = 0,1), \\
    &G_{\sigma}^{22}(\tau,\tau') = G_{\sigma}(-i \tau, -i \tau').
  \end{align}
\end{subequations}
Some of these components and their linear combinations have special notations:
\begin{subequations}
  \begin{align}
    &G_{\sigma}^{R}(t,t') = \frac{1}{2} (G_{\sigma}^{00} - G_{\sigma}^{01} + G_{\sigma}^{10} - G_{\sigma}^{11})(t,t'), \\
    &G_{\sigma}^{A}(t,t') = \frac{1}{2} (G_{\sigma}^{00} + G_{\sigma}^{01} - G_{\sigma}^{10} - G_{\sigma}^{11})(t,t'), \\
    &G_{\sigma}^{<}(t,t') = G_{\sigma}^{01}(t,t'), \\
    &G_{\sigma}^{>}(t,t') = G_{\sigma}^{10}(t,t'), \\
    &G_{\sigma}^{\leftmixing}(t,\tau') = G_{\sigma}^{02}(t, \tau'), \\
    &G_{\sigma}^{\rightmixing}(\tau,t') = G_{\sigma}^{20}(\tau,t'), \\
    &G_{\sigma}^{M}(\tau,\tau') = -i G_{\sigma}^{22}(\tau,\tau').
  \end{align}
\end{subequations}
They are called the retarded, advanced, lesser, greater, left-mixing, right-mixing, and Matsubara Green's functions, respectively.
The Matsubara Green's function depends only on the imaginary time difference $\tau - \tau'$, so we write it as $G_{\sigma}^{M}(\tau - \tau')$.

Because of the causal structure, the Green's functions satisfy the relations,
\begin{subequations}
  \begin{align}
    G_{\sigma}^{00}(t,t') &= \theta(t-t') G_{\sigma}^{>}(t,t') + \theta(t'-t) G_{\sigma}^{<}(t,t'), \\
    G_{\sigma}^{11}(t,t') &= \theta(t-t') G_{\sigma}^{<}(t,t') + \theta(t'-t) G_{\sigma}^{>}(t,t'), \\
    G_{\sigma}^{02}(t, \tau') &= G_{\sigma}^{12}(t, \tau'), \\
    G_{\sigma}^{20}(\tau, t') &= G_{\sigma}^{21}(\tau, t').
  \end{align}
  \label{eq:causal relations}
\end{subequations}
In addition to these relations, the Green's functions also satisfy the Hermiticity relations,
\begin{subequations}
  \begin{align}
    G_{\sigma}^{<}(t,t')^{*} &= - G_{\sigma}^{<}(t',t), \\
    G_{\sigma}^{>}(t,t')^{*} &= - G_{\sigma}^{>}(t',t), \\
    G_{\sigma}^{\leftmixing}(t, \tau')^{*} &= G_{\sigma}^{\rightmixing}(\beta - \tau', t).
  \end{align}
  \label{eq:hermiticity relations}
\end{subequations}
Using Eqs.~\eqref{eq:causal relations} and \eqref{eq:hermiticity relations}, we can construct all nine components of the Green's function just from 
$G_{\sigma}^{<}(t,t') \,\, (0 \le t' \le t \le t_{\mathrm{max}})$, $G_{\sigma}^{>}(t,t') \,\, (0 \le t' \le t \le t_{\mathrm{max}})$, $G_{\sigma}^{\leftmixing}(t, \tau') \,\, (0 \le t \le t_{\mathrm{max}}, 0 \le \tau' \le \beta)$ and $G_{\sigma}^{M}(\tau) \, (0 \le \tau \le \beta)$.

\subsection{Nonequilibrium DMFT for transient dynamics}
We first use the nonequilibrium DMFT formalism to analyze the transient dynamics of the model \eqref{eq:lattice model}, starting from an initial equilibrium state $\rho(t=0) \propto e^{-\beta H(0)}$ with inverse temperature $\beta$.
In nonequilibrium DMFT, the lattice model \eqref{eq:lattice model} is mapped onto a single-site impurity model defined by the action
\begin{multline}
  \mathcal{S}_{\mathrm{imp}} = -\int_{\mathcal{C}} \dd{z} \dd{z'} \sum_{\sigma} c_{\sigma}^{\dagger}(z) \Delta_{\sigma}(z,z') c_{\sigma}(z') \\
  - \int_{\mathcal{C}} \dd{z} U(z) \qty(n_{\uparrow}(z) - \frac{1}{2}) \qty(n_{\downarrow}(z) - \frac{1}{2}) \\
  + \int_{\mathcal{C}} \dd{z} \delta \mu \sum_{\sigma} n_{\sigma}(z).
  \label{eq:impurity model}
\end{multline}
Here, $\Delta_{\sigma}(z,z')$ is the hybridization function that describes the coupling between the impurity and the bath.
The action can be divided into the noninteracting part $\mathcal{S}_{0}$ and the interacting part $\mathcal{S}_{\mathrm{int}}$ as
\begin{align}
  &\mathcal{S}_{\mathrm{imp}} = \mathcal{S}_{0} + \mathcal{S}_{\mathrm{int}} + \mathrm{const}, \\ 
  &\mathcal{S}_{0} = -\int_{\mathcal{C}} \dd{z} \dd{z'} \sum_{\sigma} c_{\sigma}^{\dagger}(z) \Delta_{\sigma}(z,z') c_{\sigma}(z') \notag \\
  &\qquad + \int_{\mathcal{C}} \dd{z} \qty[ \delta \mu - U(z) \qty(\alpha - \frac{1}{2}) ] \sum_{\sigma} n_{\sigma}(z), \\
  &\mathcal{S}_{\mathrm{int}} = - \int_{\mathcal{C}} \dd{z} U(z) \qty[n_{\uparrow}(z) - \alpha] \qty[n_{\downarrow}(z) - \alpha],
\end{align}
where $\alpha$ is a parameter that can be chosen arbitrarily.
The noninteracting Green's function of the impurity model is defined as
\begin{align}
  \mathcal{G}_{\sigma}(z,z') = -i \frac{\tr [ T_{\mathcal{C}} c_{\sigma}(z) c_{\sigma}^{\dag}(z') e^{i\mathcal{S}_{0}}]}{\tr [ T_{\mathcal{C}} e^{i\mathcal{S}_{0}}]},
\end{align}
and is called the Weiss Green's function.
The Weiss Green's function satisfies the Dyson equation
\begin{multline}
  \qty[i \dv{z} + \delta \mu - U(z) \qty(\alpha - \frac{1}{2})] \mathcal{G}_{\sigma}(z,z') \\
  - \int_{\mathcal{C}} \dd{\bar{z}} \Delta_{\sigma}(z,\bar{z}) \mathcal{G}_{\sigma}(\bar{z},z') = \delta_{\mathcal{C}}(z,z'),
  \label{eq:Weiss Green's function eq}
\end{multline}
where $d/dz$ is the derivative along the contour and $\delta_{\mathcal{C}}(z,z')$ is the delta function on the contour.
Due to this relation, the Weiss Green's function contains the same information as the hybridization function.

In nonequilibrium DMFT, the hybridization function is determined in such a way that the interacting impurity Green's function
\begin{align}
  G_{\sigma}(z,z') = -i \frac{\tr [ T_{\mathcal{C}} c_{\sigma}(z) c_{\sigma}^{\dag}(z') e^{i\mathcal{S}_{\mathrm{imp}}}]}{\tr [ T_{\mathcal{C}} e^{i\mathcal{S}_{\mathrm{imp}}}]}
  \label{eq:interacting Green's function}
\end{align}
coincides with the local Green's function of the lattice model \eqref{eq:lattice model},
\begin{align}
  G_{ii\sigma}(z,z') = -i \frac{\tr [ T_{\mathcal{C}} c_{i\sigma}(z) c_{i\sigma}^{\dag}(z') e^{i \mathcal{S}_{\mathrm{lat}}}]}{\tr [ T_{\mathcal{C}} e^{i \mathcal{S}_{\mathrm{lat}}}]},
\end{align}
where $\mathcal{S}_{\mathrm{lat}}$ is the action of the lattice model \eqref{eq:lattice model},
\begin{align}
  \mathcal{S}_{\mathrm{lat}} = -\int_{\mathcal{C}} \dd{z} H(z).
\end{align}
In the case of the Bethe lattice with infinite coordination number, this solution can be obtained with a simplified self-consistency condition \cite{eckstein2009new},
\begin{align}
  \Delta_{\sigma}(z,z') = v_{\sigma}^{2} G_{\sigma}(z,z').
\end{align}
Thus, the Green's function of the original lattice model can be obtained by the following steps:
(i) Initialize the Weiss Green's function $\mathcal{G}_{\sigma}(z,z')$.
(ii) Solve the impurity model defined by $\mathcal{G}_{\sigma}(z,z')$ to obtain the interacting Green's function $G_{\sigma}(z,z')$.
(iii) Update the Weiss Green's function $\mathcal{G}_{\sigma}(z,z')$ by solving Eq.~\eqref{eq:Weiss Green's function eq} with the hybridization function $\Delta_{\sigma}(z,z') = v_{\sigma}^{2} G_{\sigma}(z,z')$.
(iv) Iterate the steps (ii) and (iii) until convergence is achieved.
After convergence, the local Green's function of the lattice model is given by $G_{ii\sigma}(z,z') = G_{\sigma}(z,z')$.

\subsection{Nonequilibrium DMFT for steady states}
\label{sec:steady-state DMFT}
In the previous section, we formulated DMFT for the transient dynamics starting from a thermal equilibrium state.
While this approach is general, simulating the time evolution until the system relaxes to a steady state is often computationally expensive.
If we are solely interested in the steady state, which is characterized by a given distribution function $f_{\mathrm{st}}(\omega)$, 
we can employ a more efficient scheme that directly targets the stationary limit \cite{kunzel2024numerically}. 

To review the formulation, let us consider the model \eqref{eq:lattice model} with $U(t) = U$.
In the steady state, time-translation invariance holds, and the Green's functions $G_{\sigma}^{R,<}(t,t')$ depend only on the time difference $t - t'$ as $G_{\sigma}^{R,<}(t,t') = G_{\sigma}^{R,<}(t - t')$.
Therefore, we can define the Fourier transforms $G_{\sigma}^{R,<}(\omega)$.
The lesser Green's function $G_{\sigma}^{<}(\omega)$ encodes the distribution of particles $f_{\mathrm{st}}(\omega)$ in the steady state through the fluctuation-dissipation-theorem-type (FDT-type) relation,
\begin{align}
	G_{\sigma}^{<}(\omega) = 2\pi i A_{\sigma}(\omega) f_{\mathrm{st}}(\omega) = -2 i f_{\mathrm{st}}(\omega) \Im G_{\sigma}^{R}(\omega).
	\label{eq:noneq_distribution_function}
\end{align}
In thermal equilibrium, $f_{\mathrm{st}}(\omega)$ becomes the Fermi--Dirac distribution function.
In a nonequilibrium setting, $f_{\mathrm{st}}(\omega)$ can be a nonthermal distribution function, e.g., a double-step function describing independent thermalization in the upper and lower Hubbard bands  \cite{kunzel2024numerically}.

Also, in a steady state, we can assume that the system loses memory of the initial state during the time evolution toward the steady state.
Thus, we can focus only on the real-time branches $\mathcal{C}_{0} \cup \mathcal{C}_{1}$ of the contour, and do not have to consider the imaginary-time branch $\mathcal{C}_{2}$ explicitly.
These properties allow us to directly obtain the steady state Green's functions $G_{\sigma}^{R,<}(\omega)$ which are consistent with a given nonequilibrium distribution function $f_{\mathrm{st}}(\omega)$, i.e., $G_{\sigma}^{R,<}(\omega)$ satisfying Eq.~\eqref{eq:noneq_distribution_function}, without simulating the transient dynamics.

Specifically, we can perform the following DMFT self-consistency loop for the Bethe lattice:
(i) Initialize the Weiss Green's function $\mathcal{G}_{\sigma}^{R,<}(\omega)$.
(ii) Perform the inverse Fourier transform to obtain $\mathcal{G}_{\sigma}^{R,<}(t)$ for $-t_{\mathrm{max}} \le t \le t_{\mathrm{max}}$.
(iii) Solve the impurity model defined by $\mathcal{G}_{\sigma}^{R,<}(t)$ to obtain the interacting Green's function $G_{\sigma}^{R,<}(t)$ for $-t_{\mathrm{max}} \le t \le t_{\mathrm{max}}$.
(iv) Perform the Fourier transform to obtain $G_{\sigma}^{R}(\omega)$.
(v) Update the retarded component of the Weiss Green's function by solving Eq.~\eqref{eq:Weiss Green's function eq} as
\begin{align}
  \mathcal{G}_{\sigma}^{R}(\omega) = \frac{1}{\omega + \delta \mu - U(\alpha - 1/2) - \Delta_{\sigma}^{R}(\omega) + i\delta}
\end{align}
with $\Delta_{\sigma}^{R}(\omega) = v_{\sigma}^{2} G_{\sigma}^{R}(\omega)$.
Then, update the lesser component of the Weiss Green's function as 
\begin{align}
  \mathcal{G}_{\sigma}^{<}(\omega) = -2i f_{\mathrm{st}}(\omega) \Im \mathcal{G}_{\sigma}^{R}(\omega),
  \label{eq:noneq_distribution_function_Weiss}
\end{align}
where $f_{\mathrm{st}}(\omega)$ is the target distribution function, which is kept fixed throughout the DMFT loop.
(vi) Repeat steps (ii)--(v) until convergence is achieved.

A few remarks are in order regarding this procedure.
First, at step (iii), there is a simplification which occurs in the steady state formalism.
For the steady-state case, to obtain $G_{\sigma}^{R,<}(t)$, it suffices to compute $G_{\sigma}^{R,<}(t_{\mathrm{max}}, t')$ for $0 \le t' \le t_{\mathrm{max}}$. 
From this information, we can reconstruct $G_{\sigma}^{R,<}(t)$ as 
\begin{align}
  G_{\sigma}^{R}(t) &=
  \begin{cases}
    0  & \text{if } -t_{\mathrm{max}} \le t < 0 \\
    G_{\sigma}^{R}(t_{\mathrm{max}}, t_{\mathrm{max}} - t) & \text{if } 0 \le t \le t_{\mathrm{max}},
  \end{cases} \\
  G_{\sigma}^{<}(t) &= 
  \begin{cases}
    -G_{\sigma}^{<}(t_{\mathrm{max}}, t_{\mathrm{max}}+t)^{*} & \text{if } -t_{\mathrm{max}} \le t < 0 \\
    G_{\sigma}^{<}(t_{\mathrm{max}}, t_{\mathrm{max}}-t) & \text{if } 0 \le t \le t_{\mathrm{max}},
  \end{cases}
\end{align}
where we use the time-translation invariance and the Hermiticity \eqref{eq:hermiticity relations}.
Since we fix the first argument of the Green's function to $t_{\mathrm{max}}$, the computational cost is smaller than that for the transient dynamics,
where we need to compute the Green's function for all $0 \le t, t' \le t_{\mathrm{max}}$.

Second, we have to be careful about the choice of the cutoff time $t_{\mathrm{max}}$, which is kept fixed throughout the DMFT self-consistency loop.
In step (iv), one needs to obtain $G_{\sigma}^{R}(\omega)$ by Fourier transforming $G_{\sigma}^{R}(t)$ on the finite time window $-t_{\mathrm{max}} \le t \le t_{\mathrm{max}}$.
If $\abs*{G_{\sigma}^{R}(t_{\mathrm{max}})}$ is not sufficiently small, this time-domain truncation induces spurious oscillations in $G_{\sigma}^{R}(\omega)$ and hence in the spectral function.
Therefore, after the convergence of the DMFT loop, one should check whether $\abs*{G_{\sigma}^{R}(t_{\mathrm{max}})}$ is small enough. Otherwise, one needs to increase $t_{\mathrm{max}}$ and repeat the DMFT calculation.

Third, at step (v), instead of imposing the nonequilibrium distribution function on the Green's function $G_{\sigma}^{<}(\omega)$ as in Eq.~\eqref{eq:noneq_distribution_function}, we impose it on the Weiss Green's function $\mathcal{G}_{\sigma}^{<}(\omega)$ as in Eq.~\eqref{eq:noneq_distribution_function_Weiss}.
Since it is not trivial whether the converged Green's function $G_{\sigma}^{<}(\omega)$ satisfies Eq.~\eqref{eq:noneq_distribution_function} or not, we need to check this condition after the convergence.

\section{Tensor-train methods for nonequilibrium DMFT} 
\label{sec:tensor-train methods}

In this section, we describe our tensor-train impurity solver for computing the nonequilibrium Green's functions of the impurity model. We first review the weak-coupling expansion of the Green's function, and then explain how the tensor cross interpolation (TCI) algorithm is used to efficiently evaluate the high-order contributions.

\subsection{Weak-coupling expansion of the impurity model}
\label{sec:weak-coupling expansion}
The most challenging part in a nonequilibrium DMFT calculation is to obtain the solution of the effective impurity model, such as the one defined by the action \eqref{eq:impurity model}.
To calculate the interacting Green's function $G_{\sigma}(z,z')$ for a given Weiss Green's function $\mathcal{G}_{\sigma}(z,z')$, the weak-coupling expansion \cite{hershfield1991probing,hershfield1992resonant,fujii2003perturbative} is one of the powerful methods.
In this method, we treat the interaction part $\mathcal{S}_{\mathrm{int}}$ as a perturbation and expand the Green's function in powers of the interaction strength $U$.
Expanding Eq.~\eqref{eq:interacting Green's function} in terms of $\mathcal{S}_{\mathrm{int}}$ and applying the Wick's theorem,
we obtain the following formulas for the partition function and Green's function:
\begin{align}
  &\frac{\mathcal{Z}}{\mathcal{Z}_{0}} = \sum_{n=0}^{\infty} \frac{i^{n}}{n!} \int_{\mathcal{C}} \dd{z_{1}} \cdots \dd{z_{n}} U(z_{1}) \cdots U(z_{n}) \notag \\
  &\hspace{2.5cm} \times \det \bm{D}_{n\uparrow}(\qty{z_{i}}) \det \bm{D}_{n\downarrow}(\qty{z_{i}}), \label{eq:partition function expansion} \\
  &G_{\sigma}(z,z') = \frac{\mathcal{Z}_{0}}{\mathcal{Z}} \sum_{n=0}^{\infty} \frac{i^{n}}{n!} \int_{\mathcal{C}} \dd{z_{1}} \cdots \dd{z_{n}} U(z_{1}) \cdots U(z_{n}) \notag \\
  &\hspace{2.5cm} \times \det \tilde{\bm{D}}_{n\sigma}(z,z',\qty{z_{i}}) \det \bm{D}_{n\bar{\sigma}}(\qty{z_{i}}), \label{eq:Green's function expansion}
\end{align}
where $\mathcal{Z}_{0}$ and $\mathcal{Z}$ are the partition functions of the noninteracting and interacting impurity models, respectively, 
and the symbol $\bar{\sigma}$ denotes the spin polarization opposite to $\sigma$.
$\bm{D}_{n\sigma}(\qty{z_{i}})$ and $\tilde{\bm{D}}_{n\sigma}(z,z',\qty{z_{i}})$ are $n \times n$ and $(n+1) \times (n+1)$ matrices defined as
\begin{align}
  &\bm{D}_{n\sigma}(\qty{z_{i}}) \notag \\
  &=
  \begin{pmatrix}
    \mathcal{G}_{\sigma}(z_{1},z_{1}^{+}) - i\alpha & \cdots & \mathcal{G}_{\sigma}(z_{1}, z_{n}) \\
    \vdots & \ddots & \vdots \\
    \mathcal{G}_{\sigma}(z_{n}, z_{1}) & \cdots & \mathcal{G}_{\sigma}(z_{n}, z_{n}^{+}) - i\alpha
  \end{pmatrix}, \\
  &\tilde{\bm{D}}_{n\sigma}(z,z',\qty{z_{i}}) \notag \\
  &=
  \begin{pmatrix}
    \mathcal{G}_{\sigma}(z,z') & \mathcal{G}_{\sigma}(z, z_{1}) & \cdots & \mathcal{G}_{\sigma}(z, z_{n}) \\
    \mathcal{G}_{\sigma}(z_{1}, z') & \mathcal{G}_{\sigma}(z_{1}, z_{1}^{+}) - i\alpha & \cdots & \mathcal{G}_{\sigma}(z_{1}, z_{n}) \\
    \vdots & \vdots & \ddots & \vdots \\
    \mathcal{G}_{\sigma}(z_{n}, z') & \mathcal{G}_{\sigma}(z_{n}, z_{1}) & \cdots & \mathcal{G}_{\sigma}(z_{n}, z_{n}^{+}) - i\alpha
  \end{pmatrix},
\end{align}
respectively, where $z_{i}^{+}$ is the time infinitesimally later than $z_{i}$ on the contour.
For practical calculations, we need to truncate the infinite series \eqref{eq:partition function expansion} and \eqref{eq:Green's function expansion} at a certain order $n = n_{\mathrm{max}}$ and approximate them as
\begin{align}
  &\frac{\mathcal{Z}}{\mathcal{Z}_{0}} \simeq \sum_{n=0}^{n_{\mathrm{max}}} \frac{i^{n}}{n!} \int_{\mathcal{C}} \dd{z_{1}} \cdots \dd{z_{n}} U(z_{1}) \cdots U(z_{n}) \notag \\
  &\hspace{2.5cm} \times \det \bm{D}_{n\uparrow}(\qty{z_{i}}) \det \bm{D}_{n\downarrow}(\qty{z_{i}}), \label{eq:partition function expansion approximation} \\
  &G_{\sigma}(z,z') \simeq \frac{\mathcal{Z}_{0}}{\mathcal{Z}} \sum_{n=0}^{n_{\mathrm{max}}} \frac{i^{n}}{n!} \int_{\mathcal{C}} \dd{z_{1}} \cdots \dd{z_{n}} U(z_{1}) \cdots U(z_{n}) \notag \\
  &\hspace{2.5cm} \times \det \tilde{\bm{D}}_{n\sigma}(z,z',\qty{z_{i}}) \det \bm{D}_{n\bar{\sigma}}(\qty{z_{i}}). \label{eq:Green's function expansion approximation}
\end{align}

\subsection{Tensor-train impurity solver}
\label{sec:tensor-train impurity solver}
To calculate the interacting Green's function $G_{\sigma}(z,z')$ using Eqs.~\eqref{eq:partition function expansion approximation} and \eqref{eq:Green's function expansion approximation}, we need to evaluate high-dimensional integrals, whose computational cost grows exponentially with the maximum order $n_{\mathrm{max}}$ when a standard quadrature method is used.
To overcome this difficulty, we apply the tensor-train method based on the tensor cross interpolation algorithm \cite{nunez2022learning,nunez2025learning}.
Our solver can be regarded as an extension of the TCI approach in Ref.~\cite{nunez2022learning} and enables us to calculate the two-time nonequilibrium Green's functions required in nonequilibrium DMFT.
For simplicity, we assume in the following that the time dependence of the interaction is given by
\begin{align}
  U(t) = 
  \begin{cases}
    0 & \text{if } t = 0 \\
    U & \text{if } t > 0.
  \end{cases}
\end{align}
Since the initial state is noninteracting in this case, the Matsubara Green's function $G_{\sigma}^{M}(\tau)$ is equal to the noninteracting one $\mathcal{G}_{\sigma}^{M}(\tau)$.
Thus, we only need to calculate $G_{\sigma}^{<}(t,t') \, (0 \le t' \le t \le t_{\mathrm{max}})$, $G_{\sigma}^{>}(t,t') \, (0 \le t' \le t \le t_{\mathrm{max}})$ and $G_{\sigma}^{\leftmixing}(t,\tau) \, (0 \le t \le t_{\mathrm{max}}, 0 \le \tau \le \beta)$ to reconstruct the full Green's function by using Eqs.~\eqref{eq:causal relations} and \eqref{eq:hermiticity relations}.
Also, since no interaction vertices are placed on the imaginary-time branch in this setup, the solver for this quench protocol can also be applied to steady-state DMFT calculations, where the imaginary-time branch is not treated explicitly.

\subsubsection{Lesser and greater Green's functions}
\label{sec:lesser and greater Green's functions}
We first discuss how to evaluate the lesser and greater Green's functions for $0 \le t' \le t \le t_{\mathrm{max}}$ using the tensor-train method.
By extracting the lesser component from Eq.~\eqref{eq:Green's function expansion approximation}, we obtain
\begin{align}
  &G_{\sigma}^{<}(t,t') \notag \\
  &\simeq \sum_{n=0}^{n_{\mathrm{max}}} \frac{(iU)^{n}}{n!} \sum_{s_{1},\dots,s_{n}=0}^{1} (-1)^{\sum_{\ell} s_{\ell}} \int_{0}^{t_{\mathrm{max}}} \dd{t_{1}} \cdots \dd{t_{n}} \notag \\
  &\hspace{1cm} \times \det \tilde{\bm{D}}_{n\sigma}^{<}(t,t',\qty{t_{i}}, \qty{s_{i}}) \det \bm{D}_{n\bar{\sigma}}(\qty{t_{i}}, \qty{s_{i}}),
  \label{eq:lesser Green's function expansion approximation}
\end{align}
where the matrices $\bm{D}_{n\bar{\sigma}}(\qty{t_{i}}, \qty{s_{i}})$ and $\tilde{\bm{D}}_{n\sigma}^{<}(t,t',\qty{t_{i}}, \qty{s_{i}})$ are defined as
\begin{align}
  &\bm{D}_{n\sigma}(\qty{t_{i}}, \qty{s_{i}}) \notag \\
  &=
  \begin{pmatrix}
    \mathcal{G}_{\sigma}^{<}(t_{1}, t_{1}) - i\alpha & \cdots & \mathcal{G}_{\sigma}^{s_{1}s_{n}}(t_{1}, t_{n}) \\
    \vdots & \ddots & \vdots \\
    \mathcal{G}_{\sigma}^{s_{n}s_{1}}(t_{n}, t_{1}) & \cdots & \mathcal{G}_{\sigma}^{<}(t_{n}, t_{n}) - i\alpha
  \end{pmatrix}, \label{eq:D matrix} \\
  &\tilde{\bm{D}}_{n\sigma}^{<}(t,t',\qty{t_{i}}, \qty{s_{i}}) \notag \\
  &=
  \begin{pmatrix}
    \mathcal{G}_{\sigma}^{<}(t,t') & \mathcal{G}_{\sigma}^{0s_{1}}(t,t_{1}) & \cdots & \mathcal{G}_{\sigma}^{0s_{n}}(t,t_{n}) \\
    \mathcal{G}_{\sigma}^{s_{1}1}(t_{1},t') & \mathcal{G}_{\sigma}^{<}(t_{1}, t_{1}) - i\alpha & \cdots & \mathcal{G}_{\sigma}^{s_{1}s_{n}}(t_{1}, t_{n}) \\
    \vdots & \vdots & \ddots & \vdots \\
    \mathcal{G}_{\sigma}^{s_{n}1}(t_{n},t') & \mathcal{G}_{\sigma}^{s_{n}s_{1}}(t_{n}, t_{1}) & \cdots & \mathcal{G}_{\sigma}^{<}(t_{n}, t_{n}) - i\alpha
  \end{pmatrix}.
\end{align}
Here, we have used $\mathcal{Z}_{0}/\mathcal{Z} = 1$ since the initial state is noninteracting.
For notational simplicity, we denote the integrand of Eq.~\eqref{eq:lesser Green's function expansion approximation} as
\begin{align}
  &Q_{n\sigma}^{<}(t,t',\qty{t_{i}}) = \sum_{s_{1},\dots,s_{n}=0}^{1} \hspace{-2mm} (-1)^{\sum_{\ell} s_{\ell}} \det \tilde{\bm{D}}_{n\sigma}^{<}(t,t',\qty{t_{i}}, \qty{s_{i}}) \notag \\
  &\hspace{4.5cm} \times \det \bm{D}_{n\bar{\sigma}}(\qty{t_{i}}, \qty{s_{i}}).
  \label{eq:integrand_lesser}
\end{align}
Because of Eq.~\eqref{eq:causal relations}, we can show that the value of the integrand $Q_{n\sigma}^{<}(t,t',\qty{t_{i}})$ is zero when $(t_{1}, \dots, t_{n}) \notin [0,t]^{n}$.
Furthermore, since the integrand $Q_{n\sigma}^{<}(t,t',\qty{t_{i}})$ is invariant under the permutation of the integration variables $t_{1}, \dots, t_{n}$, we can rewrite the integral as
\begin{align}
  G_{\sigma}^{<}(t,t') \simeq \sum_{n=0}^{n_{\mathrm{max}}} (iU)^{n} \int_{S_{n}^{0,t}} \dd{t_{1}} \cdots \dd{t_{n}} Q_{n\sigma}^{<}(t,t',\qty{t_{i}}),
  \label{eq:lesser Green's function expansion approximation simplex}
\end{align}
where $S_{n}^{a,b}$ is an $n$-dimensional region defined as
\begin{align}
  S_{n}^{a,b} = \{ (t_{1}, \dots, t_{n}) \in \mathbb{R}^{n} \mid a \le t_{n} \le \cdots \le t_{1} \le b \}.
\end{align}

Although Eq.~\eqref{eq:lesser Green's function expansion approximation simplex} provides a compact expression for the lesser Green's function, the evaluation of the high-dimensional integral using the tensor-train method remains difficult.
Since the Weiss Green's function $\mathcal{G}_{\sigma}^{00}(t,t')$ and $\mathcal{G}_{\sigma}^{11}(t,t')$ have discontinuities at $t=t'$, the integrand $Q_{n\sigma}^{<}(t,t',\qty{t_{i}})$ also has discontinuities at $t_{i} = t'$ $(1 \le i \le n)$.
Because the TCI algorithm performs poorly for discontinuous functions, we have to remove these discontinuities from the integrand to apply the TCI algorithm successfully.
To this end, we divide the region $S_{n}^{0,t}$ into $n+1$ subregions $S_{k}^{t',t} \times S_{n-k}^{0,t'}$ with $k = 0, 1, \dots, n$, as
\begin{align}
  &G_{\sigma}^{<}(t,t') \simeq \sum_{n=0}^{n_{\mathrm{max}}} (iU)^{n} \sum_{k=0}^{n} 
  \int_{S_{k}^{t',t}} \dd{t_{1}} \cdots \dd{t_{k}} \notag \\
  &\hspace{2.5cm}
  \int_{S_{n-k}^{0,t'}} \dd{t_{k+1}} \cdots \dd{t_{n}}
  Q_{n\sigma}^{<}(t,t',\qty{t_{i}}),
  \label{eq:lesser Green's function expansion final}
\end{align}
which was also used for the equilibrium TCI impurity solver in Refs.~\cite{erpenbeck2023tensor,matsuura2025tensor}.
In the domain labeled by $k$, the variable $t'$ satisfies $0 \le t_{n} \le \cdots \le t_{k+1} \le t' \le t_{k} \le \cdots \le t_{1} \le t$, and thus the integrand $Q_{n\sigma}^{<}(t,t',\qty{t_{i}})$ is smooth in this domain.

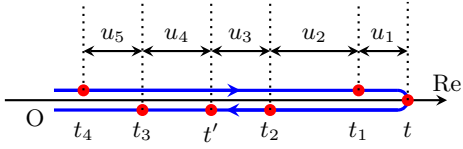
\begin{figure}
  \centering
  \begin{tikzpicture}[>=stealth, thick, scale=1.3]
    \draw[->,thick] (-0.5,0) -- (4.0,0) node[above] {Re};
  
    \draw[very thick, blue] (0,0.1) -- (3.5,0.1);
    \draw[->,very thick, blue] (0,0.1) -- (1.9,0.1);
    \node[below left] at (0,0) {O};
  
    \draw[very thick, blue] (3.5,0.1) arc[start angle=90, end angle=-90, radius=0.1];
  
    \draw[very thick, blue] (3.5,-0.1) -- (0.0,-0.1);
    \draw[->,very thick, blue] (3.5,-0.1) -- (1.75,-0.1);

    \draw[red, fill] (3.6, 0) circle (0.05);
    \draw (3.6, 0) node[below=5pt] {$t$};

    \draw[red, fill] (0.3, 0.1) circle (0.05);
    \draw[red, fill] (0.9, -0.1) circle (0.05);
    \draw[red, fill] (1.6, -0.1) circle (0.05);
    \draw[red, fill] (2.2, -0.1) circle (0.05);
    \draw[red, fill] (3.1, 0.1) circle (0.05);

    \draw (0.3, 0.0) node[below=5pt] {$t_{4}$};
    \draw (0.9, 0.0) node[below=5pt] {$t_{3}$};
    \draw (1.6, 0.0) node[below=5pt] {$t'$};
    \draw (2.2, 0.0) node[below=5pt] {$t_{2}$};
    \draw (3.1, 0.0) node[below=5pt] {$t_{1}$};

    \draw[dotted] (3.6, 0) -- (3.6, 1.0);
    \draw[dotted] (0.3, 0.1) -- (0.3, 1.0);
    \draw[dotted] (0.9, -0.1) -- (0.9, 1.0);
    \draw[dotted] (1.6, -0.1) -- (1.6, 1.0);
    \draw[dotted] (2.2, -0.1) -- (2.2, 1.0);
    \draw[dotted] (3.1, 0.1) -- (3.1, 1.0);

    \draw[<->] (0.3, 0.5) --node[midway,above]{$u_{5}$} (0.9, 0.5);
    \draw[<->] (0.9, 0.5) --node[midway,above]{$u_{4}$} (1.6, 0.5);
    \draw[<->] (1.6, 0.5) --node[midway,above]{$u_{3}$} (2.2, 0.5);
    \draw[<->] (2.2, 0.5) --node[midway,above]{$u_{2}$} (3.1, 0.5);
    \draw[<->] (3.1, 0.5) --node[midway,above]{$u_{1}$} (3.6, 0.5);

  \end{tikzpicture}
  \caption{The definition of the variables $u_{1}, \dots, u_{n+1}$. In this illustration, we take $n=4$ and $k=2$ with $(s_{1}, s_{2}, s_{3}, s_{4}) = (0,1,1,0)$.}
  \label{fig:u_variables}
\end{figure}

To evaluate the integral in each domain, we use an extension of the method proposed in Ref.~\cite{nunez2022learning}.
We first introduce new variables $u_{1}, u_{2}, \dots, u_{n+1}$ as
\begin{equation}\label{eq:u_def}
  \begin{gathered}
    u_{1} = t - t_{1}, \\
    u_{2} = t_{1} - t_{2}, \quad \dots, \quad u_{k} = t_{k-1} - t_{k}, \\
    u_{k+1} = t_{k} - t', \quad u_{k+2} = t' - t_{k+1}, \\
    u_{k+3} = t_{k+1} - t_{k+2}, \quad \dots, \quad u_{n+1} = t_{n-1} - t_{n},
  \end{gathered}
\end{equation}
and regard the integrand $Q_{n\sigma}^{<}(t,t',\qty{t_{i}})$ as a function of $t, u_{1}, u_{2}, \dots, u_{n+1}$, i.e.,
\begin{align}
  \mathcal{Q}_{nk\sigma}^{<}(t, \qty{u_{i}}) = Q_{n\sigma}^{<}(t,t',\qty{t_{i}}).
\end{align}
The definition of these new variables is illustrated in Fig.~\ref{fig:u_variables}.

Next, we discretize the variables $t, u_{1}, u_{2}, \dots, u_{n+1}$ over the range $0 \le t \le t_{\mathrm{max}}$, $0 \le u_{i} \le t_{\mathrm{max}}$ $(1 \le i \le n+1)$ with a suitable mesh,
and perform a tensor-train decomposition by the TCI algorithm for $\mathcal{Q}_{nk\sigma}^{<}(t,\qty{u_{i}})$ to obtain its tensor-train representation as
\begin{align}
  \mathcal{Q}_{nk\sigma}^{<}(t, \qty{u_{i}}) 
  &\simeq M_{0}(t) M_{1}(u_{1}) \cdots M_{n+1}(u_{n+1}) \notag \\
  &= M_{0}(t) M_{1}(t-t_{1}) \cdots M_{n+1}(t_{n-1} - t_{n}).
\end{align}
Here, the $M_{i}(u_{i})$ are matrix-valued functions, whose matrix sizes define the bond dimensions of the tensor-train representation.
These bond dimensions are determined adaptively by the TCI algorithm according to the target accuracy, and become larger when a smaller approximation error is required.
In the actual calculations, the tensor-train approximation is constructed with \texttt{TensorCrossInterpolation.jl}~\cite{nunez2025learning}, using the TCI2 algorithm based on the block-rook update.
For each TCI run, we typically perform three optimization iterations, each consisting of a two-site sweep over the tensor-train bonds followed by a global-pivot search.
For the details of the TCI algorithm, we refer the readers to Refs.~\cite{nunez2022learning,nunez2025learning,matsuura2025tensor}.

After the tensor-train decomposition, the integral over the region $S_{k}^{t',t} \times S_{n-k}^{0,t'}$ can be expressed as a recursive convolution of the matrices $M_i$, 
\begin{align}
  &\int_{S_{k}^{t',t}} \dd{t_{1}} \cdots \dd{t_{k}} \int_{S_{n-k}^{0,t'}} \dd{t_{k+1}} \cdots \dd{t_{n}} Q_{n\sigma}^{<}(t,t',\qty{t_{i}}) \notag \\
  &\hspace{1cm} = M_{0}(t) (M_{1} * \cdots * M_{k+1})(t-t') \notag \\
  &\hspace{3cm} (M_{k+2} * \cdots * M_{n+1} * I)(t'),
  \label{eq:recursive convolution}
\end{align}
where the symbol $*$ denotes the convolution defined as
\begin{align}
  (A * B)(t) = \int_{0}^{t} \dd{\bar{t}} A(t-\bar{t}) B(\bar{t}),
\end{align}
and $I(t)$ is the identity function defined as $I(t) = 1$.
Since the cost of performing each convolution is small, this procedure allows us to efficiently evaluate the original high-dimensional integral.
The details of the discretization and the method to efficiently evaluate the recursive convolution are explained in Appendix~\ref{sec:recursive convolution}.
For a fixed Weiss Green's function, the TCI calculations for different pairs $(n,k)$ in Eq.~\eqref{eq:lesser Green's function expansion final} are independent.
In practice, we therefore perform separate TCI runs for each pair $(n,k)$ and parallelize these runs over message passing interface (MPI) processes.

Since we assume that $u_{i}$ takes values in the range $0 \le u_{i} \le t_{\mathrm{max}}$ for some sets of $(t, u_{1}, \dots, u_{n+1})$, the 
corresponding $t_{i}$ can lie outside of $[0, t_{\mathrm{max}}]$.
Since the Weiss Green's function $\mathcal{G}_{\sigma}^{s_{i}s_{j}}(t_{i}, t_{j})$ is defined only for $(t_{i},t_{j}) \in [0,t_{\mathrm{max}}]^{2}$, 
$\mathcal{Q}_{nk\sigma}^{<}(t, \qty{u_{i}})$ is not defined for such sets of $(t, u_{1}, \dots, u_{n+1})$.
To address this problem, we extrapolate the Weiss Green's function from $[0,t_{\mathrm{max}}]^{2}$ to $(-\infty, t_{\mathrm{max}}]^{2}$ using the method explained in Appendix~\ref{sec:extrapolation}.
The function $\mathcal{Q}_{nk\sigma}^{<}(t, \qty{u_{i}})$ is then defined for all sets of $(t, u_{1}, \dots, u_{n+1})$ with $0 \le t \le t_{\mathrm{max}}$ and $0 \le u_{i} \le t_{\mathrm{max}}$, and therefore, we can directly apply the TCI algorithm.

During the construction of the tensor-train decomposition, the TCI algorithm repeatedly samples the function $\mathcal{Q}_{nk\sigma}^{<}(t, \qty{u_{i}})$.
As is evident from Eq.~\eqref{eq:integrand_lesser}, $\mathcal{Q}_{nk\sigma}^{<}(t,\qty{u_{i}})$ includes the summation over $2^{n}$ configurations of the Keldysh indices $\qty{s_{i}}$ and the computation of two determinants, 
and thus, the computational cost of a single evaluation scales as $\mathcal{O}(2^{n} n^{3})$.
Although we can reduce this cost to $\order{2^{n}}$ by using the fast summation technique introduced in Ref.~\cite{nunez2022learning}, the exponential scaling with respect to $n$ remains.
Typically, a single evaluation of $\mathcal{Q}_{nk\sigma}^{<}(t, \qty{u_{i}})$ takes around $100$\,{\textmu}s for $n = 10$ and $100$\,ms for $n=20$.
Due to this scaling, the computational cost increases rapidly as we increase the maximum order $n_{\mathrm{max}}$.
However, even with this limitation, we can reach a maximum order $n_{\mathrm{max}} \simeq 20$ in practice, which enables us to access timescales comparable to, or in some cases even longer than those reachable with the CT-QMC solver.

One possible way to mitigate this problem is to use the tensor-train method also for the summation over $\qty{s_{i}}$, i.e., to add legs corresponding to $\qty{s_{i}}$ to the tensor and perform the tensor-train decomposition for it.
However, we found that the convergence with respect to the bond dimension is quite slow for this enlarged tensor, so that this approach is not practical. We leave the development of more efficient methods to handle the summation over the Keldysh indices $\qty{s_{i}}$ as future work, and use the method with explicit summation over $\qty{s_{i}}$ in the following sections.

So far, we have discussed how to evaluate the lesser Green's function $G_{\sigma}^{<}(t,t')$ for $0 \le t' \le t \le t_{\mathrm{max}}$.
The greater Green's function $G_{\sigma}^{>}(t,t')$ for $0 \le t' \le t \le t_{\mathrm{max}}$ can be evaluated in the same way as the lesser component.
To be precise, we can show that the greater Green's function $G_{\sigma}^{>}(t,t')$ can be expressed as
\begin{align}
  &G_{\sigma}^{>}(t,t') \simeq \sum_{n=0}^{n_{\mathrm{max}}} (iU)^{n} \sum_{k=0}^{n} 
  \int_{S_{k}^{t',t}} \dd{t_{1}} \cdots \dd{t_{k}} \notag \\
  &\hspace{2.5cm}
  \int_{S_{n-k}^{0,t'}} \dd{t_{k+1}} \cdots \dd{t_{n}}
  Q_{n\sigma}^{>}(t,t',\qty{t_{i}}).
  \label{eq:greater Green's function expansion approximation simplex}
\end{align}
Here, the integrand $Q_{n\sigma}^{>}(t,t',\qty{t_{i}})$ is defined as
\begin{align}
  &Q_{n\sigma}^{>}(t,t',\qty{t_{i}}) = \sum_{s_{1},\cdots,s_{n}=0}^{1} \hspace{-2mm} (-1)^{\sum_{\ell} s_{\ell}} \det \tilde{\bm{D}}_{n\sigma}^{>}(t,t',\qty{t_{i}}, \qty{s_{i}}) \notag \\
  &\hspace{4.5cm} \times \det \bm{D}_{n\bar{\sigma}}(\qty{t_{i}}, \qty{s_{i}}),
\end{align}
where the matrix $\tilde{\bm{D}}_{n\sigma}^{>}(t,t',\qty{t_{i}}, \qty{s_{i}})$ is given by
\begin{align}
  &\tilde{\bm{D}}_{n\sigma}^{>}(t,t',\qty{t_{i}}, \qty{s_{i}}) \notag \\
  &=
  \begin{pmatrix}
    \mathcal{G}_{\sigma}^{>}(t,t') & \mathcal{G}_{\sigma}^{1s_{1}}(t,t_{1}) & \cdots & \mathcal{G}_{\sigma}^{1s_{n}}(t,t_{n}) \\
    \mathcal{G}_{\sigma}^{s_{1}0}(t_{1},t') & \mathcal{G}_{\sigma}^{<}(t_{1}, t_{1}) - i\alpha & \cdots & \mathcal{G}_{\sigma}^{s_{1}s_{n}}(t_{1}, t_{n}) \\
    \vdots & \vdots & \ddots & \vdots \\
    \mathcal{G}_{\sigma}^{s_{n}0}(t_{n},t') & \mathcal{G}_{\sigma}^{s_{n}s_{1}}(t_{n}, t_{1}) & \cdots & \mathcal{G}_{\sigma}^{<}(t_{n}, t_{n}) - i\alpha
  \end{pmatrix}.
\end{align}
The high-dimensional integral in Eq.~\eqref{eq:greater Green's function expansion approximation simplex} can be evaluated by exactly the same method as explained for the lesser component.

In general, we have to compute the lesser and greater Green's functions by independent TCI runs.
However, in the particle-hole symmetric case (i.e., $\delta \mu = 0$), the lesser and greater Green's functions are related through
\begin{align}
  G_{\sigma}^{>}(t,t') = -G_{\sigma}^{<}(t',t),
\end{align}
and thus calculating the lesser component is sufficient.

\subsubsection{Left-mixing Green's function}
In the case of an interaction quench from the noninteracting state, 
the evaluation of the left-mixing Green's function does not require high-dimensional integrals \cite{eckstein2010interaction}.
To see this, we start from the right Dyson equation for the model \eqref{eq:impurity model},
\begin{align}
  &\qty(-i \dv{z'} + \delta \mu) G_{\sigma}(z,z') 
  - \int_{\mathcal{C}} \dd{\bar{z}} G_{\sigma}(z,\bar{z}) \Delta_{\sigma}(\bar{z}, z') \notag \\
  &\hspace{2cm} -\int_{\mathcal{C}} \dd{\bar{z}} G_{\sigma}(z,\bar{z}) \Sigma_{\sigma}(\bar{z}, z') = \delta_{\mathcal{C}}(z,z'),
\end{align}
where $\Sigma_{\sigma}(z,z')$ is the self-energy of the impurity model.
If we extract the left-mixing component from this equation using the Langreth rule \cite{langreth1976}, we obtain
\begin{align}
  &\qty(\dv{\tau'} + \delta \mu) G_{\sigma}^{\leftmixing}(t,\tau')
  - \int_{0}^{\beta} \dd{\bar{\tau}} G_{\sigma}^{\leftmixing}(t,\bar{\tau}) \Delta_{\sigma}^{M}(\bar{\tau}, \tau') \notag \\
  &\hspace{3.3cm} -\int_{0}^{\beta} \dd{\bar{\tau}} G_{\sigma}^{\leftmixing}(t,\bar{\tau}) \Sigma_{\sigma}^{M}(\bar{\tau}, \tau') \notag \\
  &= \int_{0}^{t} \dd{\bar{t}} G_{\sigma}^{R}(t,\bar{t}) \Delta_{\sigma}^{\leftmixing}(\bar{t}, \tau') + \int_{0}^{t} \dd{\bar{t}} G_{\sigma}^{R}(t,\bar{t}) \Sigma_{\sigma}^{\leftmixing}(\bar{t}, \tau').
\end{align}
Since the initial state is noninteracting, we have $\Sigma_{\sigma}^{\leftmixing}(t, \tau') = \Sigma_{\sigma}^{M}(\tau) = 0$,
and this condition simplifies the above equation to
\begin{multline}
  \qty(\dv{\tau'} + \delta \mu) G_{\sigma}^{\leftmixing}(t, \tau')
  - \int_{0}^{\beta} \dd{\bar{\tau}} G_{\sigma}^{\leftmixing}(t, \bar{\tau}) \Delta_{\sigma}^{M}(\bar{\tau}, \tau') \\
  = \int_{0}^{t} \dd{\bar{t}} G_{\sigma}^{R}(t,\bar{t}) \Delta_{\sigma}^{\leftmixing}(\bar{t}, \tau').
\end{multline}
If we perform the Fourier transform with respect to $\tau'$, we obtain
\begin{align}
  &(i\omega_{n} + \delta \mu) G_{\sigma}^{\leftmixing}(t, i\omega_{n}) - G_{\sigma}^{\leftmixing}(t, \beta) - G_{\sigma}^{\leftmixing}(t, 0) \notag \\
  &\hspace{4cm} - \Delta_{\sigma}^{M}(i\omega_{n}) G_{\sigma}^{\leftmixing}(t, i\omega_{n}) \notag \\
  &\hspace{3.3cm} = \int_{0}^{t} \dd{\bar{t}} G_{\sigma}^{R}(t,\bar{t}) \Delta_{\sigma}^{\leftmixing}(\bar{t}, i\omega_{n}).
\end{align}
Here, $\omega_{n} = (2n+1)\pi/\beta$ is the fermionic Matsubara frequency, 
and $\Delta_{\sigma}^{M}(i\omega_{n})$, $\Delta_{\sigma}^{\leftmixing}(t, i\omega_{n})$, and $G_{\sigma}^{\leftmixing}(t, i\omega_{n})$ are defined as
\begin{subequations}
  \begin{align}
    &\Delta_{\sigma}^{M}(i\omega_{n}) = \int_{0}^{\beta} \dd{\tau} \Delta_{\sigma}^{M}(\tau) e^{+i \omega_{n} \tau}, \\
    &\Delta_{\sigma}^{\leftmixing}(t, i\omega_{n}) = \int_{0}^{\beta} \dd{\tau'} \Delta_{\sigma}^{\leftmixing}(t, \tau') e^{-i \omega_{n} \tau'}, \\
    &G_{\sigma}^{\leftmixing}(t, i\omega_{n}) = \int_{0}^{\beta} \dd{\tau'} G_{\sigma}^{\leftmixing}(t, \tau') e^{-i \omega_{n} \tau'}.
  \end{align}
\end{subequations}
Using the relations $G_{\sigma}^{\leftmixing}(t, \beta) + G_{\sigma}^{\leftmixing}(t,0) = -G_{\sigma}^{R}(t,0)$ and $\mathcal{G}_{\sigma}^{M}(i\omega_{n}) = [i\omega_{n} + \delta\mu - \Delta_{\sigma}^{M}(i\omega_{n})]^{-1}$, 
we finally obtain
\begin{align}
  &G_{\sigma}^{\leftmixing}(t,i\omega_{n}) \notag \\
  &\quad = \mathcal{G}_{\sigma}^{M}(i\omega_{n}) \qty[-G_{\sigma}^{R}(t,0) + \int_{0}^{t} \dd{\bar{t}} G_{\sigma}^{R}(t,\bar{t}) \Delta_{\sigma}^{\leftmixing}(\bar{t}, i\omega_{n})].
\end{align}
Performing the inverse Fourier transform, we arrive at
\begin{align}
  G_{\sigma}^{\leftmixing}(t, \tau')
  &= -G_{\sigma}^{R}(t,0) \mathcal{G}_{\sigma}^{M}(-\tau') \notag \\
  &\hspace{0.8cm} + \int_{0}^{t} \dd{\bar{t}} \int_{0}^{\beta} \dd{\bar{\tau}} G_{\sigma}^{R}(t,\bar{t}) \Delta_{\sigma}^{\leftmixing}(\bar{t}, \bar{\tau}) \mathcal{G}_{\sigma}^{M}(\bar{\tau} - \tau') \notag \\
  &= G_{\sigma}^{R}(t,0) \mathcal{G}_{\sigma}^{M}(\beta - \tau') \notag \\
  &\hspace{0.8cm} + \int_{0}^{t} \dd{\bar{t}} \int_{0}^{\beta} \dd{\bar{\tau}} G_{\sigma}^{R}(t,\bar{t}) \Delta_{\sigma}^{\leftmixing}(\bar{t}, \bar{\tau}) \mathcal{G}_{\sigma}^{M}(\bar{\tau} - \tau').
  \label{eq:left-mixing Green's function}
\end{align}
Since the retarded Green's function can be calculated from the lesser and greater Green's functions, 
\begin{align}
  G_{\sigma}^{R}(t,t') = \theta(t-t') \qty[G_{\sigma}^{>}(t,t') - G_{\sigma}^{<}(t,t')],
\end{align}
the left-mixing Green's function can be evaluated by performing the two-dimensional integral in Eq.~\eqref{eq:left-mixing Green's function} using a standard quadrature method, rather than using the TCI algorithm.

\subsection{Rough estimate of the maximum order}
In practical calculations, we have to choose the maximum order $n_{\mathrm{max}}$ in such a way that the truncation error becomes sufficiently small.
In this section, we roughly estimate the maximum order $n_{\mathrm{max}}$ required to obtain the Green's function with a given accuracy, and discuss how $n_{\mathrm{max}}$ depends on the parameters of the model.

To this end, we consider the atomic limit of the impurity model, where the action \eqref{eq:impurity model} reduces to
\begin{multline}
  \mathcal{S}_{\mathrm{imp}} = -\int_{\mathcal{C}} \dd{z} U(z) \qty(n_{\uparrow}(z) - \frac{1}{2}) \qty(n_{\downarrow}(z) - \frac{1}{2}) \\
  + \int_{\mathcal{C}} \dd{z} \delta \mu \sum_{\sigma} n_{\sigma}(z).
\end{multline}
In this limit, we can analytically calculate the Green's function as
\begin{subequations}
  \begin{align}
    &G_{\sigma}^{<}(t,t') = i f e^{i\delta\mu (t-t')} \qty[(1-f) e^{i \frac{U}{2} (t-t')} + f e^{-i \frac{U}{2} (t-t')}], \\
    &G_{\sigma}^{>}(t,t') = -i (1-f) e^{i\delta\mu (t-t')} \notag \\
    &\hspace{2.5cm} \times \qty[(1-f) e^{i \frac{U}{2} (t-t')} + f e^{-i \frac{U}{2} (t-t')}],
  \end{align}
\end{subequations}
where $f = 1/(e^{-\beta \delta \mu} + 1)$.
Expanding these Green's functions with respect to $U$, we obtain
\begin{subequations}
  \begin{align}
    &G_{\sigma}^{<}(t,t') = i f e^{i\delta\mu (t-t')} \notag \\
    &\hspace{2.0cm} \times \sum_{n=0}^{\infty} \frac{(iU)^{n}}{n!} \qty(\frac{t-t'}{2})^{n} \qty[(1-f) + (-1)^{n} f], \\
    &G_{\sigma}^{>}(t,t') = -i (1-f) e^{i\delta\mu (t-t')} \notag \\
    &\hspace{2.0cm} \times \sum_{n=0}^{\infty} \frac{(iU)^{n}}{n!} \qty(\frac{t-t'}{2})^{n} \qty[(1-f) + (-1)^{n} f].
  \end{align}
\end{subequations}
Since the absolute values of the lesser and greater Green's functions are of order one, the time difference satisfies $|t-t'| \le t_{\max}$, and $|(1-f)+(-1)^n f| \le 1$, we have to include terms up to the order $n_{\max}$ satisfying
\begin{align}
    \abs{\frac{(iU)^{n_{\mathrm{max}}}}{n_{\mathrm{max}}!} \qty(\frac{t_{\mathrm{max}}}{2})^{n_{\mathrm{max}}}} \ll 1.
\end{align}
To estimate the required order $n_{\mathrm{max}}$, we first consider the condition
\begin{align}
    \abs{\frac{(iU)^{n}}{n!} \qty(\frac{t_{\mathrm{max}}}{2})^{n}} = 1.
    \label{eq:estimate of n order one}
\end{align}
Applying Stirling's formula to Eq.~\eqref{eq:estimate of n order one} gives $n \simeq e|U|t_{\max}/2$.
Since the left-hand side of Eq.~\eqref{eq:estimate of n order one} decreases rapidly with increasing $n$ beyond this order, the required cutoff order in the atomic limit can be roughly estimated as
\begin{align}
    n_{\mathrm{max}} \sim \frac{e}{2}\abs{U}t_{\mathrm{max}}.
    \label{eq:estimate of n_max}
\end{align}
For the parameter regime where the hopping cannot be neglected, the required maximum order would be smaller than this estimate.
This is because the Weiss Green's function typically decays with respect to the time difference $\abs{t-t'}$, and its value becomes smaller than the one in the atomic limit.

Although this is just a rough estimate, it provides a useful guideline to choose the maximum order $n_{\mathrm{max}}$ in the practical calculation as we see below.
A similar dependence of the perturbation order on the interaction strength and $t_\text{max}$ has also been observed in real-time weak-coupling CT-QMC studies \cite{werner2009diagrammatic}.

Based on the above scaling of $n_{\mathrm{max}}$, the computational cost of the TCI solver can be estimated as follows.
Since the recursive convolution described in Appendix~\ref{sec:recursive convolution} is much cheaper than constructing the tensor-train representation of $\mathcal{Q}_{nk\sigma}^{<}(t, \qty{u_i})$, the overall computational cost of the TCI solver is dominated by the latter step.
To construct the tensor-train representation of $\mathcal{Q}_{nk\sigma}^{<}(t, \qty{u_{i}})$, we need to evaluate it $\order{n \chi^2 d}$ times, where $\chi$ is the bond dimension and $d$ is the number of discretization points for each variable.
Since the computational cost of a single evaluation of $\mathcal{Q}_{nk\sigma}^{<}(t, \qty{u_{i}})$ scales as $\order{2^{n}}$, the total computational cost to obtain a tensor-train representation is estimated as $\order{2^{n} n \chi^2 d}$.
To calculate the $n$th order contribution, we need to construct the tensor-train representation for $\mathcal{Q}_{nk\sigma}^{<}(t, \qty{u_{i}})$ for all $k=0,1,\dots,n$.
Therefore, the total computational cost to obtain the Green's function up to the order $n_{\mathrm{max}}$ is estimated as $\order{2^{n_{\mathrm{max}}} n_{\mathrm{max}}^2 \chi^2 d}$.
Using the estimate of $n_{\mathrm{max}}$ in Eq.~\eqref{eq:estimate of n_max}, we finally estimate the computational cost as
$\order{2^{\abs{U} t_{\mathrm{max}}} (Ut_{\mathrm{max}})^2 \chi^2 d}$.

\section{Benchmark against an exactly solvable impurity model} 
\label{sec:exactly solvable impurity model}

In this section, we benchmark the nonequilibrium TCI impurity solver against an exactly solvable impurity model.
After introducing the model and its exact solution, we examine the convergence of the TCI results with respect to the maximum expansion order and the bond dimension, both with and without particle-hole-symmetry.
We also quantify the phase cancellation in the TCI integration and discuss the origin of the low-rank structure of the integrand.

\subsection{Model and exact solutions}
To benchmark the nonequilibrium TCI impurity solver, we first consider an interaction quench of an impurity model for which an accurate reference solution is available.
Specifically, we study the impurity model \eqref{eq:impurity model} with $\Delta_{\downarrow}(z,z')=0$, which appears when we solve the Falicov--Kimball model within nonequilibrium DMFT \cite{freericks2003exact}.
With this setup, the down-spin occupation is conserved, and the up-spin Green's function reduces to a statistical mixture of two conditional propagators,
\begin{align}
  G_{\uparrow}(z,z') = (1 - \ev*{n_{\downarrow}}) G_{0}(z,z') + \ev*{n_{\downarrow}} G_{1}(z,z').
\end{align}
Here, $\ev{n_{\downarrow}}$ is the down-spin electron density in the initial state, which is given by
\begin{align}
  \ev*{n_{\downarrow}} = \frac{1}{e^{-\beta \delta \mu} + 1}
\end{align}
for the quench from the noninteracting state we are considering.
The propagators $G_{0}(z,z')$ and $G_{1}(z,z')$ are the solutions of the following equations:
\begin{subequations}
  \begin{align}
    &\qty(i \dv{z} + \delta\mu + \frac{U(z)}{2}) G_{0}(z,z') \notag \\
    &\hspace{2.0cm} - \int_{\mathcal{C}} \dd{\bar{z}} \Delta_{\uparrow}(z,\bar{z}) G_{0}(\bar{z},z') = \delta_{\mathcal{C}}(z,z'), \\
    &\qty(i \dv{z} + \delta\mu - \frac{U(z)}{2}) G_{1}(z,z') \notag \\
    &\hspace{2.0cm} - \int_{\mathcal{C}} \dd{\bar{z}} \Delta_{\uparrow}(z,\bar{z}) G_{1}(\bar{z},z') = \delta_{\mathcal{C}}(z,z').
  \end{align}
  \label{eq:impurity_model_G0_G1}
\end{subequations}
Equations~\eqref{eq:impurity_model_G0_G1} are Volterra integro-differential equations of the second kind \cite{Brunner1986},
for which efficient numerical algorithms are known \cite{Linz1985,Brunner1986,Press1992,Bonitz2000,stan2009time}.
Therefore, we can use the numerical solution of Eqs.~\eqref{eq:impurity_model_G0_G1} as a reference result and benchmark the output of the nonequilibrium TCI solver against it.

For all benchmarks in this section, we set the hybridization function to
\begin{subequations}
  \begin{align}
    \Delta_{\uparrow}^{R}(t,t') &= -i \theta(t-t') \int_{-2v}^{2v} \dd{\omega} \frac{\sqrt{4v^{2} - \omega^{2}}}{2\pi} e^{-i (\omega-\delta\mu) (t-t')}, \\
		\Delta_{\uparrow}^{\leftmixing}(t, \tau') &= i \int_{-2v}^{2v} \dd{\omega} \frac{\sqrt{4v^{2} - \omega^{2}}}{2\pi} \frac{e^{(\omega-\delta\mu) \tau'}}{e^{\beta (\omega-\delta\mu)} + 1} e^{-i(\omega-\delta\mu) t}, \\
		\Delta_{\uparrow}^{<}(t,t') &= i \int_{-2v}^{2v} \dd{\omega} \frac{\sqrt{4v^{2} - \omega^{2}}}{2\pi} \frac{1}{e^{\beta (\omega-\delta\mu)} + 1} e^{-i (\omega-\delta\mu) (t-t')}.
  \end{align}
  \label{eq:hybridization_function_bethe}
\end{subequations}
In Bethe lattice calculations with a semicircular density of states, these functions are often used as an initial guess for the nonequilibrium DMFT loop.
In this section, unless stated otherwise, we initialize the system in the noninteracting thermal equilibrium state with inverse temperature $\beta = 5/v$, and perform the interaction quench from $U=0$ to $U=3v$ at $t=0$.

\subsection{Particle-hole-symmetric case}
We begin with the particle-hole-symmetric case (i.e., $\delta \mu = 0$).
Here, we set $t_{\max}=5/v$ and $\alpha = 0.5$.
In the presence of particle-hole symmetry, the choice of $\alpha = 0.5$ removes all odd-order contributions in the weak-coupling expansion \eqref{eq:lesser Green's function expansion approximation simplex}, reducing the computational cost.
For the TCI algorithm, we typically use the maximum bond dimension $\chi=40$.

\begin{figure}[ht]
  \centering
  \includegraphics[width=0.4\textwidth]{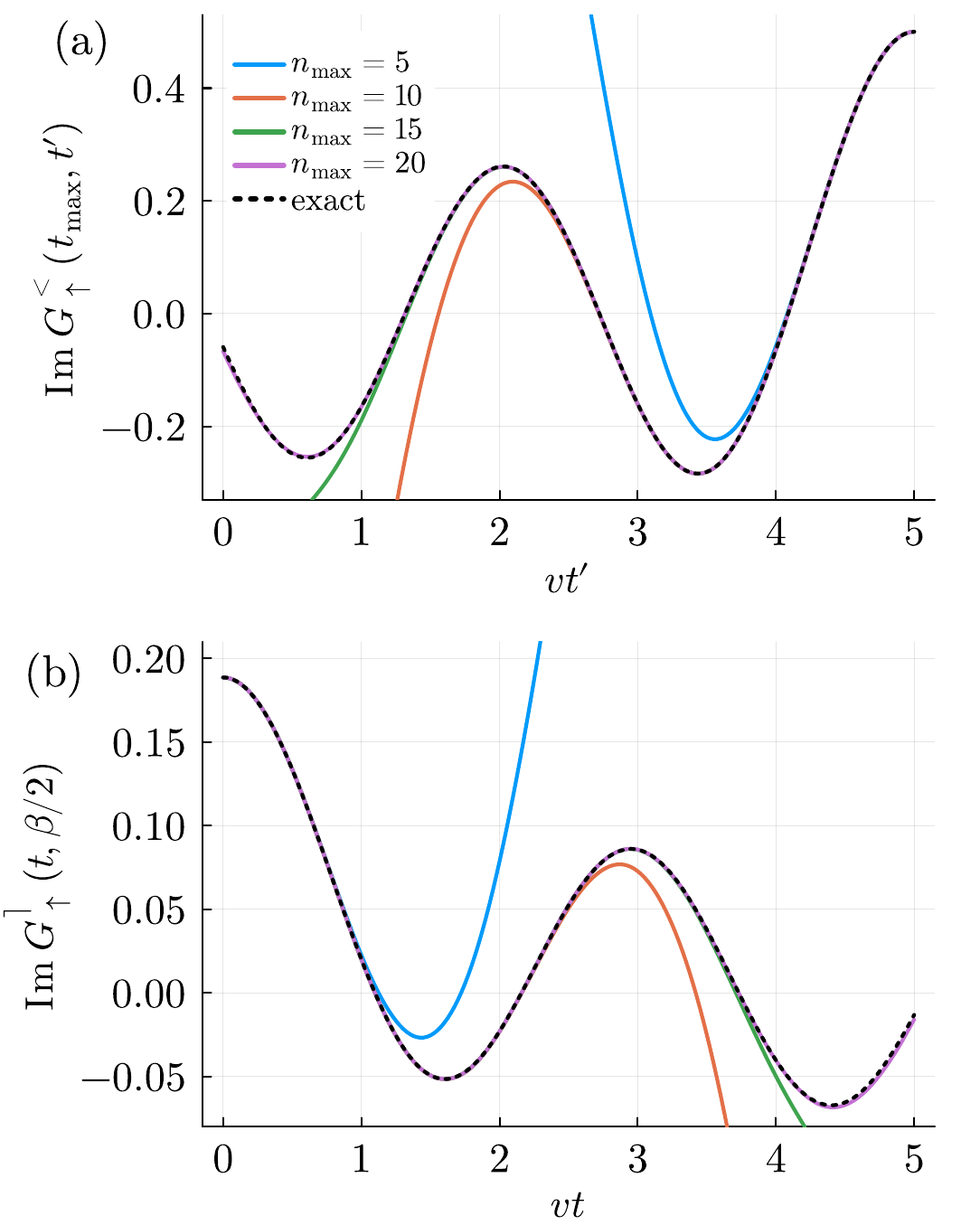}
  \caption{Results obtained with the weak-coupling nonequilibrium TCI impurity solver for the impurity model \eqref{eq:impurity model} with $\Delta_{\downarrow}(z,z') = 0$ and $\Delta_{\uparrow}^{R}(t,t')$, $\Delta_{\uparrow}^{\leftmixing}(t,\tau')$, $\Delta_{\uparrow}^{<}(t,t')$ given by Eq.~\eqref{eq:hybridization_function_bethe}.
  We set $\delta\mu=0$ (particle-hole symmetric system).
  The one-dimensional cuts of (a) $\Im G_{\uparrow}^{<}(t,t')$ at $t = t_{\mathrm{max}}$ and (b) $\Im G_{\uparrow}^{\leftmixing}(t,\tau')$ at $\tau' = \beta/2$ with several $n_{\mathrm{max}}$ are shown together with the exact results.
	The parameters are set to $\beta = 5/v$, $U = 3v$, $t_{\mathrm{max}} = 5/v$, $\alpha = 0.5$ and $\chi = 40$.}
  \label{fig:FK_quench_long}
\end{figure}

Figure~\ref{fig:FK_quench_long} shows one-dimensional cuts of the lesser and left-mixing Green's functions obtained by the TCI solver with several maximum orders $n_{\mathrm{max}}$, together with the exact reference data.
The TCI results converge systematically to the exact solution as $n_{\mathrm{max}}$ is increased.
For $n_{\mathrm{max}} = 20$, the difference from the exact curves is almost indistinguishable on the scale of Fig.~\ref{fig:FK_quench_long}.
This is consistent with the rough estimate of the maximum order \eqref{eq:estimate of n_max}, which gives $n_{\mathrm{max}} \simeq 20.4$ for this parameter set.
The agreement between the TCI results and the exact results suggests that the 20-dimensional integrals can be accurately evaluated by the tensor-train method with a relatively low bond dimension $\chi=40$.
This implies that the integrand of the weak-coupling expansion of the nonequilibrium Green's function has a low-rank structure.

\begin{figure}[t]
  \centering
  \includegraphics[width=0.48\textwidth]{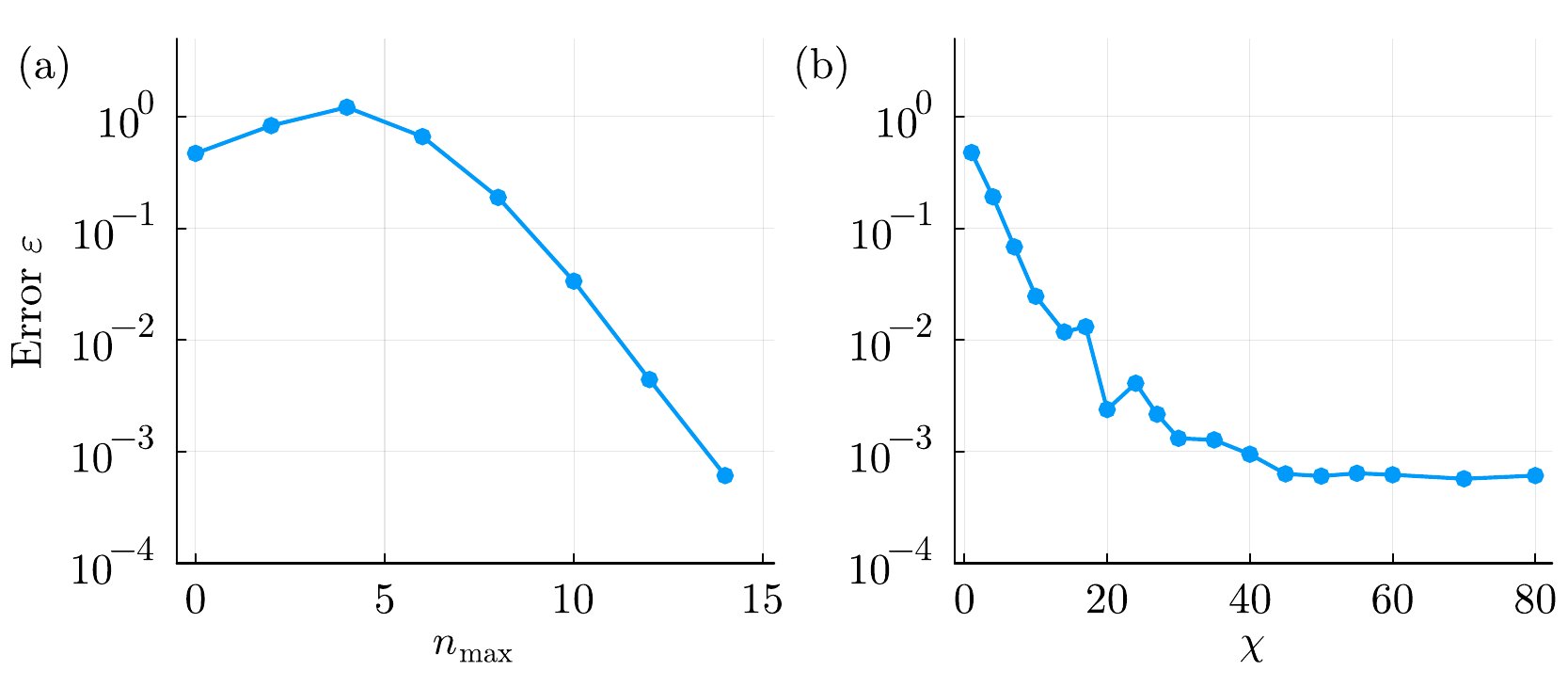}
  \caption{Convergence of the error of the lesser Green's function with respect to (a) the maximum order $n_{\mathrm{max}}$ and (b) the bond dimension $\chi$ for the exactly solvable impurity model [Eq.~\eqref{eq:impurity model} with $\Delta_\downarrow(z,z')=0$] with particle-hole symmetry $(\delta\mu = 0)$.
  The parameters are $\beta = 5/v$, $U=3v$, $t_{\mathrm{max}} = 3/v$ and $\alpha = 0.5$.
  In panel (a), the maximum bond dimension is fixed to $\chi=80$, and in panel (b), the maximum order is fixed to $n_{\mathrm{max}} = 14$.}
  \label{fig:FK_quench_error} 
\end{figure}

A more quantitative analysis of the convergence is presented in Fig.~\ref{fig:FK_quench_error}, where we set $t_{\mathrm{max}} = 3/v$ and $\alpha = 0.5$.
In Fig.~\ref{fig:FK_quench_error}(a), we show how the error of the lesser Green's function
\begin{align}
  \varepsilon = \max_{i,j=0,\dots,N_{t}} \abs{G_{\uparrow}^{<,\mathrm{TCI}}(i\Delta t, j\Delta t) - G_{\uparrow}^{<,\mathrm{exact}}(i\Delta t, j\Delta t)}
  \label{eq:error of lesser Green's function}
\end{align}
decreases with increasing maximum order $n_{\mathrm{max}}$.
Here, $N_{t}= 200$ is the number of time steps, and $\Delta t = t_{\mathrm{max}}/N_{t}$ is the time step.
To isolate the truncation error with respect to $n_{\mathrm{max}}$, we set $\chi = 80$ so that the error due to the finite bond dimension is negligible.
The error drops rapidly with increasing $n_{\mathrm{max}}$ when $n_{\mathrm{max}} \gtrsim 4$, and falls below $10^{-3}$ for $n_{\mathrm{max}} = 14$.
Figure~\ref{fig:FK_quench_error}(b) shows the same error as a function of the bond dimension $\chi$ at a fixed maximum order $n_{\mathrm{max}} = 14$.
In the range $0 \le \chi \le 30$, the error decreases approximately exponentially with increasing $\chi$ and then saturates below $10^{-3}$ for $\chi \gtrsim 50$.
Hence, the 14-dimensional integral can be evaluated accurately with a bond dimension $\chi \simeq 50$.

\subsection{Particle-hole-asymmetric case}
We next consider the case without particle-hole symmetry, where odd perturbation orders contribute in contrast with the particle-hole-symmetric case.
For conventional CT-QMC solvers, this further aggravates the sign problem, substantially increasing the computational cost required to reach a given simulation time. In practice, the maximum accessible time can be reduced by roughly a factor of two upon moving away from half filling.
Since the tensor-train approach is much less sensitive to the sign problem, one may expect that it can also handle the particle-hole-asymmetric case.

\begin{figure}[ht]
  \centering
  \includegraphics[width=0.4\textwidth]{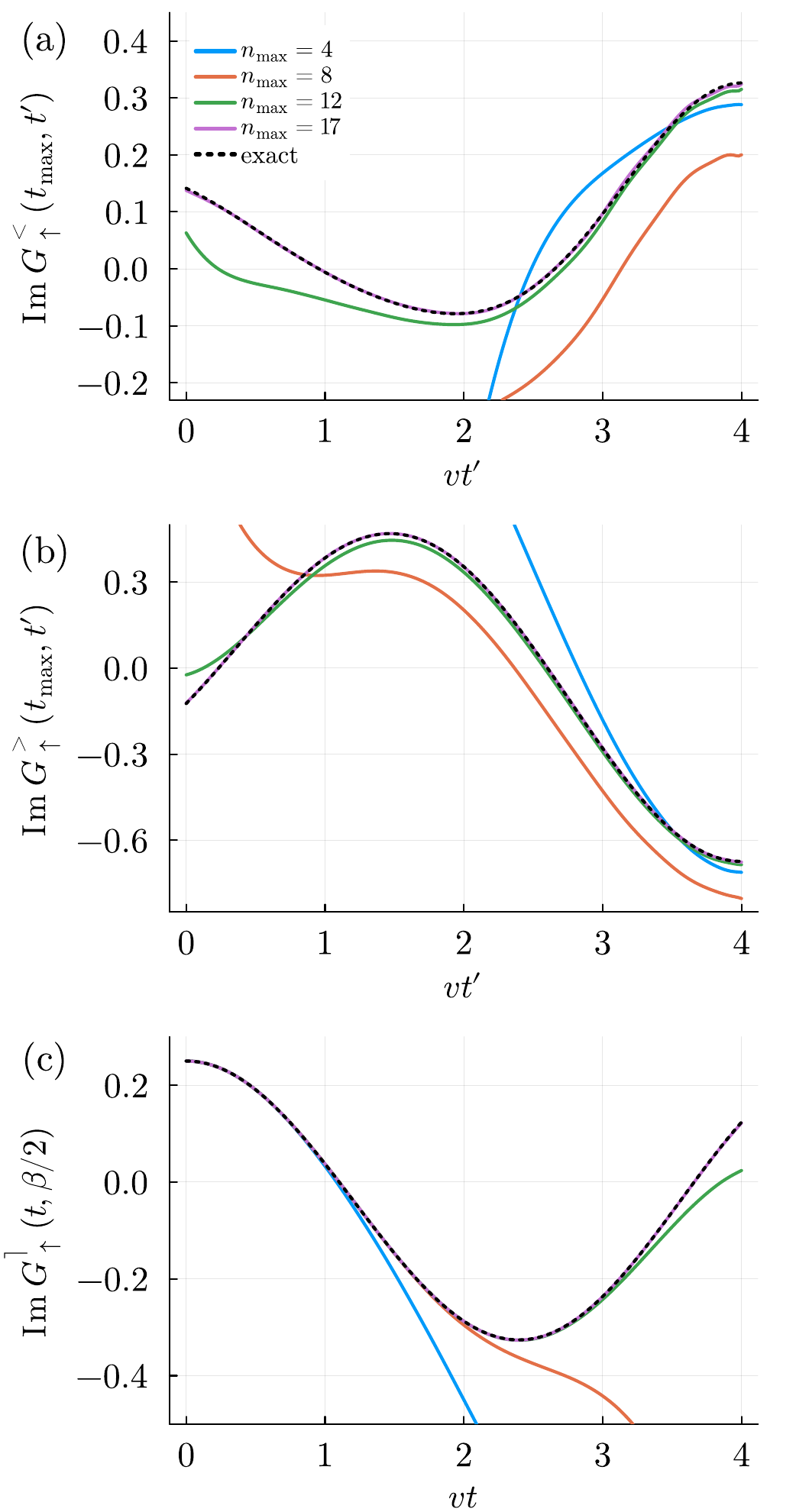}
  \caption{Results of the weak-coupling nonequilibrium TCI impurity solver for the impurity model \eqref{eq:impurity model} with $\Delta_{\downarrow}(z,z') = 0$ and $\Delta_{\uparrow}^{R}(t,t')$, $\Delta_{\uparrow}^{\leftmixing}(t,\tau')$, $\Delta_{\uparrow}^{<}(t,t')$ given by Eq.~\eqref{eq:hybridization_function_bethe}.
  The chemical-potential shift is set to $\delta\mu = 0.825 v$, where the particle-hole symmetry is broken.
  We plot the one-dimensional cuts of (a) $\Im G_{\uparrow}^{<}(t,t')$ at $t = t_{\mathrm{max}}$, (b) $\Im G_{\uparrow}^{>}(t,t')$ at $t = t_{\mathrm{max}}$, and (c) $\Im G_{\uparrow}^{\leftmixing}(t,\tau')$ at $\tau' = \beta/2$ with several $n_{\mathrm{max}}$ together with the exact results.
	The parameters are set to $\beta = 5/v$, $U = 3v$, $t_{\mathrm{max}} = 4/v$, $\alpha = 0.5$ and $\chi = 60$.}
  \label{fig:FK_quench_off_halffill_long}
\end{figure}

To demonstrate this, we set $\delta\mu = 0.825 v$, yielding an initial up-spin occupation $\langle n_\uparrow\rangle \simeq 0.75$ (i.e., 3/4-filling).
Figure~\ref{fig:FK_quench_off_halffill_long} shows one-dimensional cuts of the lesser, greater, and left-mixing Green's functions obtained by setting $t_{\mathrm{max}} = 4/v$, $\alpha = 0.5$ and $\chi = 60$.
Here we also plot the greater Green's function since it has to be computed independently when the particle-hole symmetry is broken.
For $n_{\mathrm{max}} = 17$, good agreement with the exact result is obtained for all three components, demonstrating that the tensor-train method can efficiently approximate the integrand of the weak-coupling expansion even in the absence of particle-hole symmetry.
Because this agreement is achieved already with a moderate bond dimension $\chi = 60$, the TCI solver can be numerically controlled in the particle-hole-asymmetric case with an only moderately larger bond dimension than in the particle-hole-symmetric case.
For the present benchmark problem, the additional computational cost arises mainly from the inclusion of odd perturbation orders and the independent evaluation of the greater Green's function, and remains modest compared with that of CT-QMC solvers.

\begin{figure}[ht]
  \centering
  \includegraphics[width=0.48\textwidth]{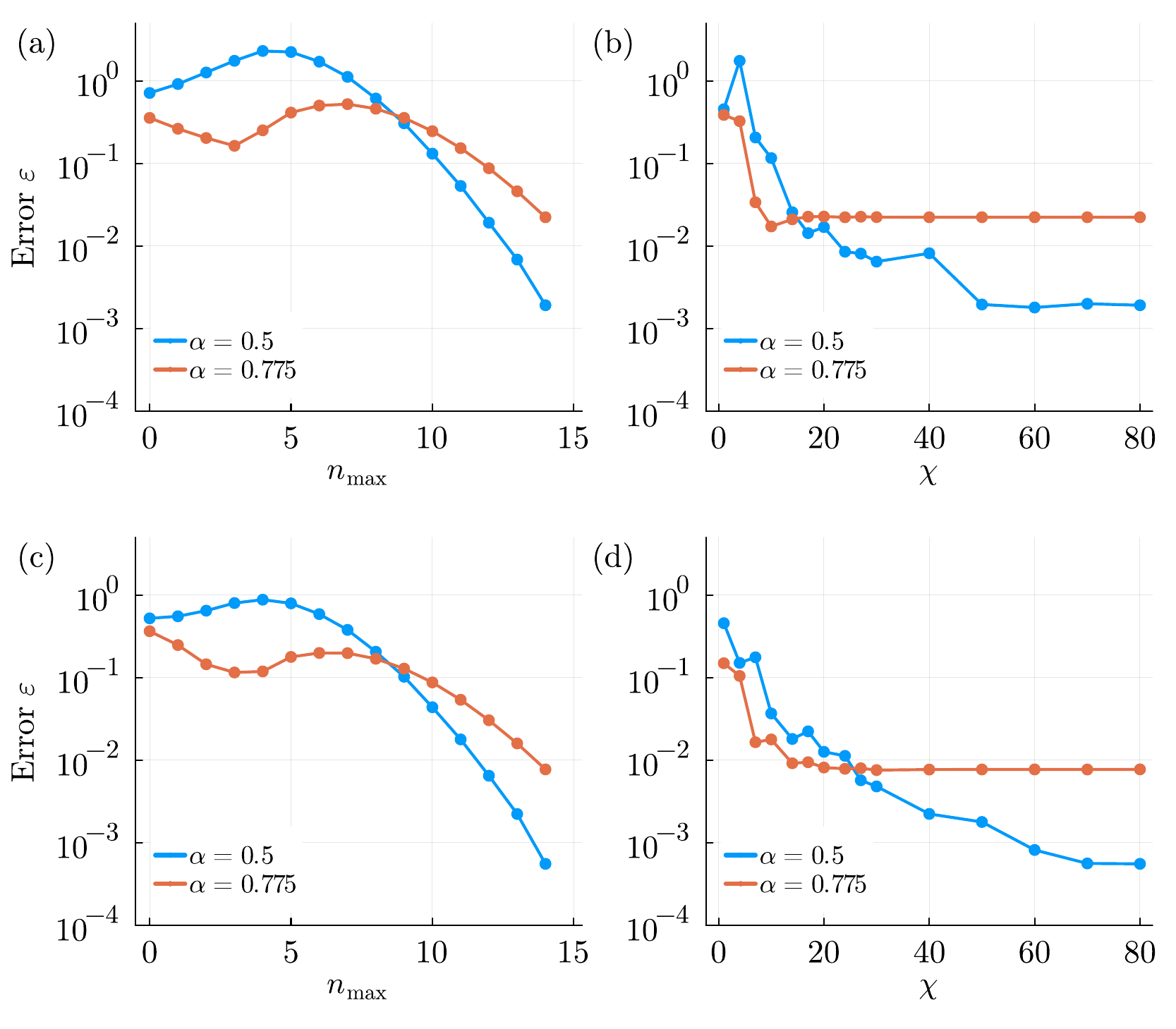}
  \caption{Convergence of the error of (a),(b) the lesser Green's function and (c),(d) the greater Green's function for the exactly solvable impurity model [Eq.~\eqref{eq:impurity model} with $\Delta_{\downarrow}(z,z') = 0$] without particle-hole symmetry $(\delta\mu = 0.825 v)$.
  In panels (a,c), the error is plotted as a function of the maximum order $n_{\mathrm{max}}$, and in panels (b,d), the error is plotted as a function of the maximum bond dimension $\chi$.
  The parameters are $\beta = 5/v$, $U=3v$, and $t_{\mathrm{max}} = 3/v$.
  In panels (a,c), the bond dimension is fixed to $\chi=80$, and in panels (b,d), the maximum order is fixed to $n_{\mathrm{max}} = 14$.}
  \label{fig:FK_quench_off_halffill_error}
\end{figure}

Figure~\ref{fig:FK_quench_off_halffill_error} provides details of the convergence of the lesser and greater Green's functions in the system without particle-hole symmetry.
For this analysis, we set $t_{\mathrm{max}} = 3/v$, and show the error of the lesser and greater Green's functions as a function of $n_{\mathrm{max}}$ and $\chi$.
The error for the greater component is defined analogously to Eq.~\eqref{eq:error of lesser Green's function}, with $G^{<}$ being replaced by $G^{>}$.
For the system away from half-filling, there is no obvious optimal choice of the shift parameter $\alpha$, so we compare the results of $\alpha=0.5$ and $\alpha = 0.775 \simeq 0.5 + \delta\mu / U$.
The choice $\alpha=0.5 + \delta\mu / U$ makes the lesser and greater components of the Weiss Green's function for the down-spin electron time-independent, which may simplify the structure of the integrand.

Figure~\ref{fig:FK_quench_off_halffill_error}(a) and (c) show the errors of the lesser and greater Green's functions as a function of $n_{\mathrm{max}}$ at fixed $\chi = 80$,
while Fig.~\ref{fig:FK_quench_off_halffill_error}(b) and (d) show the corresponding errors as a function of $\chi$ at fixed $n_{\mathrm{max}} = 14$.
For both the lesser and greater components, the errors decrease more rapidly with $n_{\mathrm{max}}$ for $\alpha = 0.5$ than for $\alpha = 0.775$.
At $n_{\mathrm{max}} = 14$, the error for $\alpha = 0.5$ is about an order of magnitude smaller than that for $\alpha = 0.775$.
At fixed $n_{\mathrm{max}} = 14$, however, the error reaches its saturation value at smaller bond dimensions for $\alpha = 0.775$ ($\chi \gtrsim 30$) than for $\alpha = 0.5$ ($\chi \gtrsim 70$).
Thus, tuning $\alpha$ involves a trade-off between convergence in $n_{\mathrm{max}}$ and $\chi$.
In the present implementation, the dominant computational cost scales with the perturbation order $n$ as $\order{2^{n}}$, while the cost associated with the bond dimension $\chi$ is subdominant.
It is therefore more important in practice to reduce the required $n_{\mathrm{max}}$ than to reduce the bond dimension.
From this viewpoint, $\alpha = 0.5$ is the more efficient choice than $\alpha = 0.775$.
For the nonequilibrium DMFT calculations in the following sections, we therefore set $\alpha = 0.5$.

\subsection{Sign cancellation in the TCI integration}
The TCI approach also allows us to quantify the sign (or more accurately phase) cancellations in the integral over the time variables.
To this end, we calculate the integral of the absolute value of the integrand in the weak-coupling expansion
\begin{align}
  &F_{\sigma}^{X}(t,t') = \sum_{n=0}^{n_{\mathrm{max}}} \abs{U}^{n} \sum_{k=0}^{n} \int_{S_{k}^{t',t}} \dd{t_{1}} \cdots \dd{t_{k}} \notag \\
  &\hspace{2.5cm}
  \int_{S_{n-k}^{0,t'}} \dd{t_{k+1}} \cdots \dd{t_{n}} \abs{Q_{n\sigma}^{X}(t,t', \qty{t_{i}})},
\end{align}
with the same methods as for the original Green's functions $G_{\sigma}^{X}(t,t')$, where $X$ stands for $<$ or $>$.
The ratio of the original Green's function to this integral of the absolute value,
\begin{align}
  S_{\sigma}^{X}(t,t') = \frac{\abs{G_{\sigma}^{X}(t,t')}}{F_{\sigma}^{X}(t,t')},
  \label{eq:average_sign_like}
\end{align}
is a measure of the severity of the sign cancellation during the integration.
We note that $S_{\sigma}^{X}(t,t')$ is not identical to the average sign in CT-QMC, since the summation over the Keldysh indices has already been performed in $Q_{n\sigma}^{X}(t,t', \qty{t_{i}})$.
Rather, it measures the sign or phase cancellation in the continuous-time integrals that are actually interpolated by the TCI algorithm.

\begin{figure*}[ht]
  \centering
  \includegraphics[width=\textwidth]{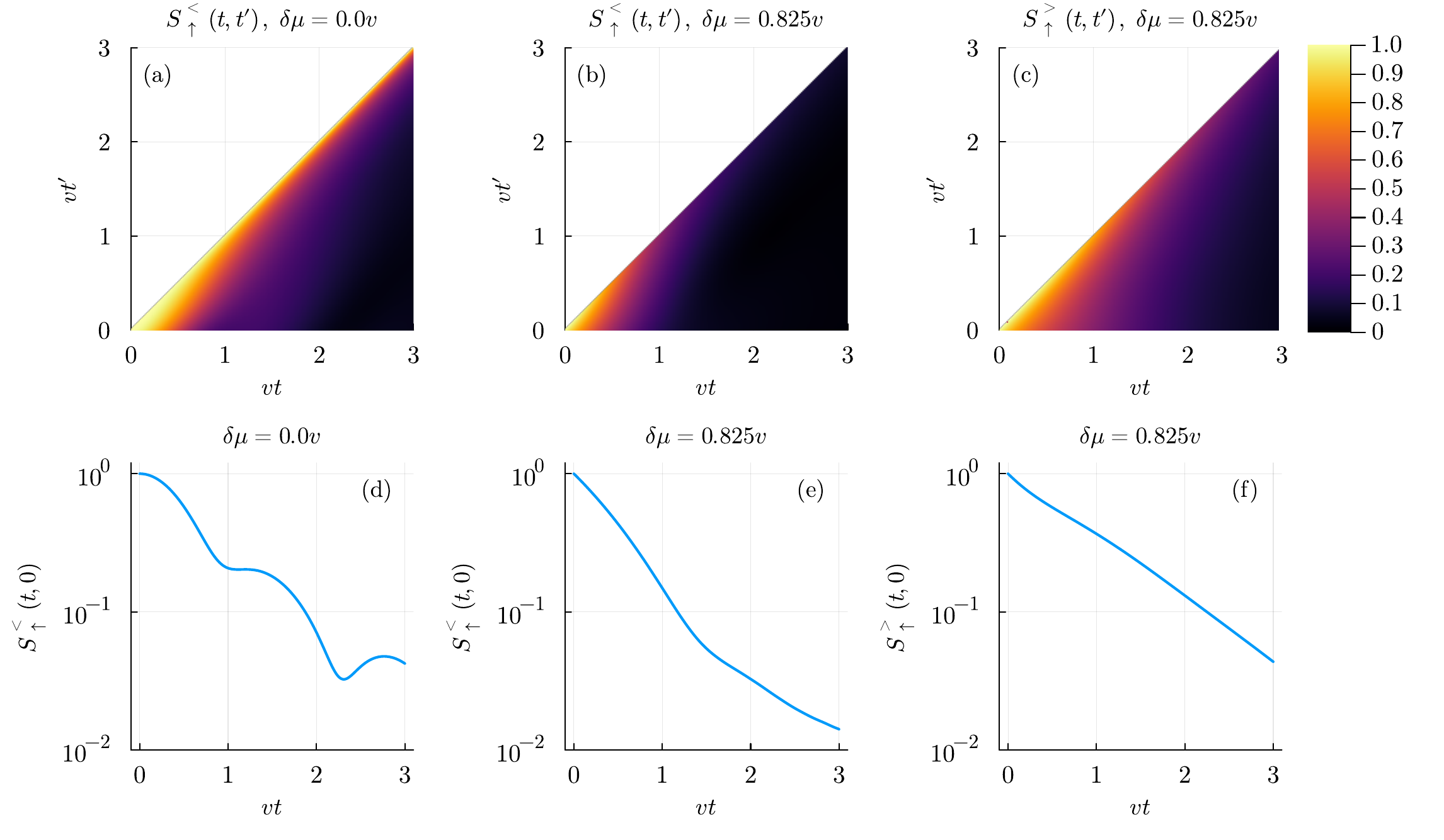}
  \caption{Average-sign quantity $S_{\uparrow}^{X}(t,t')$ defined in Eq.~\eqref{eq:average_sign_like} for the exactly solvable impurity model [Eq.~\eqref{eq:impurity model} with $\Delta_\downarrow(z,z')=0$].
    In panel (a), we show the lesser component $S_{\uparrow}^{<}(t,t')$ for the half-filled case ($\delta\mu = 0$), while in panels (b) and (c), we show the lesser and greater components $S_{\uparrow}^{<}(t,t')$ and $S_{\uparrow}^{>}(t,t')$ for the 3/4-filled case ($\delta\mu = 0.825 v$).
    In panels (d)--(f), we plot the one-dimensional cuts of (a)--(c) at $t'=0$.
    The parameters are $\beta = 5/v$, $U=3v$, $t_{\mathrm{max}} = 3/v$, and $\alpha = 0.5$.
    Both the denominator and numerator of Eq.~\eqref{eq:average_sign_like} are evaluated by the TCI method with $n_{\mathrm{max}} = 14$ and $\chi = 80$.}
  \label{fig:FK_quench_sign_cancellation}
\end{figure*}

Figure~\ref{fig:FK_quench_sign_cancellation} shows $S_{\uparrow}^{X}(t,t')$ ($0 \le t' \le t \le 3/v$) for the exactly solvable impurity model [Eq.~\eqref{eq:impurity model} with $\Delta_\downarrow(z,z')=0$].
Figure~\ref{fig:FK_quench_sign_cancellation}(a) shows the lesser component for the half-filled case, while Figs.~\ref{fig:FK_quench_sign_cancellation}(b) and \ref{fig:FK_quench_sign_cancellation}(c) show the lesser and greater components for the 3/4-filled case, respectively.
In Figs.~\ref{fig:FK_quench_sign_cancellation}(d)--\ref{fig:FK_quench_sign_cancellation}(f), we plot one-dimensional cuts of the heat maps in Figs.~\ref{fig:FK_quench_sign_cancellation}(a)--\ref{fig:FK_quench_sign_cancellation}(c), at $t'=0$.
In all cases, $S_{\uparrow}^{X}(t,t')$ decreases approximately exponentially as the time separation $t-t'$ increases, indicating that the sign cancellation becomes stronger at longer time separation.
At half filling, $S_{\uparrow}^{<}(t,t)$ is fixed to unity, which means that no cancellation occurs along the diagonal line $t = t'$.
Once particle-hole symmetry is broken, however, $S_{\uparrow}^{<}(t,t')$ and $S_{\uparrow}^{>}(t,t')$ are already smaller than unity even at $t=t'$, showing that the cancellation is stronger than in the particle-hole-symmetric case.
This trend is especially apparent for the lesser component shown in Fig.~\ref{fig:FK_quench_sign_cancellation}(b).
In the parameter range shown in Fig.~\ref{fig:FK_quench_sign_cancellation}, the minimum value is about $4\times 10^{-2}$ at half filling and about $1 \times 10^{-2}$ away from half filling.

Although the tensor-train method avoids the stochastic sign problem associated with Monte Carlo sampling, extremely small values of $S_{\uparrow}^{X}(t,t')$ can still lead to numerical difficulties.
When the sign or phase cancellation is too strong, even a small approximation error in the tensor-train representation can lead to a large relative error in the final Green's function, so that a more accurate tensor-train approximation and possibly a larger bond dimension are required.
In the present benchmark, where $S_{\uparrow}^{X}(t,t')$ reaches values of order $10^{-2}$, the convergence results in Figs.~\ref{fig:FK_quench_error} and \ref{fig:FK_quench_off_halffill_error} show that the Green's functions can be evaluated accurately with moderate bond dimensions.
These results suggest that, for this exactly solvable model, the weak-coupling expansion can be evaluated accurately by TCI even when the average sign is as small as $10^{-2}$.

\subsection{Low-rank structure of the integrand}
For the exactly solvable model, the origin of the low-rank structure of the integrand for $\alpha = 0.5 + \delta\mu/U$ can be understood as follows.
Since the lesser and greater components of the Weiss Green's function for the spin-down electrons are constant for this choice of $\alpha$, the determinant of the matrix $\bm{D}_{n\downarrow}(\qty{t_{i}}, \qty{s_{i}})$ defined in Eq.~\eqref{eq:D matrix} becomes constant and independent of $\qty{t_{i}}$ or $\qty{s_{i}}$.
Therefore, the nontrivial part of the integrand comes only from the first determinant $\det \tilde{\bm{D}}_{n \uparrow}^{<}(t,t',\qty{t_{i}},\qty{s_{i}})$.
If we naively expand this determinant, we obtain $(n+1)!$ terms.
However, not all of these terms contribute to the Green's function. 
As in the usual linked-cluster argument, only the connected terms contribute to the Green's function.
Here, ``connected'' means that there exists a path from $t$ to $t'$ that passes through each of $t_1,\ldots,t_n$ exactly once.

By taking this into account, we can reduce the number of terms contributing to the determinant from $(n+1)!$ to $n!$ as
\begin{widetext}
  \begin{align}
    &G_{\uparrow}^{<}(t,t') \propto \sum_{n=0}^{\infty} (iU)^{n} \sum_{k=0}^{n} \sum_{s_{1}, \dots, s_{n}=0}^{1} (-1)^{\sum_{\ell} s_{\ell}} \int_{S_{k}^{t',t}} \dd{t_{1}} \cdots \dd{t_{k}} \int_{S_{n-k}^{0,t'}} \dd{t_{k+1}} \cdots \dd{t_{n}} \notag \\
    &\hspace{4cm} \sum_{\sigma \in S_{n}} \mathcal{G}_{\uparrow}^{0 s_{\sigma(1)}}(t, t_{\sigma(1)}) \mathcal{G}_{\uparrow}^{s_{\sigma(1)} s_{\sigma(2)}}(t_{\sigma(1)}, t_{\sigma(2)}) \cdots \mathcal{G}_{\uparrow}^{s_{\sigma(n)} 1}(t_{\sigma(n)}, t').
    \label{eq:Falicov-Kimball lesser expansion linked cluster}
  \end{align}
Since the expression above is a convolution of the Weiss Green's functions $\mathcal{G}_{\uparrow}^{s_{i} s_{j}}(t_{i}, t_{j})$, we can use the Langreth rules to rewrite the integral in Eq.~\eqref{eq:Falicov-Kimball lesser expansion linked cluster} as
  \begin{align}
    &G_{\uparrow}^{<}(t,t') \propto \sum_{n=0}^{\infty} (iU)^{n} \sum_{k=0}^{n} \int_{S_{k}^{t',t}} \dd{t_{1}} \cdots \dd{t_{k}} \int_{S_{n-k}^{0,t'}} \dd{t_{k+1}} \cdots \dd{t_{n}} \notag \\
    &\hspace{2.5cm}  \sum_{\sigma \in S_{n}} \sum_{m=0}^{n} \qty(\prod_{j=0}^{m-1} \mathcal{G}_{\uparrow}^{R}(t_{\sigma(j)}, t_{\sigma(j+1)})) \mathcal{G}_{\uparrow}^{<}(t_{\sigma(m)}, t_{\sigma(m+1)}) \qty(\prod_{j=m+1}^{n} \mathcal{G}_{\uparrow}^{A}(t_{\sigma(j)}, t_{\sigma(j+1)})).
    \label{eq:Falicov-Kimball lesser expansion linked cluster Langreth}
  \end{align}
\end{widetext}
Here, we have defined $t_{\sigma(0)} = t$ and $t_{\sigma(n+1)} = t'$ for notational simplicity.
Because $\mathcal{G}_{\uparrow}^{R}(t_{i}, t_{j})$ and $\mathcal{G}_{\uparrow}^{A}(t_{i}, t_{j})$ are nonzero only when $t_{i} > t_{j}$ and $t_{i} < t_{j}$, respectively, only a small number of permutations $\sigma$ contribute to the summation.

As an example, let us consider the contribution with fixed $n$ and $k=n$, where the vertices are ordered as $t' \le t_n \le \cdots \le t_1 \le t$.
In this case, the integrand in Eq.~\eqref{eq:Falicov-Kimball lesser expansion linked cluster Langreth} becomes nonzero only when the permutation is $\sigma = \mathrm{id}$, and its contribution becomes
\begin{align}
	(iU)^{n} \int_{S_{n}^{t',t}} \dd{t_{1}} \cdots \dd{t_{n}} \mathcal{G}_{\uparrow}^{R}(t,t_{1}) \cdots \mathcal{G}_{\uparrow}^{R}(t_{n-1}, t_{n}) \mathcal{G}_{\uparrow}^{<}(t_{n}, t').
\end{align}
If the Weiss Green's function $\mathcal{G}_{\uparrow}^{R,<}(t_{i}, t_{j})$ is time translationally invariant, i.e., it depends only on the time difference $t_{i} - t_{j}$ as $\mathcal{G}_{\uparrow}^{R,<}(t_{i}, t_{j}) = \mathcal{G}_{\uparrow}^{R,<}(t_{i} - t_{j})$, the integrand has a rank-1 tensor-train representation with respect to the variables $u_{i}$:
\begin{align}
	\qty(iU)^{n} \mathcal{G}_{\uparrow}^{R}(u_{1}) \cdots \mathcal{G}_{\uparrow}^{R}(u_{n}) \mathcal{G}_{\uparrow}^{<}(u_{n+1}).
\end{align}
Although the Weiss Green's function in the present case is not time translationally invariant due to the extrapolation, it is expected that the integrand still has a low-rank tensor-train representation because the deviation from the time translationally invariant function is not so large.

The discussion above is specific to the case with $k = n$.
For $k \neq n$, the summation does not collapse to a single contribution.
However, the causality still eliminates many contributions and restricts the structure of the integrand.
Thus, the number of allowed terms is much smaller than the total number of permutations $(n+1)!$, and this should lead to a low-rank tensor-train representation of the integrand.

The above findings may also be related to the fact that the summation over $\qty{s_{i}}$ cannot be efficiently handled by the tensor-train method, as mentioned in Sec.~\ref{sec:lesser and greater Green's functions}.
The summation over the Keldysh indices $\qty{s_{i}}$ is essential because this summation eliminates all the disconnected contributions and retains only the connected contributions.
Moreover, once the summation over $\qty{s_{i}}$ is carried out, the resulting expression can be recast in terms of retarded, lesser and advanced components with the Langreth rules, so that causality eliminates many terms.
Without this summation, the integrand would retain a large number of contributions from the determinant expansion, and would generally not admit a compact low-rank tensor-train representation.

\section{Interaction quenches in the Hubbard model} 
\label{sec:Hubbard quench}

Having benchmarked the nonequilibrium TCI impurity solver against the exactly solvable impurity model in Sec.~\ref{sec:exactly solvable impurity model}, we now apply it within nonequilibrium DMFT to interaction quenches in the Hubbard model on the Bethe lattice with bandwidth $4v$.
We first examine the half-filled case, where previous CT-QMC results are available for comparison, and then extend the analysis to doped systems away from half filling.
We finally analyze the relaxation behavior and discuss how the fast thermalization and dynamical transition observed at half filling \cite{eckstein2009thermalization} are modified by doping.

\subsection{Half-filled case}
\begin{figure*}[ht]
  \centering
  \includegraphics[width=0.8\textwidth]{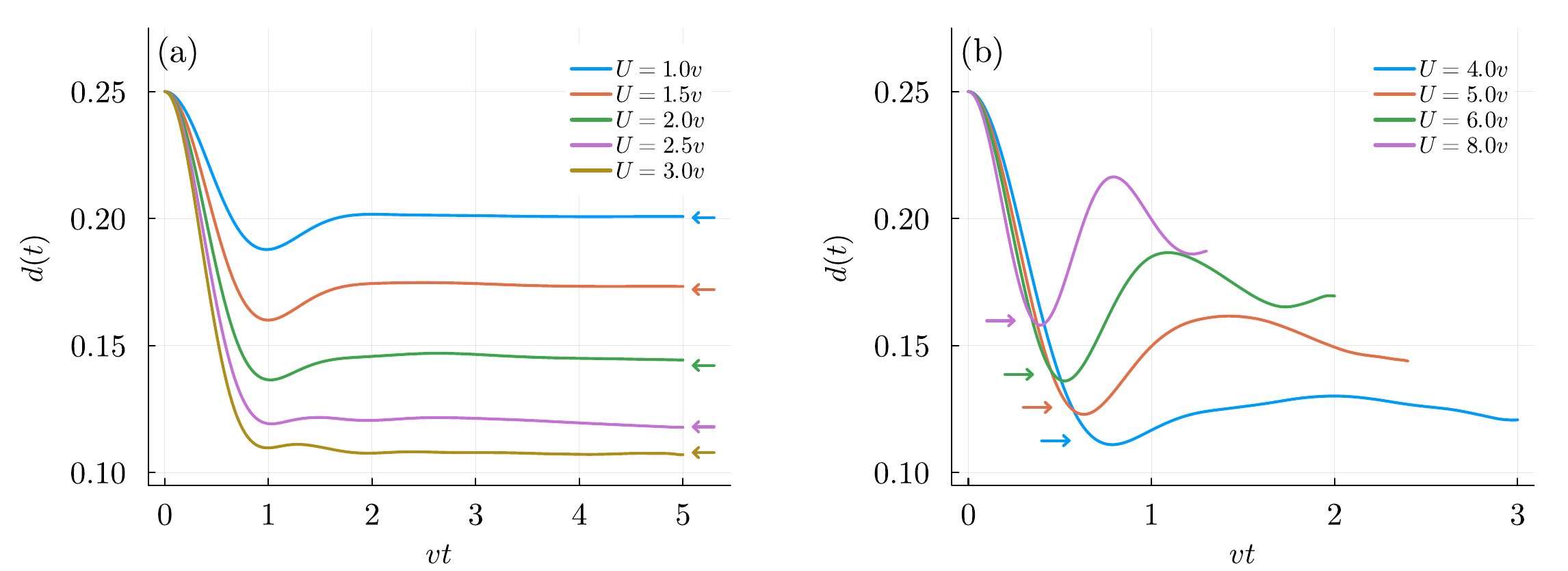}
  \caption{Time evolution of the double occupancy $d(t) = \ev*{n_{i\uparrow}(t) n_{i\downarrow}(t)}$ after interaction quenches in the half-filled Hubbard model $(\delta\mu = 0)$, calculated with nonequilibrium DMFT using the TCI solver.
  The initial state is noninteracting with $\beta = 5/v$.
  The arrows indicate the thermal values of the double occupancy expected to be reached in the limit $t \rightarrow \infty$.}
  \label{fig:Hubbard_quench_d}
\end{figure*}

We begin with the half-filled case, $\delta\mu = 0$.
To obtain the results, we iterate the nonequilibrium DMFT loop using the TCI impurity solver until self-consistency is reached.
In each DMFT iteration, we set the maximum bond dimension to $\chi = 40$ and the maximum order $n_{\mathrm{max}}$ based on the rough estimate \eqref{eq:estimate of n_max}.
For the analysis with $U = 3v$ and $t_{\mathrm{max}} = 5/v$, we set the largest order $n_{\mathrm{max}} = 20$.
The parameter $\alpha$ is set to $\alpha = 0.5$.

After convergence, we evaluate the double occupancy $d(t)= \ev*{n_{i\uparrow}(t) n_{i\downarrow}(t)}$ using
\begin{align}
	d(t) &= \alpha n - \alpha^2 \notag \\
  &\hspace{0.3cm} - \sum_{n=0}^{\infty} (iU)^{n} \sum_{s_{1}, \dots, s_{n}=0}^{1} (-1)^{\sum_{\ell} s_{\ell}} \int_{S_{n}^{0,t}} \dd{t_{1}} \cdots \dd{t_{n}} \notag \\ 
	&\hspace{1.3cm} \det \bm{E}_{n \uparrow}^{<}(t, \qty{t_{i}}, \qty{s_{i}}) \det \bm{E}_{n \downarrow}^{<}(t, \qty{t_{i}}, \qty{s_{i}}),
	\label{eq:nonequilibrium_double_occupancy}
\end{align}
where the matrix $\bm{E}_{n \sigma}^{<}(t, \qty{t_{i}}, \qty{s_{i}})$ is defined as
\begin{align}
	&\bm{E}_{n \sigma}^{<}(t, \qty{t_{i}}, \qty{s_{i}}) \notag \\
  &=
	\begin{pmatrix}
		\mathcal{G}_{\sigma}^{<}(t,t) - i\alpha & \mathcal{G}_{\sigma}^{0s_{1}}(t,t_{1}) & \cdots & \mathcal{G}_{\sigma}^{0s_{n}}(t,t_{n}) \\
		\mathcal{G}_{\sigma}^{s_{1}1}(t_{1},t) & \mathcal{G}_{\sigma}^{<}(t_{1},t_{1}) - i\alpha & \cdots & \mathcal{G}_{\sigma}^{s_{1}s_{n}}(t_{1},t_{n}) \\
		\vdots & \vdots & \ddots & \vdots \\
		\mathcal{G}_{\sigma}^{s_{n}1}(t_{n},t) & \mathcal{G}_{\sigma}^{s_{n}s_{1}}(t_{n},t_{1}) & \cdots & \mathcal{G}_{\sigma}^{<}(t_{n},t_{n}) - i\alpha
	\end{pmatrix},
\end{align}
and $n$ in the first line is the electron density.
Equation~\eqref{eq:nonequilibrium_double_occupancy} follows from Wick's theorem applied to the quantity $\ev*{[n_{i\uparrow}(t) - \alpha] [n_{i\downarrow}(t) - \alpha]}$.
Since the integral in Eq.~\eqref{eq:nonequilibrium_double_occupancy} is over the simplex $S_{n}^{0,t}$, we can evaluate it using the recursive convolution method explained in Sec.~\ref{sec:lesser and greater Green's functions}.
Since the expansion up to around 20th order can be evaluated accurately by our TCI impurity solver, the parameter regime we can access is given by 
\begin{align}
  \abs{U} t_{\mathrm{max}} \lesssim 14.7,
\end{align}
where we use the estimate \eqref{eq:estimate of n_max} with $n_{\mathrm{max}} = 20$.

In Figs.~\ref{fig:Hubbard_quench_d}(a) and \ref{fig:Hubbard_quench_d}(b), we show the time evolution of the double occupancy after the interaction quench from $U=0$ and $\beta=5/v$ to the weak-coupling ($U\le 3v$) and the strong-coupling ($U> 3v$) regime, respectively.
We can see that the results obtained from the TCI solver are in good agreement with the nonequilibrium DMFT results of Ref.~\cite{eckstein2009thermalization}, which were obtained using the CT-QMC impurity solver.
This demonstrates that the TCI solver can reproduce the CT-QMC results over a comparable time range in the particle-hole-symmetric case.

\begin{figure}[ht]
  \centering
  \includegraphics[width=0.4\textwidth]{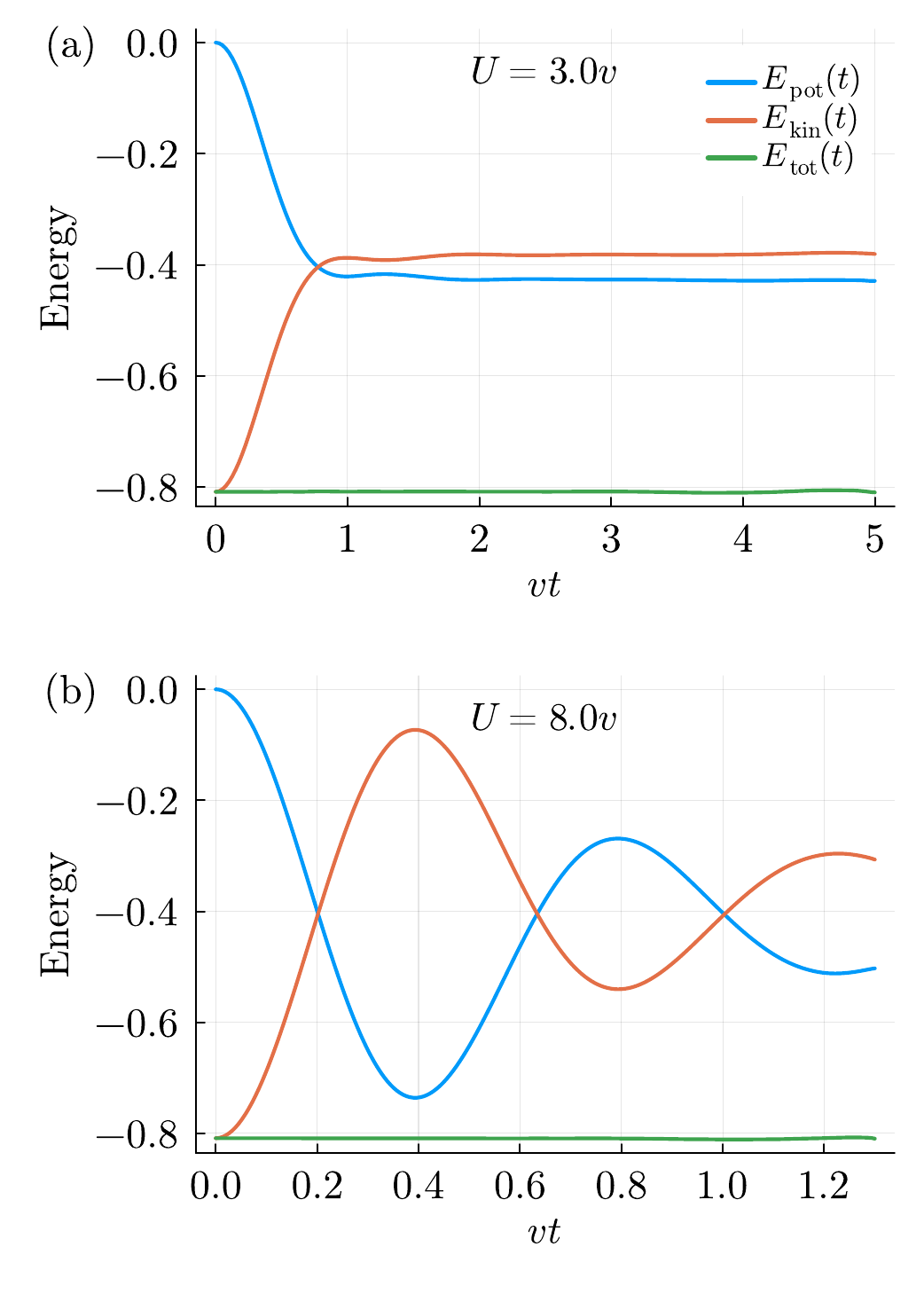}
  \caption{Time evolution of the potential, kinetic, and total energies after interaction quenches in the half-filled Hubbard model ($\delta\mu = 0$), calculated with nonequilibrium DMFT using the TCI solver.
  We plot the results for quenches to (a) the weak-coupling regime ($U = 3.0v$) and (b) the strong-coupling regime ($U = 8.0v$).
  The initial state is noninteracting with $\beta = 5/v$.}
  \label{fig:Hubbard_quench_energy_conservation}
\end{figure}

In addition to the double occupancy, we also calculate the time evolution of the potential, kinetic, and total energies per site defined by
\begin{align}
	&E_{\mathrm{pot}}(t) = \frac{1}{N} \sum_{i=1}^{N} U(t) \ev{\qty(n_{i\uparrow}(t) - \frac{1}{2}) \qty(n_{i\downarrow}(t) - \frac{1}{2})}, \\
	&E_{\mathrm{kin}}(t) = \frac{1}{N} \sum_{\expval{i,j},\sigma} \frac{v_{\sigma}}{\sqrt{Z}} \ev*{c_{i\sigma}^{\dag}(t) c_{j\sigma}(t)}, \\
	&E_{\mathrm{tot}}(t) = E_{\mathrm{pot}}(t) + E_{\mathrm{kin}}(t),
\end{align}
where $N$ is the number of lattice sites.
These quantities can be computed from
\begin{align}
	&E_{\mathrm{pot}}(t) = U \qty(d(t) - \frac{n}{2} + \frac{1}{4}), \\
	&E_{\mathrm{kin}}(t) = -i \sum_{\sigma} \qty(\Delta_{\sigma} * G_{ii\sigma})^{<}(t,t).
\end{align}
The results for $U = 3v$ and $U = 8v$ are shown in Fig.~\ref{fig:Hubbard_quench_energy_conservation}.
Since the system is isolated and the Hamiltonian is time independent after the quench, the total energy must be conserved.
Whether this energy conservation is satisfied or not provides a nontrivial consistency check for the accuracy of the nonequilibrium DMFT calculation.
Figure~\ref{fig:Hubbard_quench_energy_conservation} shows that this condition is satisfied in both the weak-coupling and strong-coupling regimes, confirming that the TCI impurity solver provides an accurate solution over a wide range of interaction strengths.

\subsection{3/4-filled case}
We next consider the 3/4-filled case with $\delta\mu=0.825v$. Away from half filling, the sign problem becomes substantially more severe for CT-QMC solvers, and, to our knowledge, nonequilibrium DMFT studies of interaction quenches in the doped Hubbard model are not available. By contrast, the benchmark in Sec.~\ref{sec:exactly solvable impurity model} demonstrated that the TCI solver can handle particle-hole asymmetric impurity problems with little additional computational cost.
This makes the TCI method particularly promising for nonequilibrium DMFT away from half filling.
Guided by the benchmark in Sec.~\ref{sec:exactly solvable impurity model}, we set $\alpha = 0.5$ here.

\begin{figure*}[ht]
  \centering
  \includegraphics[width=0.8\textwidth]{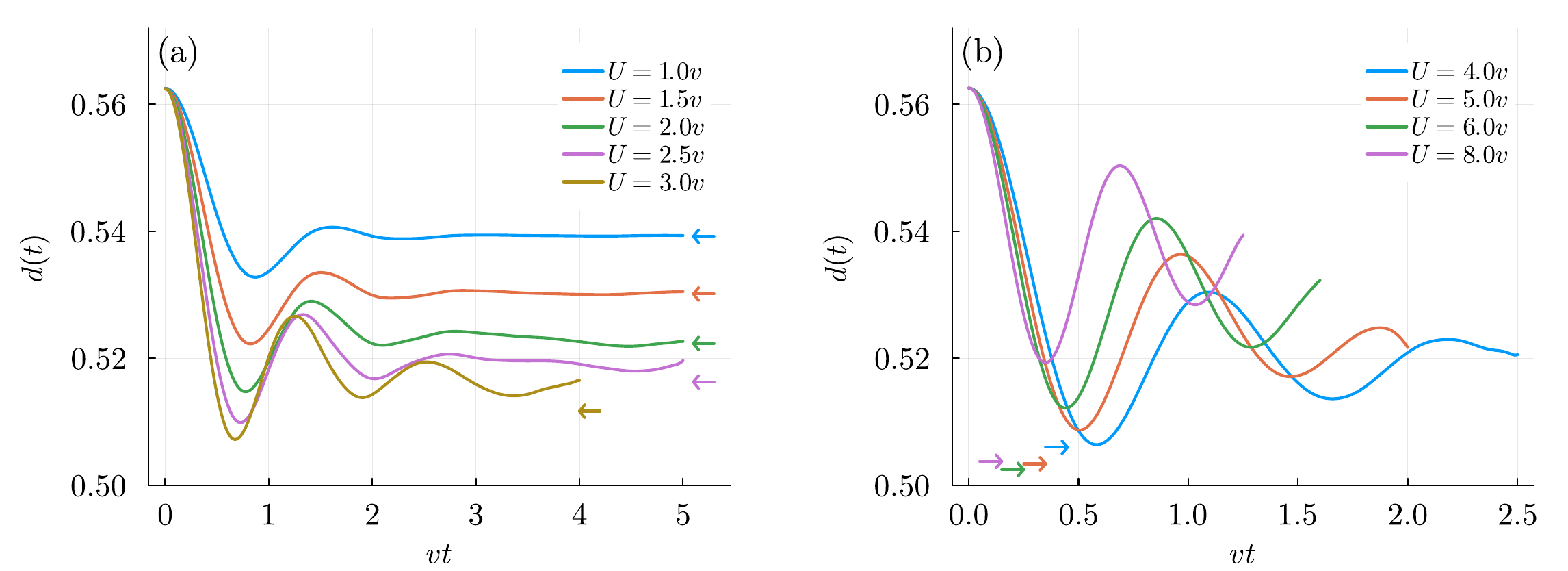}
  \caption{Time evolution of the double occupancy $d(t) = \ev*{n_{i\uparrow}(t) n_{i\downarrow}(t)}$ after the interaction quench for the Hubbard model at 3/4-filling $(\delta\mu = 0.825 v)$.
  The initial state is noninteracting with $\beta = 5/v$.
  The arrows indicate the thermal values of the double occupancy expected to be reached in the limit $t \rightarrow \infty$.}
  \label{fig:Hubbard_quench_off_halffill_d}
\end{figure*}

Figure~\ref{fig:Hubbard_quench_off_halffill_d} shows the time evolution of the double occupancy after interaction quenches from the noninteracting state with $\beta = 5/v$.
The calculation remains well controlled in the particle-hole-asymmetric case and reaches time scales comparable to those obtained at half filling. This demonstrates that nonequilibrium DMFT remains feasible away from half filling within the present TCI framework, and suggests access to longer simulation times than are presently achievable with CT-QMC-based approaches.

\begin{figure}[ht]
  \centering
  \includegraphics[width=0.4\textwidth]{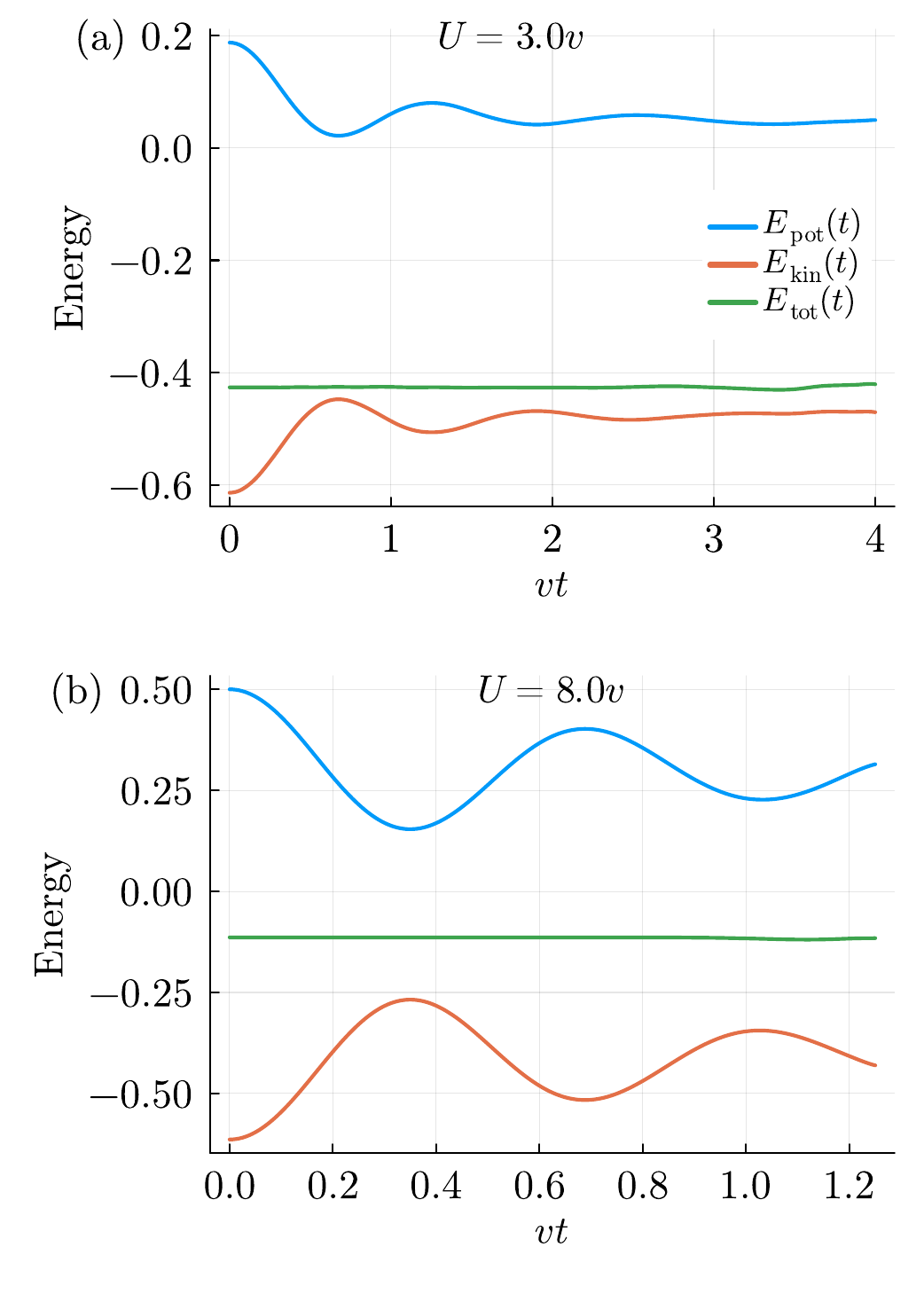}
  \caption{Time evolution of the potential, kinetic, and total energies after interaction quenches in the Hubbard model at 3/4-filling ($\delta\mu = 0.825 v$).
  We plot the results for quenches to (a) the weak-coupling regime ($U = 3.0v$) and (b) the strong-coupling regime ($U = 8.0v$).
  The initial state is noninteracting with $\beta = 5/v$.}
  \label{fig:Hubbard_quench_energy_conservation_off_halffill}
\end{figure}

The time evolution of the potential, kinetic, and total energies after the quenches to $U=3.0v$ and $8.0v$ is shown in Fig.~\ref{fig:Hubbard_quench_energy_conservation_off_halffill}.
The total energy is conserved for both the weak-coupling and strong-coupling cases.
We have also confirmed that the particle number remains constant during the time evolution, although the data are not shown.
These conservation laws provide stringent internal checks on the accuracy of the calculation.
Taken together, these results indicate that, unlike in CT-QMC, particle-hole asymmetry introduces only a constant-factor overhead in the present TCI approach, arising mainly from odd perturbation orders and the independent evaluation of the greater component of the Green's function.

\begin{figure*}[ht]
  \centering
  \includegraphics[width=\textwidth]{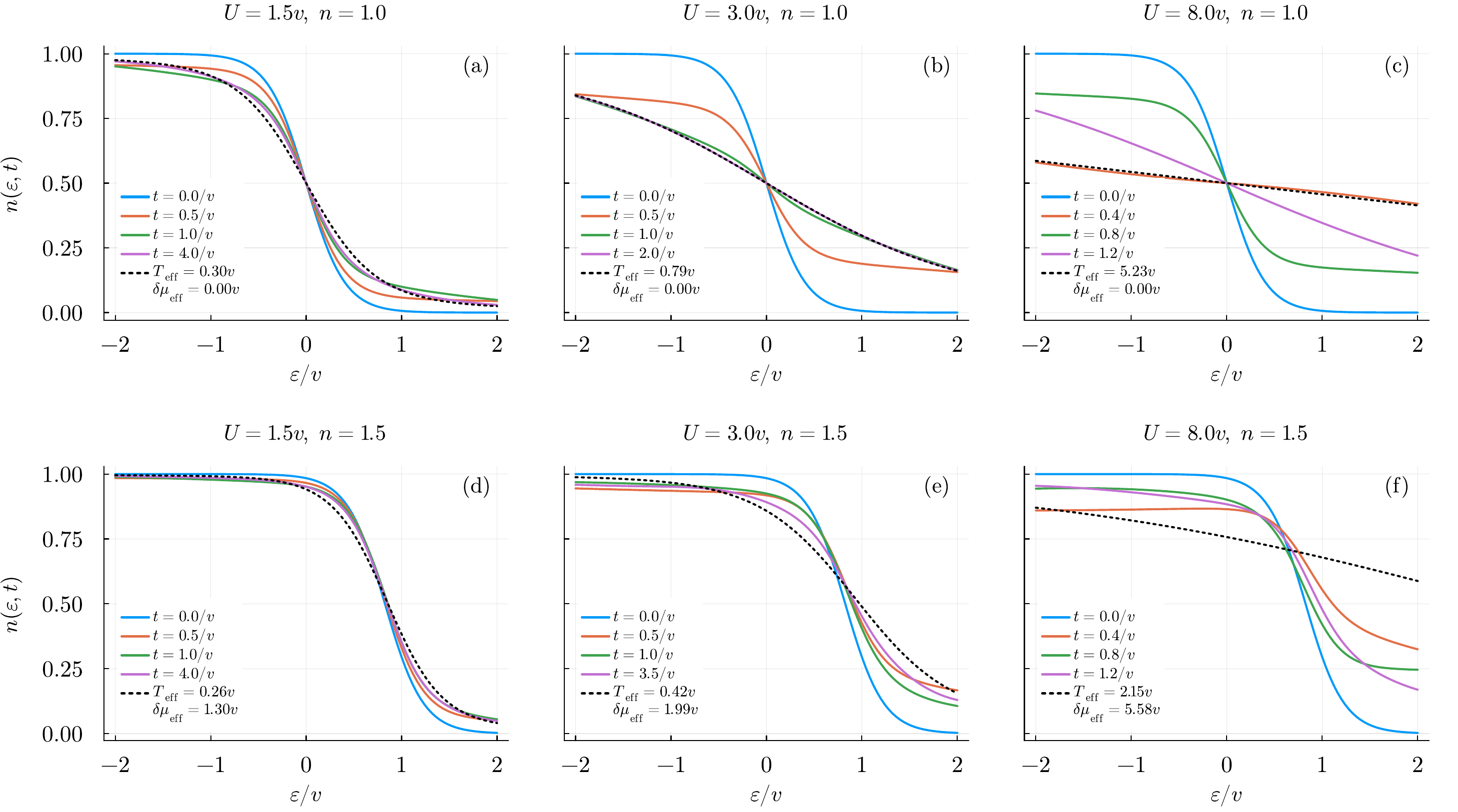}
  \caption{Time evolution of the distribution function $n(\varepsilon,t) = -i G_{\varepsilon \sigma}^{<}(t,t)$ after interaction  quenches in the Hubbard model at (a)--(c) half-filling ($\delta\mu = 0$) and (d)--(f) 3/4-filling ($\delta\mu = 0.825 v$).
  The dashed lines show the thermal distribution function $n_{\mathrm{th}}(\varepsilon)$ with effective temperature $T_{\mathrm{eff}}$ and effective chemical-potential shift $\delta \mu_{\mathrm{eff}}$, which are determined so as to match the total energy after the quench and the particle number. 
  In the half-filled case, the system quickly relaxes to a thermal distribution around $U = 3v$, while in the 3/4-filled case, there is no clear fast thermalization behavior for the considered values of $U$.}
  \label{fig:Hubbard_quench_off_halffill_distribution}
\end{figure*}

\subsection{Relaxation toward thermal equilibrium}

\begin{figure*}[ht]
  \centering
  \includegraphics[width=\textwidth]{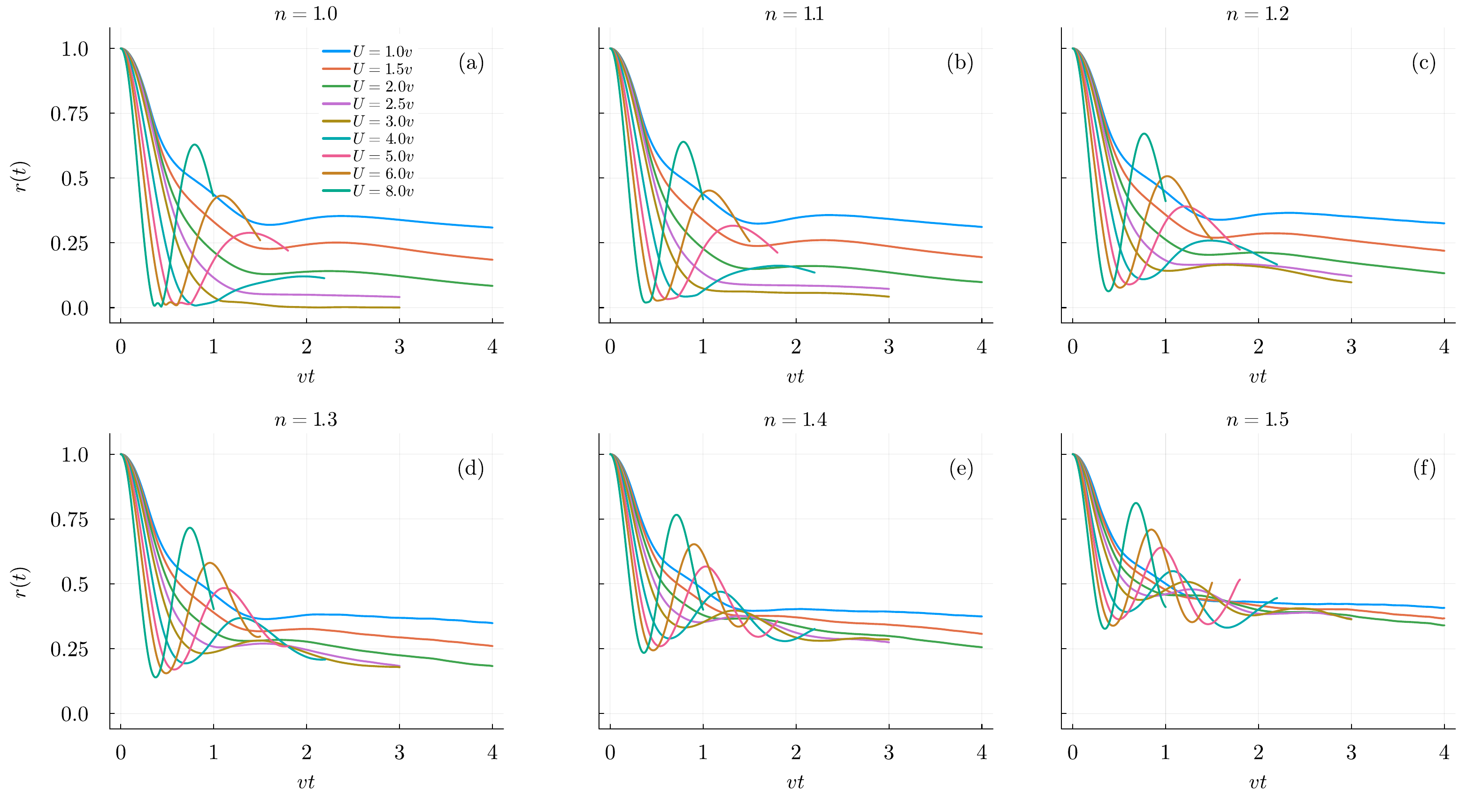}
  \caption{Time evolution of the normalized deviation of the nonequilibrium distribution function from the corresponding thermal distribution,
    $r(t)=D(t)/D(0)$, after an interaction quench in the Hubbard model.
    The panels (a)--(f) show the results for $n=1.0$, $1.1$, $1.2$, $1.3$, $1.4$, and $1.5$, respectively.
    The fast thermalization behavior can only be observed at half filling with $U=3.0v$.}
  \label{fig:Hubbard_quench_distribution_diff_time_evolution}
\end{figure*}

In this section, we discuss the thermalization behavior after the interaction quench.
We focus on the double occupancy $d(t)$ and the distribution function $n(\varepsilon,t) = -i G_{\varepsilon\sigma}^{<}(t,t)$, where $G_{\varepsilon\sigma}^{<}(t,t')$ is the lesser component of the lattice Green's function $G_{\varepsilon\sigma}(z,z')$, which satisfies the Dyson equation
\begin{align}
  &\qty(i \dv{z} - \varepsilon + \delta\mu - U(z) \qty(\alpha - \frac{1}{2})) G_{\varepsilon \sigma}(z,z') \notag \\
  &\hspace{1.5cm} - \int_{\mathcal{C}} \dd{\bar{z}} \Sigma_{\sigma}(z,\bar{z}) G_{\varepsilon \sigma}(\bar{z}, z') = \delta_{\mathcal{C}}(z,z').
\end{align}
The dynamics of the double occupancy has already been shown in Figs.~\ref{fig:Hubbard_quench_d} and \ref{fig:Hubbard_quench_off_halffill_d}.
In these figures, the arrows indicate the values of the double occupancy corresponding to the thermal states with effective temperatures $T_{\mathrm{eff}}$ and effective chemical-potential shifts $\delta\mu_{\mathrm{eff}}$, which are chosen such that the total energy and particle number of the thermal states agree with those after the quench.
To determine $T_{\mathrm{eff}}$ and $\delta\mu_{\mathrm{eff}}$, we performed  equilibrium DMFT calculations, and calculated the total energy and particle number for the Hubbard model with the TCI impurity solver developed in Ref.~\cite{matsuura2025tensor}.
In Fig.~\ref{fig:Hubbard_quench_off_halffill_distribution}, we show the time evolution of the distribution function $n(\varepsilon, t)$ for representative values of the interaction strength $U$, together with the thermal distribution function $n_{\mathrm{th}}(\varepsilon)$ with the effective temperature $T_{\mathrm{eff}}$ and effective chemical-potential shift $\delta\mu_{\mathrm{eff}}$.

In the particle-hole symmetric case, which has already been studied in Refs.~\cite{eckstein2009thermalization,eckstein2010interaction}, 
there is a dynamical transition at around $U = U_{c}^{\mathrm{dyn}} \simeq 3v$, and the relaxation behavior differs qualitatively between the weak-coupling regime $U < U_{c}^{\mathrm{dyn}}$ and the strong-coupling regime $U > U_{c}^{\mathrm{dyn}}$.
For the double occupancy, the weak-coupling case shows an overshoot beyond the thermal value, followed by relaxation toward it, whereas in the strong-coupling case the double occupancy oscillates around a nonthermal value with a period of about $2\pi/U$.
A similar distinction is seen in the distribution function.
In the weak-coupling regime, it settles on a short time scale into a nearly stationary form, but this stationary distribution still differs visibly from the thermal distribution $n_{\mathrm{th}}(\varepsilon)$.
This is consistent with the prethermalization scenario discussed in Ref.~\cite{moeckel2008interaction}, where the weak-coupling quench leads first to an intermediate quasistationary regime with a nonequilibrium distribution, while full thermalization of the distribution function occurs only on longer time scales.
In the strong-coupling regime, the distribution function oscillates around a nonthermal form with a period of approximately $2\pi/U$, similar to the behavior of the double occupancy.
Between the weak- and strong-coupling regimes, especially at $U = 3v \simeq U_{c}^{\mathrm{dyn}}$, the system relaxes rapidly to a thermal distribution within a short time of order $t \simeq 2.0/v$.

In the 3/4-filled case, by contrast, no clear transition between the weak-coupling and strong-coupling behavior is observed.
The double occupancy exhibits oscillatory behavior not only in the strong-coupling regime but already in the weak-coupling regime, for example at $U = 1.5v$.
Also, the distribution function does not show rapid thermalization for any of the interaction strengths studied here.

To characterize the approach to thermal equilibrium more quantitatively, we introduce the indicator
\begin{align}
  D(t) = \int_{-2v}^{2v} \dd{\varepsilon} \rho(\varepsilon) \abs{n(\varepsilon, t) - n_{\mathrm{th}}(\varepsilon)},
\end{align}
where $\rho(\varepsilon)$ is the density of states for the Bethe lattice:
\begin{align}
    \rho(\varepsilon) = \frac{\sqrt{4v^2 - \varepsilon^2}}{2\pi v^{2}}.
\end{align}
This quantity measures the distance between the instantaneous distribution function and the corresponding thermal distribution.
In the following, we use the normalized deviation
\begin{align}
    r(t) = D(t)/D(0)
\end{align}
to compare different fillings and interaction strengths.

Figure~\ref{fig:Hubbard_quench_distribution_diff_time_evolution} shows the time evolution of $r(t)$ for several fillings and interaction strengths.
A distinctive feature of the half-filled case is that $r(t)$ can reach zero, within numerical accuracy, at certain interaction strengths and times.
For $U=3v$, $r(t)$ rapidly decreases to zero and remains close to zero within the accessible time window, indicating fast relaxation of the distribution function.
For stronger interactions ($U\ge 4v$), $r(t)$ also reaches nearly zero at certain times but subsequently increases again, showing a revival of the nonthermal distribution.
When the filling is shifted away from half filling, however, this zero-touching behavior is lost.
As we increase the doping, the curves of $r(t)$ are gradually lifted away from zero, and $r(t)$ remains nonzero for all interaction strengths and times considered here.
This means that, away from the particle-hole symmetric point, no fast thermalization behavior can be observed.

\begin{figure}[ht]
  \centering
  \includegraphics[width=0.45\textwidth]{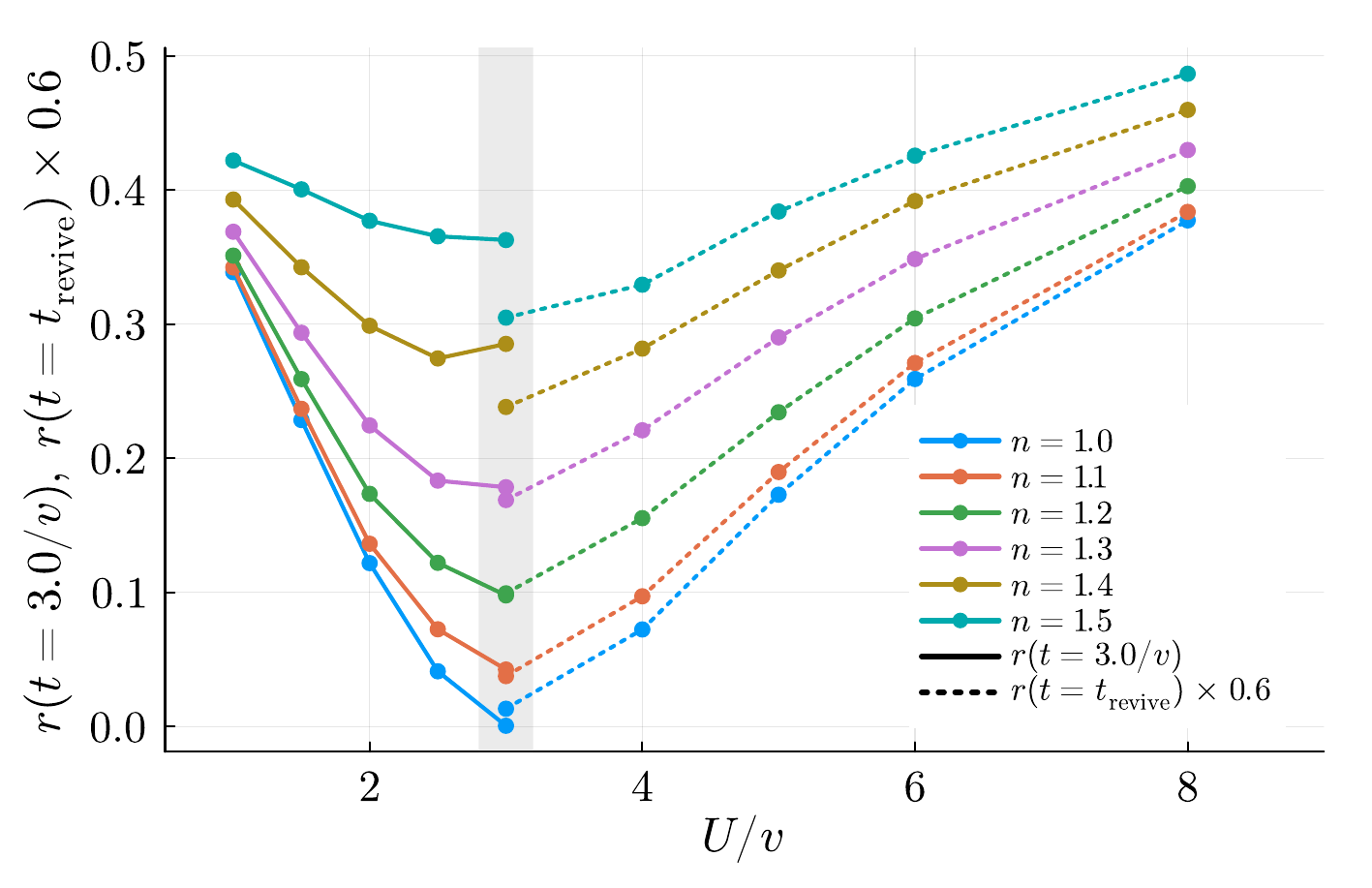}
  \caption{
  $r(t)=D(t)/D(0)$ as a function of $U/v$ for several fillings $n$. Here, $r(t)$ measures the distance between the distribution function $n(\varepsilon,t)$ and the thermal distribution $n_{\mathrm{th}}(\varepsilon)$ at time $t$.
  In the weak-coupling regime, solid lines show $r(t)$ at $t=3.0/v$, whereas in the strong-coupling regime, dotted lines show $r(t)$ at the time $t = t_{\mathrm{revive}}$ when $r(t)$ reaches its first revival maximum after the quench.
  For presentation purposes, the values of $r(t)$ on the strong-coupling side are multiplied by $0.6$.}
  \label{fig:Hubbard_quench_distribution_diff_int_vs_U}
\end{figure}

The disappearance of fast thermalization upon doping can also be clearly seen in Fig.~\ref{fig:Hubbard_quench_distribution_diff_int_vs_U}, where we plot $r(t)$ at representative times as a function of the interaction strength for several fillings.
In the weak-coupling regime with $U \le 3v$ (solid lines), we evaluate $r(t)$ at a fixed time $t = 3.0/v$, while in the strong-coupling regime with $U \ge 3v$ (dotted lines), where the dynamics is oscillatory, we evaluate $r(t)$ at the time when $r(t)$ reaches its first revival maximum after the quench.
For example, in the half-filled case with $U = 8.0v$, we use $t = t_{\mathrm{revive}} \simeq 0.85/v$, as seen in Fig.~\ref{fig:Hubbard_quench_distribution_diff_time_evolution}.
For the half-filled case, $r(t)$ becomes very small as $U=3v$ is approached from both the weak- and strong-coupling sides, reflecting the dynamical transition near $U=3v$.
By contrast, away from half filling, the dip in $r(t)$ around $U\simeq 3v$ becomes progressively shallower with increasing doping.
When the density reaches $n = 1.5$, the dip is essentially wiped out, and no clear signature of the dynamical transition remains.

\begin{figure}[t]
    \centering
    \includegraphics[width=0.49\textwidth]{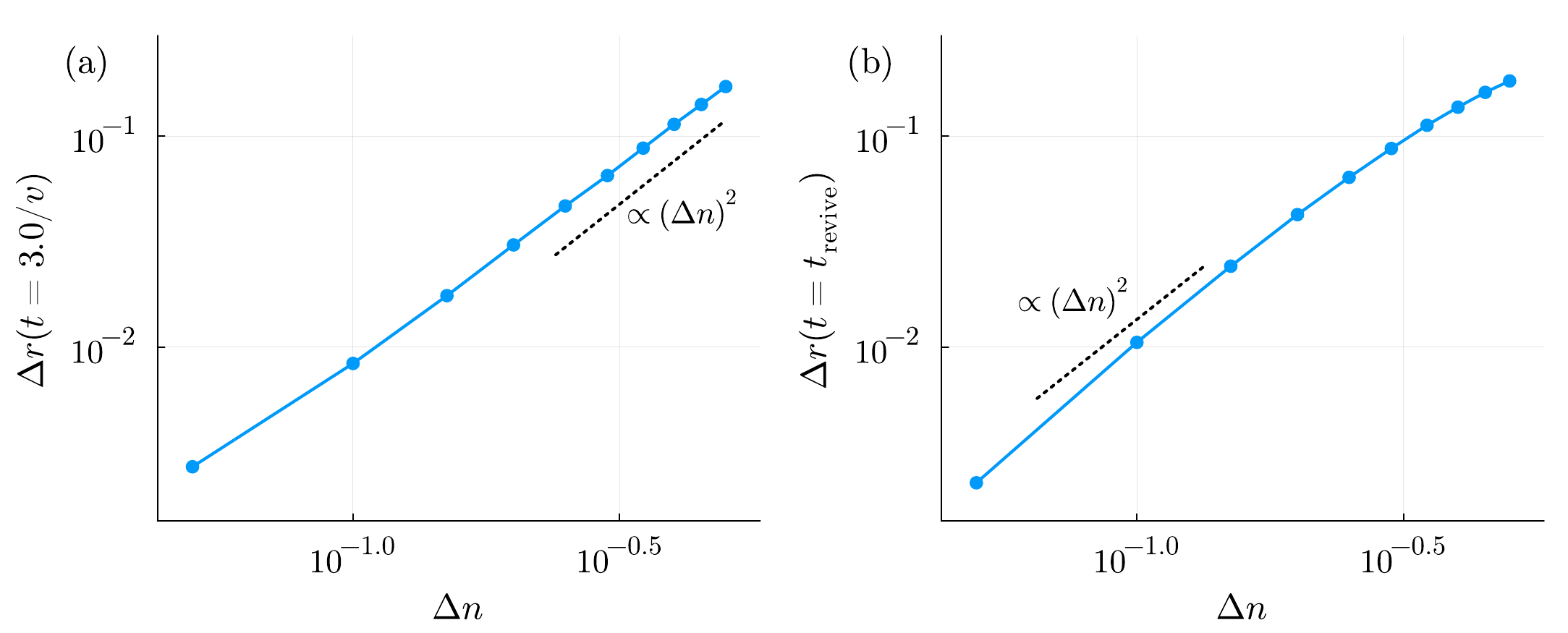}
    \caption{Doping dependence of the deviation of $r(t)$ from its half-filled value.
    We plot $\Delta r(t) = \left. r(t)\right|_{n} - \left. r(t)\right|_{n=1}$ as a function of $\Delta n = n-1$ on a log-log scale.
    Panel (a) shows the weak-coupling case $U=1.5v$ with $t=3.0/v$, and panel (b) shows the strong-coupling case $U=8.0v$ with $t=t_{\mathrm{revive}}$.
    The dashed lines, proportional to $(\Delta n)^2$, are guides to the eye.
    }
    \label{fig:Hubbard_quench_Delta_r_doping_dependence}
\end{figure}

To further quantify how the half-filling behavior is modified by doping, we define 
\begin{align}
    \Delta r(t) = r(t) |_{n} - r(t) |_{n=1},
\end{align}
which is plotted as a function of $\Delta n = n - 1$ for $U = 1.5v$ and $U = 8.0v$ in Fig.~\ref{fig:Hubbard_quench_Delta_r_doping_dependence}.
Here, we fix $t = 3.0/v$ for $U = 1.5v$ and $t = t_{\mathrm{revive}}$ for $U = 8.0v$, following the same convention as in Fig.~\ref{fig:Hubbard_quench_distribution_diff_int_vs_U}.
For both $U=1.5v$ and $U=8.0v$, the data approximately follow a quadratic dependence, $\Delta r \propto (\Delta n)^2$, in the density range studied here.
This indicates that the deviation from the half-filling result grows roughly quadratically with doping, and as a consequence, the characteristic behavior associated with fast thermalization and the dynamical transition is gradually lost as $\Delta n$ increases.

This doping-induced smearing of the dynamical transition is qualitatively consistent with the behavior found in the time-dependent Gutzwiller approximation discussed in Refs.~\cite{schiro2010time,schiro2011quantum}.
Within that approximation, for the half-filled case, physical quantities such as the double occupancy usually show oscillatory behavior after the quench.
However, at the special value of $U = U_{c} / 2$, where $U_c$ is the critical interaction strength for the Mott transition in the equilibrium phase, the double occupancy exponentially decays and the system relaxes to a steady state within a short time.
Also, the long-time limit of the double occupancy shows a singular behavior at $U = U_{c}/2$.
Thus, the time-dependent Gutzwiller approximation predicts a dynamical transition at around $U = U_{c}^{\mathrm{dyn}} = U_{c}/2$ for the half-filled case.
When the particle-hole symmetry is absent, however, this sharp dynamical transition is known to disappear and is replaced by a crossover in the Gutzwiller approach \cite{schiro2011quantum}.
So far, nonequilibrium DMFT with a CT-QMC impurity solver has been applied only to the half-filled case \cite{eckstein2009thermalization}, where the dynamical transition was demonstrated beyond the level of the time-dependent Gutzwiller approximation.
Our results for the particle-hole asymmetric case indicate that the absence of a sharp dynamical transition away from particle-hole symmetry is also a finding which is valid beyond the time-dependent Gutzwiller approximation.

It is also worth noting that a similar tendency has been reported for interaction quenches in the one-dimensional Hubbard model~\cite{hamerla2013dynamical}.
In that work, the half-filled system exhibits a dynamical transition after the interaction quench, while this transition is smeared out away from half filling.
Although the one-dimensional integrable Hubbard model and the infinite-dimensional DMFT setup studied here are not directly comparable, doping has a qualitatively similar effect in both cases.

While the dynamical transition only exists in the half-filled model, it is interesting to note that irrespective of doping, the $r(t)$ curves in Fig.~\ref{fig:Hubbard_quench_distribution_diff_time_evolution} exhibit a qualitative change with increasing $U$. In particular, well-defined $2\pi/U$ oscillations appear for $U\gtrsim 3v$. Since the critical interaction $U \simeq 3v$ in the half-filled system is linked to the Mott transition \cite{schiro2010time}, this observation implies that Mott physics to some degree controls the dynamics even away from half-filling. The situation is reminiscent of the Hund metal phenomenon in multi-orbital Hubbard systems \cite{Georges2013}, where a crossover line within the doped metal phase separates a Fermi liquid regime from a Hund metal regime which is influenced by Mott physics \cite{Werner2008}. The single-band Hubbard model considered here does not show a prominent spin-freezing effect or Hund metal crossover, but the post-quench dynamics shows atomic-like collapse-and-revival behavior for $U\gtrsim 3v$.

\section{Steady-state DMFT for the Hubbard model} 
\label{sec:Hubbard steady state}
In this section, we apply the nonequilibrium TCI impurity solver to compute the steady-state spectral function of the Hubbard model on the Bethe lattice (bandwidth $4v$)
within the formalism introduced in Sec.~\ref{sec:steady-state DMFT}.
In this steady-state approach, the distribution function $f_{\mathrm{st}}(\omega)$ is specified as an input, and the spectral function $A_{\sigma}(\omega)$ consistent with $f_{\mathrm{st}}(\omega)$ is calculated directly.
Here, we choose $f_{\mathrm{st}}(\omega)$ to be the Fermi--Dirac distribution function
\begin{align}
  f_{\mathrm{st}}(\omega) = \frac{1}{e^{\beta \omega} + 1},
\end{align}
and discuss the thermal equilibrium state.

\begin{figure}[t]
  \centering
  \includegraphics[width=0.42\textwidth]{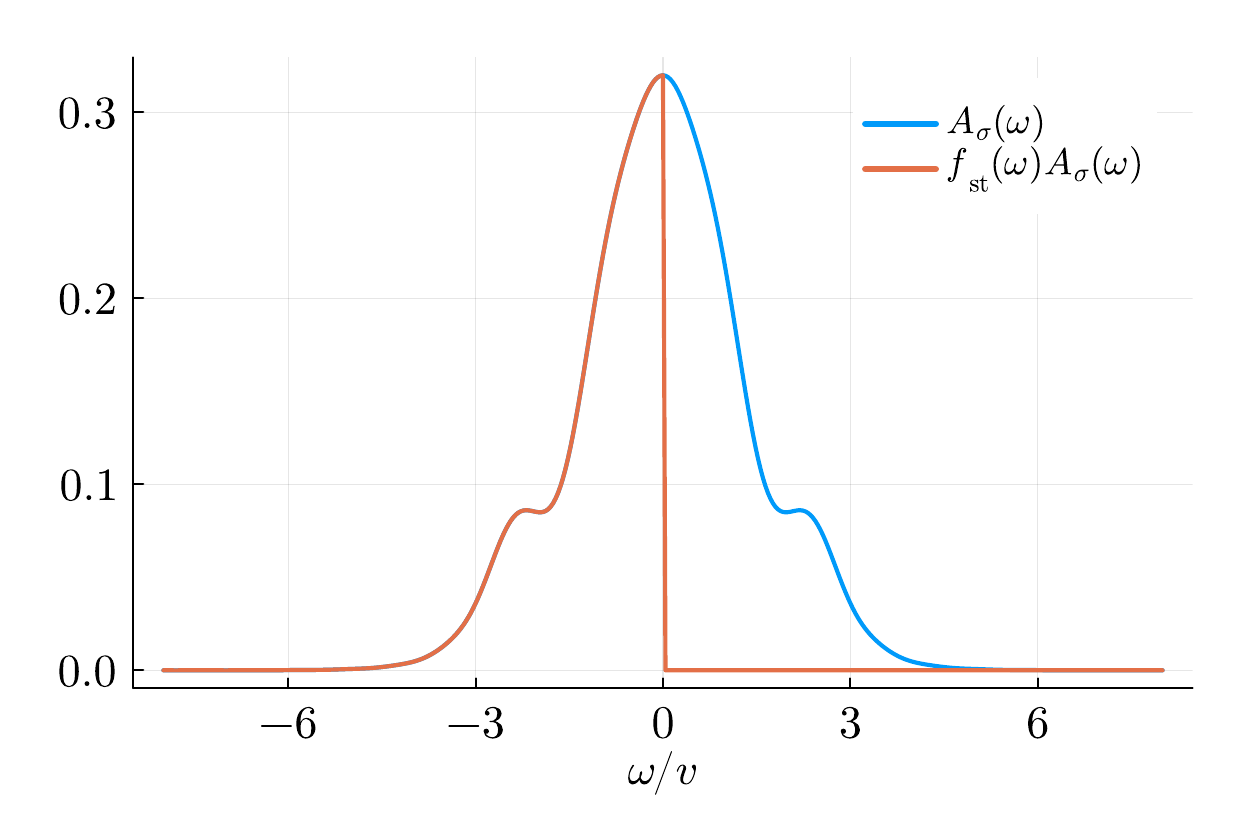}
  \caption{Spectral function $A_{\sigma}(\omega)$ and the occupation function $f_{\mathrm{st}}(\omega) A_{\sigma}(\omega)$ of the Hubbard model obtained by the steady-state DMFT calculation.
  The distribution function $f_{\mathrm{st}}(\omega)$ is set to the Fermi--Dirac distribution function with $T = 0$.
  The other parameters are $U = 2v$, and $t_{\mathrm{max}} = 8/v$.
  The maximum bond dimension is set to $\chi = 40$, and the maximum order is set to $n_{\mathrm{max}} = 18$.
  }
  \label{fig:Hubbard_steady_state_spectral_function}
\end{figure}

Figure~\ref{fig:Hubbard_steady_state_spectral_function} shows the spectral function $A_{\sigma}(\omega)$ and the occupation function $f_{\mathrm{st}}(\omega) A_{\sigma}(\omega)$ obtained by the steady-state DMFT calculation for the half-filled Hubbard model with $T = 0$, and $U = 2v$.
The calculation is performed with $t_{\mathrm{max}} = 8/v$, $\alpha = 0.5$, $\chi = 40$, and $n_{\mathrm{max}} = 18$.
Before Fourier transforming $G_{\sigma}^{R}(t)$, we extrapolate it beyond the cutoff time $t_{\max}$ using linear prediction in order to reduce finite-time truncation artifacts in the resulting spectral function.
Since $U = 2v$ corresponds to half the bandwidth, the system is in the metallic phase, and the spectral function exhibits a quasiparticle peak at the Fermi level $\omega = 0$.
In addition, 
satellite features appear near $\omega \simeq \pm 2v$, which may be precursors of the Hubbard bands. 
This result is in good quantitative agreement with the spectra reported in Refs.~\cite{ganahl2015efficient,nayak2025steady}.

\begin{figure}[t]
  \centering
  \includegraphics[width=0.375\textwidth]{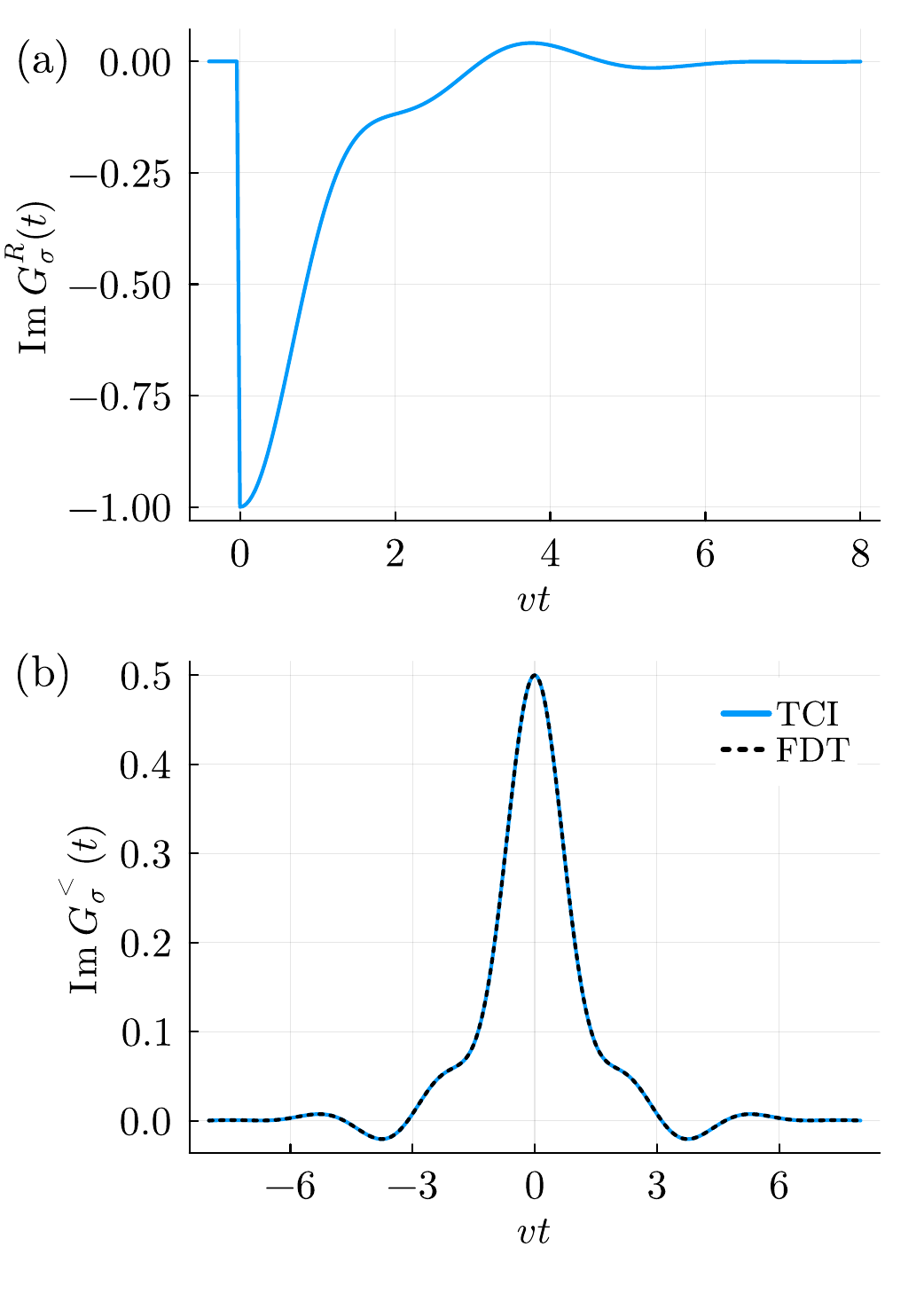}
  \caption{Green's function of the Hubbard model in thermal equilibrium with $U = 2v$ and $T = 0$.
  (a) The imaginary part of the retarded Green's function in the time domain.
  (b) The imaginary part of the lesser Green's function in the time domain.
  The solid line represents the result obtained by the TCI impurity solver, and the dashed line represents the result obtained by the fluctuation-dissipation relation, $G_{\sigma}^{<}(\omega) = -2i f_{\mathrm{st}}(\omega) \Im G_{\sigma}^{R}(\omega)$.
  }
  \label{fig:Hubbard_steady_state_green_function}
\end{figure}

As discussed in Sec.~\ref{sec:steady-state DMFT}, we have to check whether $\abs{G_{\sigma}^{R}(t_{\mathrm{max}})}$ is sufficiently small, and whether the FDT-type relation \eqref{eq:noneq_distribution_function} is satisfied.
In Fig.~\ref{fig:Hubbard_steady_state_green_function} (a), we show the imaginary part of the retarded Green's function in the time domain.
We can see that the value of $\abs{G_{\sigma}^{R}(t_{\mathrm{max}})}$ at $t = t_{\mathrm{max}} = 8/v$ is sufficiently small (around $10^{-3}$), indicating that the maximum time is long enough to obtain reliable results.
Figure~\ref{fig:Hubbard_steady_state_green_function} (b) shows the imaginary part of the lesser Green's function in the time domain obtained by the TCI impurity solver, together with the results obtained by Fourier transforming the FDT-type relation as
\begin{align}
  G_{\sigma}^{<}(t) = \int_{-\infty}^{\infty} \frac{\dd{\omega}}{2\pi} e^{-i\omega t} [-2i f_{\mathrm{st}}(\omega) \Im G_{\sigma}^{R}(\omega)].
\end{align}
We can see that the result of the TCI solver (solid lines) matches well with that obtained by the FDT-type relation (dashed line), indicating that the steady-state DMFT calculation is consistent.

Let us finally discuss the parameter regime in which the nonequilibrium TCI impurity solver combined with the steady-state formalism yields reliable results.
As indicated in Fig.~\ref{fig:Hubbard_steady_state_green_function} (a), we typically need a cutoff time of $t_{\mathrm{max}} \simeq 8/v$ so that the retarded Green's function decays sufficiently within the simulated time window.
In practice, our implementation reaches a maximum order of $n_{\mathrm{max}} \simeq 20$.
Using the rough estimate in Eq.~\eqref{eq:estimate of n_max}, this translates into an estimated interaction range $U \lesssim 2v$ for which accurate results can be expected.

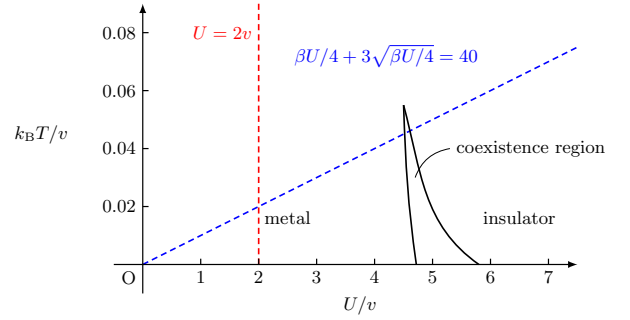
\begin{figure}[t]
	\centering
	\resizebox{0.45\textwidth}{!}{%
	\begin{tikzpicture}
			\draw[thick, blue, densely dashed] (0, 0) -- (7.5, 7.5/100 * 50);
			\draw (2.5, 3.6) node[right, blue]{$\beta U/4 + 3\sqrt{\beta U / 4} = 40$};

			\draw[thick, red, densely dashed] (2.0, 0.0) -- (2.0, 4.5);
			\draw[red] (2.0, 4.0) node[left]{$U = 2v$};
	
			\draw[->, semithick, >=latex] (-0.5, 0) -- (7.5, 0);  
			\draw[->, semithick, >=latex] (0, -0.5) -- (0, 4.5);  
			\draw (0,0) node[below left]{O};
			\draw (3.75, -0.7) node{$U/v$};
			\draw (-1.75, 2.25) node{$k_{\mathrm{B}}T / v$};
	
			\foreach \i in {1, 2, 3, 4, 5, 6, 7} {
					\draw[semithick] (\i, 0.1) -- (\i, 0.0) node[below] {\i};  
			}
			\foreach \i/\label in {1/0.02, 2/0.04, 3/0.06, 4/0.08} {
				\draw[semithick] (0.1, \i) -- (0.0, \i) node[left] {\label};   
			}
	
			\draw[thick] (4.5, 2.75) to [out=-88,in=98] (4.72, 0.0);
			\draw[thick] (4.5, 2.75) to [out=-75, in=145] (5.8, 0.0);
	
			\draw (6.5, 0.8) node{insulator};
			\draw (2.5, 0.8) node{metal};
			\draw (4.7, 1.5) to [out=70, in=190] (5.3, 2.0) node[right]{coexistence region};
	\end{tikzpicture}
	}
	\caption{
	Equilibrium DMFT phase diagram for the Hubbard model on the Bethe lattice. 
	The red dashed line marks the boundary of the region $U\lesssim 2v$, which is roughly the region that can be explored within steady-state DMFT using the nonequilibrium weak-coupling TCI impurity solver.
	The blue dashed line is a rough estimate of the boundary above which the equilibrium weak-coupling TCI solver works efficiently \cite{matsuura2025tensor}.
	}
	\label{fig:phase_diagram}
\end{figure}

Figure~\ref{fig:phase_diagram} summarizes the parameter regime accessible to the nonequilibrium TCI impurity solver.
The blue line is the boundary for the equilibrium TCI solver discussed in Ref.~\cite{matsuura2025tensor}, and is shown for comparison.
Unlike the equilibrium solver, whose required maximum order increases at low temperatures, the nonequilibrium solver remains effective down to low temperatures because the required order is essentially insensitive to temperature in this setup.
Moreover, it directly provides real-frequency spectra, eliminating the need for numerically uncontrolled analytic continuation, which cannot be avoided in the case of equilibrium imaginary-time solvers.

Although we have focused here on the Fermi--Dirac distribution, the steady-state formalism itself can also be used with nonthermal input distributions, such as a double-step distribution, as discussed in Sec.~\ref{sec:steady-state DMFT}.
In the weak-coupling metallic regime accessible with the present solver, however, we find that even when the DMFT loop converges, the resulting Green's functions do not satisfy the FDT-type relation~\eqref{eq:noneq_distribution_function} with the imposed distribution.
This suggests that such nonthermal distributions are short-lived in this regime and are not realized as long-lived stationary states.
Long-lived nonequilibrium distributions of this type are more naturally stabilized in the strongly correlated regime, where doublon--hole recombination is suppressed by the Mott gap.
Since this regime is outside the efficiently treatable range of the present weak-coupling solver, such states may be better explored by TCI impurity solvers based on strong-coupling expansions~\cite{eckstein2024solving,kim2025strong,geng2025third,geng2025photoinduced,geng2026high}.\\

\section{Summary and outlook}
\label{sec:discussions}
In this work, we developed a nonequilibrium TCI impurity solver based on the weak-coupling expansion on the Kadanoff--Baym contour and applied it to nonequilibrium DMFT for both transient and steady-state calculations.
The benchmarks against an exactly solvable impurity problem showed systematic convergence with increasing expansion order and for  moderate bond dimensions, indicating that the nonequilibrium integrand admits an efficient low-rank tensor-train representation.

For the Hubbard model, the method reproduced the known half-filled quench dynamics and the rapid thermalization near the dynamical transition point around $U = 3v$.
Moreover, it remained well controlled in the 3/4-filled case, where the rapid-thermalization regime observed at half filling evolves into a much broader crossover and no comparably fast thermalization behavior is found.
The present calculations thus provide numerically exact evidence that the dynamical transition at half filling evolves into a smooth crossover upon doping. 
This crossover marks a qualitative change from a relatively smooth relaxation on the weak-$U$ side to an atomic-like oscillatory post-quench dynamics on the strong-$U$ side.
To obtain these results, the development of the TCI solver was crucial, since particle-hole-asymmetric impurity problems are known to be substantially more difficult for CT-QMC solvers, due to the severe sign problem.
In the steady-state formalism, the TCI solver can directly provide real-frequency spectra at zero temperature, without any analytic continuation.
These results establish the present TCI impurity solver as a useful approach for nonequilibrium DMFT.

On the other hand, the present method still has some limitations.
Since the summation over the Keldysh indices is performed explicitly, the computational cost grows exponentially with the perturbation order.
In the current implementation, the maximum reachable time in transient calculations is limited to approximately $|U|t_{\max}\lesssim 14.7$.
In the steady-state formalism, the need for a sufficiently long real-time cutoff currently restricts accurate calculations to $U \lesssim 2v$.
Thus, although the method opens a window to deal with the weak-to-intermediate-coupling regime and systems away from half filling, it is not yet capable of reaching the long-time dynamics deep in the strongly correlated regime.

To further clarify the advantages and limitations of the present method, it is useful to compare the TCI solver not only with CT-QMC but also with other impurity solvers for nonequilibrium DMFT.
We begin with weak-coupling impurity solvers based on low-order expansions of the self-energy \cite{eckstein2010interaction,tsuji2013nonequilibrium}.
The second-order perturbation theory for the self-energy is computationally cheap and can describe the dynamics reasonably well in the weak-coupling regime, but its reliability already deteriorates at moderate interaction strength around $U \simeq 2v$.
One may go beyond the second order perturbation by including higher-order diagrams, but the convergence of such a series is not straightforward and does not necessarily lead to systematic improvements.
By contrast, the TCI impurity solver evaluates the weak-coupling expansion of the Green's function itself, whose convergence radius is infinite for a fixed maximum time $t_{\mathrm{max}}$ \cite{bertrand2019reconstructing}.
This makes the method systematically improvable by increasing the maximum order and allows us to obtain reliable results even in the intermediate-to-strong-coupling regime, at least for short times.

From a complementary perspective, one may also compare the TCI solver with strong-coupling impurity solvers such as the noncrossing approximation (NCA) and its extensions \cite{eckstein2010nonequilibrium}.
These methods are reliable in the strongly correlated regime and converge rapidly toward the exact result as the expansion order is increased.
Although high-order calculations are typically expensive because of the large number of diagrams and the associated high-dimensional integrals, recent work has shown that the TCI approach can alleviate this computational burden \cite{eckstein2024solving,kim2025strong,geng2025third,geng2025photoinduced,geng2026high}.
The accuracy of the strong-coupling approach, however, generally deteriorates in the weak-coupling regime, and may even suffer from misleading convergence \cite{geng2025third}.
In this sense, the present weak-coupling TCI solver and the strong-coupling family should be regarded as complementary rather than competing approaches.

There are also approaches that do not rely on perturbative expansions.
Representative examples are Hamiltonian-based impurity solvers, such as exact diagonalization (ED) \cite{gramsch2013hamiltonian,arrigoni2013nonequilibrium} and density matrix renormalization group (DMRG) solvers \cite{wolf2014solving,balzer2015nonthermal,ganahl2015efficient}.
Recently developed influence-functional (IF) approaches \cite{thoenniss2023nonequilibrium,thoenniss2023efficient,ng2023real,chen2024grassmann,nayak2025steady} are also promising methods that belong to this category.
These methods are attractive because they do not rely on an expansion around either the noninteracting or the atomic limit and can therefore be applied, in principle, over a broad interaction range.
Their practical computational cost, however, generally increases with the maximum simulation time $t_{\max}$, due to the increasing number of bath sites required in ED, the buildup of entanglement in DMRG, and the larger grids and bond dimensions needed in IF methods.
According to the existing literature, in the weak-to-intermediate-coupling regime $U \lesssim 3v$, the maximum time scales reached by these methods are roughly comparable to those reached by the present TCI solver.
Because these methods remain computationally demanding even in the weak-coupling regime, the present TCI solver may provide a more efficient route to the same time scales in that regime.

To further extend the applicability of the present TCI solver, the most important next step is to compress the information carried by the Keldysh indices within the tensor-train representation.
If this becomes possible, the explicit summation over the Keldysh indices would no longer be necessary, and the associated exponential computational cost could be reduced to a polynomial scaling.
This would make it possible to investigate longer-time dynamics of strongly correlated systems in a systematic way.
We have tested a simple strategy in which the Keldysh indices are treated as additional tensor indices, but this leads to a large increase in bond dimension and does not significantly improve the accessible time scales.
More suitable variable transformations that disentangle the correlations associated with the Keldysh indices, or possibly entirely different strategies, may therefore be necessary.
Such advances would enable us to follow transient dynamics over longer times and investigate nontrivial thermalization processes.
They could also extend the parameter range accessible to accurate steady-state calculations using either the formulation adopted in this work or the order-by-order DMFT algorithm of Bertrand {\it et al.}~\cite{bertrand2025algorithm}, with TCI replacing the quasi-Monte Carlo impurity solver used in the latter.

Another important direction is the extension to cluster impurity solvers, since the present weak-coupling solver is naturally suited for cluster impurity problems. 
In that case, one must treat not only the time integrations but also the summation over cluster-site indices, for which the sign problem is known to become even more severe in conventional approaches.
If the dependence on these site indices can also be compressed efficiently in the tensor-train framework, the TCI approach may open a route to nonequilibrium cluster DMFT \cite{tsuji2014nonequilibrium} and thereby to the study of nonequilibrium dynamics with nonlocal correlations.

Overall, the present work establishes the nonequilibrium TCI impurity solver as a promising route to nonequilibrium DMFT in parameter regimes that are difficult to access with existing approaches, while also identifying clear directions for further methodological developments.

\section*{Acknowledgments}
We thank M. Eckstein, O. Parcollet, K. Inayoshi, L. Geng and Y. Yu for fruitful discussions.
This work is supported by JST FOREST (Grant No.~JPMJFR2131) and JSPS KAKENHI (Grants No.~JP24H00191, No.~JP25H01246 and No.~JP25H01251).
S.M. is also supported by JSPS KAKENHI (Grant No.~JP26KJ1014), the Forefront Physics and Mathematics Program to Drive Transformation (FoPM), a World-Leading Innovative Graduate Study (WINGS) Program, and the JSR Fellowship, the University of Tokyo.
H.S. was supported by JSPS KAKENHI (Grant No.~22KK0226) and JST FOREST (Grant No.~JPMJFR2232), Japan. 
P.W. acknowledges support by the Swiss National Science Foundation via Grant No.~2000--1-240023.
The computations for this work have been done using the facilities of the Supercomputer Center, the Institute for Solid State Physics, the University of Tokyo (2025-Ba-0039), 
and the implementation of the TCI algorithm is based on \texttt{TensorCrossInterpolation.jl} \cite{nunez2025learning}.

\appendix
\onecolumngrid
\section{Evaluation of the recursive convolution}
\label{sec:recursive convolution}

In this appendix, we discuss the recursive convolution \eqref{eq:recursive convolution} used in the main text.
By combining the TCI approach and the Chebyshev polynomial expansion, we can evaluate recursive convolutions efficiently as tensor contractions.
The method described below is based on the approach mentioned in Ref.~\cite{nunez2022learning}.
Since the implementation details are not spelled out in Ref.~\cite{nunez2022learning}, we provide here a self-contained description of the recursive-convolution procedure used in our calculations.

\subsection{Chebyshev polynomial expansion}
First, we briefly review the Chebyshev polynomial expansion.
Let us consider the matrix-valued function $A(t)$ defined on the interval $0 \le t \le t_{\mathrm{max}}$, and expand it in terms of the Chebyshev polynomials $T_k(x) = \cos(k \arccos x)$.
Since the Chebyshev polynomials are defined on the interval $-1 \le x \le 1$, it is convenient to introduce the rescaled Chebyshev polynomials $\mathcal{T}_{k}(t)$ as
\begin{align}
	\mathcal{T}_{k}(t) = T_k \qty( \frac{2t}{t_{\mathrm{max}}} - 1 ).
\end{align}
If we expand $A(t)$ in terms of $\mathcal{T}_k(t)$ up to order $N_{\mathrm{ch}}-1$ as
\begin{align}
	A(t) \simeq \sum_{k=0}^{N_{\mathrm{ch}}-1} \tilde{A}_{k} \mathcal{T}_k(t),
\end{align}
the optimal expansion coefficients $\tilde{A}_k$ are given by
\begin{align}
	\tilde{A}_{k} = 
	\begin{cases}
		\displaystyle\frac{2}{N_{\mathrm{ch}}} \displaystyle\sum_{\ell=0}^{N_{\mathrm{ch}}-1} A(t^{(\ell)}) \mathcal{T}_k(t^{(\ell)}) &  \text{if } k = 1, \dots, N_{\mathrm{ch}}-1 \\[15pt]
		\displaystyle\frac{1}{N_{\mathrm{ch}}} \displaystyle\sum_{\ell=0}^{N_{\mathrm{ch}}-1} A(t^{(\ell)}) & \text{if } k = 0,
	\end{cases}
	\label{eq:chebyshev_expansion_coefficients}
\end{align}
where $t^{(\ell)} = \frac{t_{\mathrm{max}}}{2} \qty[ 1 + \cos \frac{\pi (\ell + 1/2)}{N_{\mathrm{ch}}} ]$ are the Chebyshev nodes.
This relation can be derived from the orthogonality relation of the Chebyshev polynomials.
If we introduce the diagrammatic representation of $A(t^{(\ell)})$ and $\tilde{A}_k$ as
\begin{align}
	A(t^{(\ell)}) \; \; = \; \;
	\begin{tikzpicture}[baseline=-2.0ex]
		\node[draw, fill, blue!70!gray, minimum width=0.6cm, minimum height=0.6cm, inner sep=0pt] (A) at (0,0) {\textcolor{white}{$A$}};
		\draw[thick] (A.south) -- ++(0,-0.25) node[below]{$\ell$};
		\draw[thick] (A.west) -- ++(-0.25,0) node[left, inner sep=0pt]{};
		\draw[thick] (A.east) -- ++(0.25,0) node[right, inner sep=0pt]{};
	\end{tikzpicture}, \hspace{1cm}
	\tilde{A}_{k} \; \; = \; \;
	\begin{tikzpicture}[baseline=-2.0ex]
		\node[draw, fill, green!50!gray, minimum width=0.6cm, minimum height=0.6cm, inner sep=0pt] (A) at (0,0) {\textcolor{white}{$\tilde{A}$}};
		\draw[thick] (A.south) -- ++(0,-0.25) node[below]{$k$};
		\draw[thick] (A.west) -- ++(-0.25,0) node[left, inner sep=0pt]{};
		\draw[thick] (A.east) -- ++(0.25,0) node[right, inner sep=0pt]{};
	\end{tikzpicture},
\end{align}
the relation \eqref{eq:chebyshev_expansion_coefficients} between the values of $A(t)$ at the Chebyshev nodes $A(t^{(\ell)})$ and the expansion coefficients $\tilde{A}_k$ can be 
diagrammatically represented as
\begin{align}
	\tilde{A}_{k} \; \; = \; \;
  \begin{tikzpicture}[baseline=-2.0ex]
    \node[draw, fill, green!50!gray, minimum width=0.6cm, minimum height=0.6cm, inner sep=0pt] (A) at (0,0) {\textcolor{white}{$\tilde{A}$}};
    \draw[thick] (A.south) -- ++(0,-0.25) node[below]{$k$};
    \draw[thick] (A.west) -- ++(-0.25,0) node[left, inner sep=0pt]{};
    \draw[thick] (A.east) -- ++(0.25,0) node[right, inner sep=0pt]{};
  \end{tikzpicture}
  \; \; = \; \;
  \begin{tikzpicture}[baseline=-3.5ex]
    \node[draw, fill, blue!70!gray, minimum width=0.6cm, minimum height=0.6cm, inner sep=0pt] (A) at (0,0) {\textcolor{white}{$A$}};
    \node[circle, draw, fill, black, minimum size=0.2cm, inner sep=0pt] (R) at (0,-0.65) {};
    \draw[thick] (A.south) -- (R.north);
    \draw[thick] (R.south) -- ++(0,-0.25) node[below]{$k$};
    \draw[thick] (A.west) -- ++(-0.25,0) node[left, inner sep=0pt]{};
    \draw[thick] (A.east) -- ++(0.25,0) node[right, inner sep=0pt]{};
  \end{tikzpicture},
	\label{eq:chebyshev_expansion_conversion}
\end{align}
where the black dot represents the transformation matrix
\begin{align}
  \begin{tikzpicture}[baseline=-1.0ex]
    \node[circle, draw, fill, black, minimum size=0.2cm, inner sep=0pt] (R) at (0,0) {};
    \draw[thick] (R.north) -- ++(0,0.25) node[above]{$\ell$};
    \draw[thick] (R.south) -- ++(0,-0.25) node[below]{$k$};
  \end{tikzpicture}
  \; \; = \; \;
  \begin{cases}
    \dfrac{2}{N_{\mathrm{ch}}} \mathcal{T}_{k}(t^{(\ell)}) & \text{if} \;\; k = 1, \cdots, N_{\mathrm{ch}}-1 \\[10pt]
    \dfrac{1}{N_{\mathrm{ch}}} & \text{if} \;\; k = 0.
  \end{cases}
\end{align}
In the following, we use the blue and green boxes to represent the values of functions at the Chebyshev nodes and the Chebyshev expansion coefficients, respectively.

\subsection{Computation of the convolution of two functions}
\label{sec:convolution_two_functions}
Let us consider the convolution of two matrix-valued functions $A(t)$ and $B(t)$ defined on the interval $0 \le t \le t_{\mathrm{max}}$,
\begin{align}
	C(t) = (A * B)(t) = \int_{0}^{t} \dd{t'} \, A(t - t') B(t').
\end{align}
Here we assume that we know the values of $A(t)$ and $B(t)$ at the Chebyshev nodes $t^{(\ell)}$ $(\ell=0,1,\dots,N_{\mathrm{ch}}-1)$.
Then, we can expand $A(t)$ and $B(t)$ in terms of the rescaled Chebyshev polynomials $\mathcal{T}_k(t)$ as
\begin{align}
	A(t) \simeq \sum_{k=0}^{N_{\mathrm{ch}}-1} \tilde{A}_{k} \mathcal{T}_k(t), \quad B(t) \simeq \sum_{k=0}^{N_{\mathrm{ch}}-1} \tilde{B}_{k} \mathcal{T}_k(t),
\end{align}
where the coefficients $\tilde{A}_k$ and $\tilde{B}_k$ can be obtained from \eqref{eq:chebyshev_expansion_conversion}.
Then the convolution $C(t)$ can be expressed as
\begin{align}
	C(t) \simeq \sum_{k=0}^{N_{\mathrm{ch}}-1} \sum_{k'=0}^{N_{\mathrm{ch}}-1} \tilde{A}_{k} \tilde{B}_{k'} \int_{0}^{t} \dd{t'} \, \mathcal{T}_k(t - t') \mathcal{T}_{k'}(t').
\end{align}
By introducing a diagrammatic representation of the convolution of the rescaled Chebyshev polynomials as
\begin{align}
  \int_{0}^{t^{(\ell)}} \dd{t'} \mathcal{T}_{k}(t^{(\ell)}-t') \mathcal{T}_{k'}(t')
  \;\; = \;\;
  \begin{tikzpicture}[baseline=-0.8ex]
    \node[draw, fill, gray, minimum width=1.8cm, minimum height=0.6cm, inner sep=0pt] (TT) at (0,0) {\textcolor{white}{$\mathcal{T}*\mathcal{T}$}};
    \coordinate (leftbase)  at ($ (TT.north west)!0.2!(TT.north east)$);
    \coordinate (rightbase) at ($ (TT.north west)!0.8!(TT.north east)$);
    \coordinate (leftleg)  at ($ (leftbase)  + (0,0.25) $);
    \coordinate (rightleg) at ($ (rightbase) + (0,0.25) $);
    \draw[thick] (leftbase)  -- (leftleg) node[above]{$k$};
    \draw[thick] (rightbase) -- (rightleg) node[above]{$k'$};
    \draw[thick] (TT.south) -- ++(0,-0.25) node[below]{$\ell$};
  \end{tikzpicture}
	\; \; ,
\end{align}
we can obtain the values of the convolution $C(t)$ at the Chebyshev nodes $t^{(\ell)}$ by performing the tensor contraction
\begin{align}
  C(t^{(\ell)}) \;\;=\;\;
  \begin{tikzpicture}[baseline=-1.5ex]
    \node[draw, fill, blue!70!gray, minimum width=0.6cm, minimum height=0.6cm, inner sep=0pt] (A) at (0,0) {\textcolor{white}{$C$}};
    \draw[thick] (A.south) -- ++(0,-0.25) node[below]{$\ell$};
    \draw[thick] (A.west) -- ++(-0.25,0) node[left, inner sep=0pt]{};
    \draw[thick] (A.east) -- ++(0.25,0) node[right, inner sep=0pt]{};
  \end{tikzpicture}
  \;\;=\;\;
  \begin{tikzpicture}[baseline=0.0ex]
    \node[draw, fill, gray, minimum width=1.8cm, minimum height=0.6cm, inner sep=0pt] (TT) at (0,0) {\textcolor{white}{$\mathcal{T}*\mathcal{T}$}};
    \coordinate (leftbase)  at ($ (TT.north west)!0.166!(TT.north east)$);
    \coordinate (rightbase) at ($ (TT.north west)!0.833!(TT.north east)$);
    \node[draw, fill, green!50!gray, minimum width=0.6cm, minimum height=0.6cm, inner sep=0pt] (A) at ($ (leftbase) + (0,0.6) $) {\textcolor{white}{$\tilde{A}$}};
    \node[draw, fill, green!50!gray, minimum width=0.6cm, minimum height=0.6cm, inner sep=0pt] (B) at ($ (rightbase) + (0,0.6) $) {\textcolor{white}{$\tilde{B}$}};
    \draw[thick] (leftbase)  -- (A.south);
    \draw[thick] (rightbase) -- (B.south);
    \draw[thick] (A.east) -- (B.west);
    \draw[thick] (TT.south) -- ++(0,-0.25) node[below]{$\ell$};
    \draw[thick] (A.west) -- ++(-0.25,0) node[left, inner sep=0pt]{};
    \draw[thick] (B.east) -- ++(0.25,0) node[right, inner sep=0pt]{};
  \end{tikzpicture}
  \;\; = \;\;
  \begin{tikzpicture}[baseline=0.5ex]
    \node[draw, fill, gray, minimum width=1.8cm, minimum height=0.6cm, inner sep=0pt] (TT) at (0,0) {\textcolor{white}{$\mathcal{T}*\mathcal{T}$}};
    \coordinate (leftbase)  at ($ (TT.north west)!0.166!(TT.north east)$);
    \coordinate (rightbase) at ($ (TT.north west)!0.833!(TT.north east)$);
    \node[circle, draw, fill, black, minimum size=0.2cm, inner sep=0pt] (R1) at ($ (leftbase) + (0,0.25) $) {};
    \node[circle, draw, fill, black, minimum size=0.2cm, inner sep=0pt] (R2) at ($ (rightbase) + (0,0.25) $) {};
    \node[draw, fill, blue!70!gray, minimum width=0.6cm, minimum height=0.6cm, inner sep=0pt] (A) at ($ (R1.north) + (0,0.45) $) {\textcolor{white}{$A$}};
    \node[draw, fill, blue!70!gray, minimum width=0.6cm, minimum height=0.6cm, inner sep=0pt] (B) at ($ (R2.north) + (0,0.45) $) {\textcolor{white}{$B$}};
    \draw[thick] (leftbase)  -- (A.south);
    \draw[thick] (rightbase) -- (B.south);
    \draw[thick] (A.east) -- (B.west);
    \draw[thick] (TT.south) -- ++(0,-0.25) node[below]{$\ell$};
    \draw[thick] (A.west) -- ++(-0.25,0) node[left, inner sep=0pt]{};
    \draw[thick] (B.east) -- ++(0.25,0) node[right, inner sep=0pt]{};
  \end{tikzpicture}.
\end{align}
From the values of $C(t)$ at the Chebyshev nodes, we can also obtain the Chebyshev expansion coefficients of $C(t)$ using the relation \eqref{eq:chebyshev_expansion_coefficients},
\begin{align}
  \tilde{C}_{k} \;\;=\;\; 
  \begin{tikzpicture}[baseline=-1.5ex]
    \node[draw, fill, green!50!gray, minimum width=0.6cm, minimum height=0.6cm, inner sep=0pt] (A) at (0,0) {\textcolor{white}{$\tilde{C}$}};
    \draw[thick] (A.south) -- ++(0,-0.25) node[below]{$k$};
    \draw[thick] (A.west) -- ++(-0.25,0) node[left, inner sep=0pt]{};
    \draw[thick] (A.east) -- ++(0.25,0) node[right, inner sep=0pt]{};
  \end{tikzpicture}
  \;\;=\;\;
  \begin{tikzpicture}[baseline=-2.0ex]
    \node[draw, fill, blue!70!gray, minimum width=0.6cm, minimum height=0.6cm, inner sep=0pt] (A) at (0,0) {\textcolor{white}{$C$}};
    \node[circle, draw, fill, black, minimum size=0.2cm, inner sep=0pt] (R) at ($(A.south) + (0.0, -0.25)$) {};
    \draw[thick] (A.south) -- (R.north);
    \draw[thick] (R.south) -- ++(0,-0.15) node[below]{$k$};
    \draw[thick] (A.west) -- ++(-0.25,0) node[left, inner sep=0pt]{};
    \draw[thick] (A.east) -- ++(0.25,0) node[right, inner sep=0pt]{};
  \end{tikzpicture}
  \;\;=\;\;
  \begin{tikzpicture}[baseline=1.0ex]
    \node[draw, fill, gray, minimum width=1.8cm, minimum height=0.6cm, inner sep=0pt] (TT) at (0,0) {\textcolor{white}{$\mathcal{T}*\mathcal{T}$}};
    \coordinate (leftbase)  at ($ (TT.north west)!0.166!(TT.north east)$);
    \coordinate (rightbase) at ($ (TT.north west)!0.833!(TT.north east)$);
    \node[circle, draw, fill, black, minimum size=0.2cm, inner sep=0pt] (R1) at ($ (leftbase) + (0,0.25) $) {};
    \node[circle, draw, fill, black, minimum size=0.2cm, inner sep=0pt] (R2) at ($ (rightbase) + (0,0.25) $) {};
    \node[circle, draw, fill, black, minimum size=0.2cm, inner sep=0pt] (R3) at ($ (TT.south) + (0,-0.25)$) {};
    \node[draw, fill, blue!70!gray, minimum width=0.6cm, minimum height=0.6cm, inner sep=0pt] (A) at ($ (R1.north) + (0,0.45) $) {\textcolor{white}{$A$}};
    \node[draw, fill, blue!70!gray, minimum width=0.6cm, minimum height=0.6cm, inner sep=0pt] (B) at ($ (R2.north) + (0,0.45) $) {\textcolor{white}{$B$}};
    \draw[thick] (leftbase)  -- (A.south);
    \draw[thick] (rightbase) -- (B.south);
    \draw[thick] (A.east) -- (B.west);
    \draw[thick] (TT.south) -- (R3.north) node[below]{};
    \draw[thick] (A.west) -- ++(-0.25,0) node[left, inner sep=0pt]{};
    \draw[thick] (B.east) -- ++(0.25,0) node[right, inner sep=0pt]{};
    \draw[thick] (R3.south) -- ++(0,-0.15) node[below]{$k$};
  \end{tikzpicture}
  \;\; = \;\;
  \begin{tikzpicture}[baseline=0.8ex]
    \node[draw=black, line width=2.0pt, fill=gray, minimum width=1.8cm, minimum height=0.6cm, inner sep=0pt] (TT) at (0,0) {\textcolor{white}{$\mathcal{T}*\mathcal{T}$}};
    \coordinate (leftbase)  at ($ (TT.north west)!0.166!(TT.north east)$);
    \coordinate (rightbase) at ($ (TT.north west)!0.833!(TT.north east)$);
    \node[draw, fill, blue!70!gray, minimum width=0.6cm, minimum height=0.6cm, inner sep=0pt] (A) at ($ (leftbase) + (0,0.6) $) {\textcolor{white}{$A$}};
    \node[draw, fill, blue!70!gray, minimum width=0.6cm, minimum height=0.6cm, inner sep=0pt] (B) at ($ (rightbase) + (0,0.6) $) {\textcolor{white}{$B$}};
    \draw[thick] (leftbase)  -- (A.south);
    \draw[thick] (rightbase) -- (B.south);
    \draw[thick] (A.east) -- (B.west);
    \draw[thick] (TT.south) -- ++(0,-0.25) node[below]{$k$};
    \draw[thick] (A.west) -- ++(-0.25,0) node[left, inner sep=0pt]{};
    \draw[thick] (B.east) -- ++(0.25,0) node[right, inner sep=0pt]{};
  \end{tikzpicture},
\end{align}
where the gray box with a thick border is defined as
\begin{align}
  \begin{tikzpicture}[baseline=-0.8ex]
    \node[draw=black, line width=2.0pt, fill=gray, minimum width=1.8cm, minimum height=0.6cm, inner sep=0pt] (TT) at (0,0) {\textcolor{white}{$\mathcal{T}*\mathcal{T}$}};
    \coordinate (leftbase)  at ($ (TT.north west)!0.2!(TT.north east)$);
    \coordinate (rightbase) at ($ (TT.north west)!0.8!(TT.north east)$);
    \coordinate (leftleg)  at ($ (leftbase)  + (0,0.25) $);
    \coordinate (rightleg) at ($ (rightbase) + (0,0.25) $);
    \draw[thick] (leftbase)  -- (leftleg) node[above]{$k$};
    \draw[thick] (rightbase) -- (rightleg) node[above]{$k'$};
    \draw[thick] (TT.south) -- ++(0,-0.25) node[below]{$k''$};
  \end{tikzpicture}
	\; \; \coloneqq \; \;
	\begin{tikzpicture}[baseline=-0.75ex]
    \node[draw, fill, gray, minimum width=1.8cm, minimum height=0.6cm, inner sep=0pt] (TT) at (0,0) {\textcolor{white}{$\mathcal{T}*\mathcal{T}$}};
    \coordinate (leftbase)  at ($ (TT.north west)!0.166!(TT.north east)$);
    \coordinate (rightbase) at ($ (TT.north west)!0.833!(TT.north east)$);
    \node[circle, draw, fill, black, minimum size=0.2cm, inner sep=0pt] (R1) at ($ (leftbase) + (0,0.25) $) {};
    \node[circle, draw, fill, black, minimum size=0.2cm, inner sep=0pt] (R2) at ($ (rightbase) + (0,0.25) $) {};
    \node[circle, draw, fill, black, minimum size=0.2cm, inner sep=0pt] (R3) at ($ (TT.south) + (0,-0.25)$) {};
    \draw[thick] (TT.south) -- (R3.north) node[below]{};
    \draw[thick] (R3.south) -- ++(0,-0.15) node[below]{$k''$};
		\draw[thick] (leftbase)  -- (R1.north);
		\draw[thick] (rightbase) -- (R2.north);
		\draw[thick] (R1.north) -- ++(0,0.15) node[above]{$k$};
		\draw[thick] (R2.north) -- ++(0,0.15) node[above]{$k'$};
  \end{tikzpicture}
	\; \; .
	\label{eq:chebyshev_convolution_box}
\end{align}
Thus, we can construct an approximation to $C(t) = (A*B)(t)$ from the values of $A(t)$ and $B(t)$ at the Chebyshev nodes by performing the tensor contraction as
\begin{align}
	C(t) \simeq \sum_{k=0}^{N_{\mathrm{ch}}-1} \tilde{C}_{k} \mathcal{T}_k(t) \simeq
	\sum_{k=0}^{N_{\mathrm{ch}}-1} 
	\;
	\begin{tikzpicture}[baseline=0.8ex]
		\node[draw=black, line width=2.0pt, fill=gray, minimum width=1.8cm, minimum height=0.6cm, inner sep=0pt] (TT) at (0,0) {\textcolor{white}{$\mathcal{T}*\mathcal{T}$}};
		\coordinate (leftbase)  at ($ (TT.north west)!0.166!(TT.north east)$);
		\coordinate (rightbase) at ($ (TT.north west)!0.833!(TT.north east)$);
		\node[draw, fill, blue!70!gray, minimum width=0.6cm, minimum height=0.6cm, inner sep=0pt] (A) at ($ (leftbase) + (0,0.6) $) {\textcolor{white}{$A$}};
		\node[draw, fill, blue!70!gray, minimum width=0.6cm, minimum height=0.6cm, inner sep=0pt] (B) at ($ (rightbase) + (0,0.6) $) {\textcolor{white}{$B$}};
		\draw[thick] (leftbase)  -- (A.south);
		\draw[thick] (rightbase) -- (B.south);
		\draw[thick] (A.east) -- (B.west);
		\draw[thick] (TT.south) -- ++(0,-0.25) node[below]{$k$};
		\draw[thick] (A.west) -- ++(-0.25,0) node[left, inner sep=0pt]{};
		\draw[thick] (B.east) -- ++(0.25,0) node[right, inner sep=0pt]{};
	\end{tikzpicture}
	\;
	\mathcal{T}_k(t).
\end{align}
Note that the gray box with a thick border \eqref{eq:chebyshev_convolution_box} is determined solely by the Chebyshev polynomials and is independent of the functions $A(t)$ and $B(t)$.
Therefore, we can precompute and store these values for later use.

\subsection{Computation of the recursive convolution}
Let us consider evaluating an integral
\begin{align}
	\mathcal{I}(t) = \int_{S_{n}^{0,t}} \dd{t_{1}} \cdots \dd{t_{n}} \, F(t; t_{1}, \dots, t_{n})
\end{align}
for $0 \le t \le t_{\mathrm{max}}$.
As we discussed in Sec.~\ref{sec:lesser and greater Green's functions}, we first introduce new time difference variables $u_{1}, \dots, u_{n}$ as
\begin{align}
	u_{1} = t - t_{1}, \quad u_{2} = t_{1} - t_{2}, \quad \dots, \quad u_{n} = t_{n-1} - t_{n}
\end{align}
and regard $F(t;t_{1}, \dots, t_{n})$ as a function of $t$ and $u_{1}, \dots, u_{n}$,
\begin{align}
	\mathcal{F}(t; u_{1}, \dots, u_{n}) = F(t; t_{1}, \dots, t_{n}).
\end{align}
Then we use the TCI algorithm to approximate the function $\mathcal{F}(t; u_{1}, \dots, u_{n})$ in the tensor-train format as
\begin{align}
	\mathcal{F}(t; u_{1}, \dots, u_{n}) \simeq M_{0}(t) M_{1}(u_{1}) \cdots M_{n}(u_{n}),
	\label{eq:tensor_train_decomposition_F_continuous}
\end{align}
to rewrite the integral $\mathcal{I}(t)$ as a recursive convolution as
\begin{align}
	\mathcal{I}(t) \simeq M_{0}(t) (M_{1} * M_{2} * \cdots * M_{n} * I)(t),
\end{align}
where $I(t) = 1$.

For simplicity, Eq.~\eqref{eq:tensor_train_decomposition_F_continuous} is written as if we directly obtained a tensor-train decomposition of the continuous function $\mathcal{F}(t; u_{1}, \dots, u_{n})$.
However, in practice, we first discretize the function $\mathcal{F}(t; u_{1}, \dots, u_{n})$ and then perform the tensor-train decomposition of the discretized tensor.
In actual computations, it is convenient to discretize the variables $t, u_{1}, \dots, u_{n}$ at $N_{\mathrm{ch}}$ Chebyshev nodes defined on the interval $[0, t_{\mathrm{max}}]$, and regard $\mathcal{F}(t; u_{1}, \dots, u_{n})$ as a $(n+1)$-dimensional tensor whose local dimensions are all $N_{\mathrm{ch}}$,
\begin{align}
	[\mathcal{F}]_{\ell_{0} \ell_{1} \cdots \ell_{n}} = \mathcal{F}(t^{(\ell_{0})}; t^{(\ell_{1})}, \dots, t^{(\ell_{n})}).
\end{align}
Then we perform the tensor-train decomposition of the tensor $[\mathcal{F}]_{\ell_{0} \ell_{1} \cdots \ell_{n}}$ using the TCI algorithm to obtain
\begin{align}
  [\mathcal{F}]_{\ell_{0}\ell_{1}\cdots\ell_{n}} = \;
  \begin{tikzpicture}[baseline=-1.0ex]
    \node[draw, fill, blue!70!gray, minimum width=2.5cm, minimum height=0.6cm] (A) at (0,0) {\textcolor{white}{$\mathcal{F}$}};
    \coordinate (base1) at ($ (A.south west)!0.1!(A.south east)$);
    \coordinate (base2) at ($ (A.south west)!0.3!(A.south east)$);
    \coordinate (base3) at ($ (A.south west)!0.9!(A.south east)$);
    \draw[thick] (base1) -- ++(0,-0.3) node[below] {$\ell_{0}$};
    \draw[thick] (base2) -- ++(0,-0.3) node[below] {$\ell_{1}$};
    \draw[thick] (base3) -- ++(0,-0.3) node[below] {$\ell_{n}$};
    \node (base_dot) at ($ (A.south west)!0.6!(A.south east) + (0,-0.2)$) {$\cdots$};
  \end{tikzpicture}
  \;\;\simeq\;\;
  \begin{tikzpicture}[baseline=-1.0ex]
    \node[draw, fill, blue!70!gray, minimum width=0.6cm, minimum height=0.6cm, inner sep=0pt] (T1) at (0,0) {\textcolor{white}{$M_{0}$}};
    \node[draw, fill, blue!70!gray, minimum width=0.6cm, minimum height=0.6cm, inner sep=0pt] (T2) at (1.0,0) {\textcolor{white}{$M_{1}$}};
    \node[draw, fill, blue!70!gray, minimum width=0.6cm, minimum height=0.6cm, inner sep=0pt] (Tn) at (3.0,0) {\textcolor{white}{$M_{n}$}};
    \node[inner sep=0pt] (left_end) at ($(T1.west) + (-0.4,0)$) {$\times$};
    \node[inner sep=0pt] (right_end) at ($(Tn.east) + (0.4,0)$) {$\times$};
    \node (cdot) at ($(T2.east)!0.5!(Tn.west)$) {$\cdots$};
    \draw[thick] (left_end.center) -- (T1) -- (T2) -- ($(T2) + (0.4,0)$);
    \draw[thick] ($(Tn) + (-0.4,0.0)$) -- (Tn) -- (right_end.center);
    \draw[thick] (T1.south) -- ++(0,-0.3) node[below] {$\ell_{0}$};
    \draw[thick] (T2.south) -- ++(0,-0.3) node[below] {$\ell_{1}$};
    \draw[thick] (Tn.south) -- ++(0,-0.3) node[below] {$\ell_{n}$};
  \end{tikzpicture}
	\;\;\simeq\;\;
	M_{0}^{\ell_{0}} M_{1}^{\ell_{1}} \cdots M_{n}^{\ell_{n}}.
\end{align}
The site tensors $M_{0}^{\ell_{0}}$, $M_{1}^{\ell_{1}}$, $\dots$, $M_{n}^{\ell_{n}}$ obtained in this way can be regarded as the discretized values of the continuous functions $M_{0}(t)$, $M_{1}(u_{1})$, $\dots$, $M_{n}(u_{n})$ in Eq.~\eqref{eq:tensor_train_decomposition_F_continuous} at the Chebyshev nodes,
\begin{align}
	M_{i}^{\ell_{i}} = M_{i}(t^{(\ell_{i})}), \quad i = 0, 1, \dots, n.
\end{align}
Then we can evaluate the recursive convolution $M_{1} * M_{2} * \cdots * M_{n} * I$ by performing $n$ successive convolutions of two functions using the method explained in Sec.~\ref{sec:convolution_two_functions}.
Diagrammatically, the process of evaluating the recursive convolution can be represented as
\begin{align}
  (M_1*\cdots*M_n*I)(t)
  \;\;\simeq\;\;
  \sum_{k=0}^{N_{\mathrm{ch}}-1}
  \;
  \begin{tikzpicture}[baseline=8.0ex]
    \node[
      draw=black, line width=2.0pt, fill=gray,
      minimum width=3.6cm, minimum height=0.6cm,
      inner sep=0pt
    ] (conv1) at (0,0)
      {\textcolor{white}{$\mathcal{T}*\mathcal{T}$}};
    \draw[thick]
      (conv1.south) -- +(0.0,-0.25) node[below]{$k$};
    \coordinate (conv1left) at
      ($(conv1.north west)!0.08!(conv1.north east)$);
    \coordinate (conv1right) at
      ($(conv1.north west)!0.92!(conv1.north east)$);
    \node[
      minimum width=0.6cm, minimum height=0.6cm,
      inner sep=1pt
    ] (vdot) at ($(conv1right)+(0,0.5)$)
      {\scriptsize$\vdots$};
    \draw[thick] (conv1right) -- +(0.0,0.1);
    \node[
      draw=black, line width=2.0pt, fill=gray,
      minimum width=2.4cm, minimum height=0.6cm,
      inner sep=0pt
    ] (convn-1) at ($(vdot)+(0,0.6)$)
      {\textcolor{white}{$\mathcal{T}*\mathcal{T}$}};
    \draw[thick] (convn-1.south) -- +(0.0,-0.1);
    \coordinate (convn-1left) at
      ($(convn-1.north west)!0.125!(convn-1.north east)$);
    \coordinate (convn-1right) at
      ($(convn-1.north west)!0.875!(convn-1.north east)$);
    \node[
      draw=black, line width=2.0pt, fill=gray,
      minimum width=1.8cm, minimum height=0.6cm,
      inner sep=0pt
    ] (convn) at ($(convn-1right)+(0,0.5)$)
      {\textcolor{white}{$\mathcal{T}*\mathcal{T}$}};
    \draw[thick] (convn-1right) -- (convn.south);
    \coordinate (convnleft) at
      ($(convn.north west)!0.166!(convn.north east)$);
    \coordinate (convnright) at
      ($(convn.north west)!0.833!(convn.north east)$);
    \node[
      draw, fill, blue!70!gray,
      minimum width=0.6cm, minimum height=0.6cm,
      inner sep=0pt
    ] (Mn) at ($(convnleft)+(0,0.5)$)
      {\textcolor{white}{$M_n$}};
    \node[
      draw, fill, blue!70!gray,
      minimum width=0.6cm, minimum height=0.6cm,
      inner sep=0pt
    ] (Mn+1) at ($(convnright)+(0,0.5)$)
      {\textcolor{white}{$I$}};
    \node[
      draw, fill, blue!70!gray,
      minimum width=0.6cm, minimum height=0.6cm,
      inner sep=0pt
    ] (Mn-1) at ($(Mn)!(convn-1left)!(Mn+1)$)
      {\scalebox{0.7}[1.0]{\textcolor{white}{$M_{n-1}$}}};
    \draw[thick] (convnleft) -- (Mn.south);
    \draw[thick] (convnright) -- (Mn+1.south);
    \draw[thick] (convn-1left) -- (Mn-1.south);
    \draw[thick] (Mn-1.east) -- (Mn.west);
    \draw[thick] (Mn.east) -- (Mn+1.west);
    \node[
      draw, fill, blue!70!gray,
      minimum width=0.6cm, minimum height=0.6cm,
      inner sep=0pt
    ] (M1) at ($(Mn)!(conv1left)!(Mn+1)$)
      {\textcolor{white}{$M_1$}};
    \draw[thick] (conv1left) -- (M1.south);
    \draw[thick] (M1.west) -- +(-0.3,0.0);
    \node[
      text=black,
      minimum width=0.6cm, minimum height=0.6cm,
      inner sep=1pt
    ] (cdot) at ($(M1)!0.5!(Mn-1)$)
      {$\cdots$};
    \draw[thick] (M1.east) -- +(0.3,0.0);
    \draw[thick] (Mn-1.west) -- +(-0.3,0.0);
  \end{tikzpicture}
  \;\;
  \mathcal{T}_k(t).
\end{align}
For the calculation in the main text, we set the number of Chebyshev nodes to $N_{\mathrm{ch}} = 20$ for the transient dynamics case and $N_{\mathrm{ch}} = 30$ for the steady-state case, which is sufficient to obtain converged results.

\section{Extrapolation of the Green's function}
\label{sec:extrapolation}

As discussed in Sec.~\ref{sec:lesser and greater Green's functions},
we need to smoothly extrapolate the Weiss Green's function $\mathcal{G}_{\sigma}^{s_{i}s_{j}}(t_{i},t_{j})$ from $[0, t_{\mathrm{max}}]^2$ to $(-\infty, t_{\mathrm{max}}]^2$ in order to apply the TCI algorithm for the function $\mathcal{Q}^{<}_{nk\sigma}(t, \qty{u_{i}})$.
In this appendix, we explain how this extrapolation is implemented.

\subsection{Extrapolation of single-variable functions}
Let us start with the extrapolation of a single-variable function $f(t)$ defined on $[0, t_{\mathrm{max}}]$ to $(-\infty, t_{\mathrm{max}}]$.
We want an extrapolation procedure which ensures that the extrapolated function is differentiable at $t = 0$ and decays exponentially as $t \to -\infty$.
To this end, we introduce two functions,
\begin{align}
	h_{1}(t) = (1 - \lambda t) e^{\lambda t}, \quad h_{2}(t) = t e^{\lambda t},
\end{align}
where $\lambda > 0$ controls the decay rate.
These functions satisfy 
\begin{align}
	h_{1}(0) = 1, \quad h_{1}'(0) = 0, \quad h_{2}(0) = 0, \quad h_{2}'(0) = 1,
	\label{eq:basis_functions}
\end{align}
and decay exponentially as $t \to -\infty$.
Then, we can extrapolate the function $f(t)$ by the linear combination of these two functions as
\begin{align}
	\tilde{f}(t) = f(0^{+}) h_{1}(t) + f'(0^{+}) h_{2}(t) \quad (t < 0),
\end{align}
where $f'(0^{+})$ is the right derivative of $f(t)$ at $t = 0$.
The extrapolated function $\tilde{f}(t)$ is differentiable at $t = 0$ because of Eq.~\eqref{eq:basis_functions} and decays exponentially as $t \to -\infty$.

\subsection{Extrapolation of two-variable functions}

\begin{figure}
	\centering
	\begin{tikzpicture}[scale=0.8]
		\fill[blue!20] (0, 0) rectangle (2.0, 2.0);
		
		\fill[red!20] (-3.0, 0) rectangle (0, 2.0);
		\fill[red!20] (0, -3.0) rectangle (2.0, 0);
		\fill[red!20] (-3.0, -3.0) rectangle (0, 0);

		\draw[->,>=latex,thick] (-3, 0) -- (3, 0) node[above] {$t$};
		\draw[->,>=latex,thick] (0, -3) -- (0, 3) node[right] {$t'$};
		\node at (0,0) [below left] {O};
		
		\node at (1.0, 1.0) {$f(t,t')$};
		\node at (-1.5, 1.0) {$f_{2}(t,t')$};
		\node at (-1.5, -1.5) {$f_{3}(t,t')$};
		\node at (1.0, -1.5) {$f_{4}(t,t')$};
		\node at (2.0, 0.0) [below right] {$t_{\mathrm{max}}$};
		\node at (0.0, 2.0) [above left] {$t_{\mathrm{max}}$};
	\end{tikzpicture}
	\caption{Schematic illustration of the extrapolation of a two-variable function $f(t,t')$ defined on $[0, t_{\mathrm{max}}]^2$ to $(-\infty, t_{\mathrm{max}}]^2$. The blue region represents the original domain of $f(t,t')$, while the red regions represent the extrapolated domains with the corresponding extrapolated functions $f_{2}(t,t')$, $f_{3}(t,t')$, and $f_{4}(t,t')$.}
    \label{fig:two-variable function extrapolation}
\end{figure}
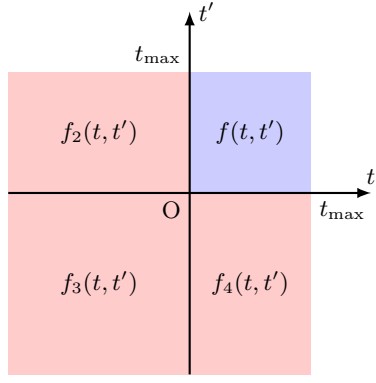

The extrapolation of a two-variable function $f(t, t')$ defined on $[0, t_{\mathrm{max}}]^2$ to $(-\infty, t_{\mathrm{max}}]^2$ can be performed in a similar manner.
The decomposition of the original and extrapolated domains is schematically shown in Fig.~\ref{fig:two-variable function extrapolation}.
We first extrapolate the function along the $t$ direction for each fixed $t' \in [0, t_{\mathrm{max}}]$ using the method explained in the previous section.
This gives us the extrapolated function
\begin{align}
	f_{2}(t,t') = f(0^{+},t') h_{1}(t) + \partial_{t} f(0^{+},t') h_{2}(t) \quad (t < 0, \; t' \in [0, t_{\mathrm{max}}]).
	\label{eq:extrapolation_t_direction}
\end{align}
We also extrapolate the function along the $t'$ direction for each fixed $t \in [0, t_{\mathrm{max}}]$ as
\begin{align}
	f_{4}(t,t') = f(t,0^{+}) h_{1}(t') + \partial_{t'} f(t,0^{+}) h_{2}(t') \quad (t' < 0, \; t \in [0, t_{\mathrm{max}}]).
	\label{eq:extrapolation_tprime_direction}
\end{align}
For the bottom-left region where both $t$ and $t'$ are negative, we can extrapolate the function either from $f_{2}(t,t')$ along the $t'$ direction or from $f_{4}(t,t')$ along the $t$ direction.
If we extrapolate from $f_{2}(t,t')$, we obtain
\begin{align}
	f_{3}(t,t') = f_{2}(t,0^{+}) h_{1}(t') + \partial_{t'} f_{2}(t,0^{+}) h_{2}(t') \quad (t < 0, \; t' < 0).
\end{align}
Inserting Eq.~\eqref{eq:extrapolation_t_direction} into this equation, we have
\begin{align}
	&f_{3}(t,t') = f(0^{+},0^{+}) h_{1}(t) h_{1}(t') + \partial_{t} f(0^{+},0^{+}) h_{2}(t) h_{1}(t') \notag \\
	&\hspace{2.5cm} + \partial_{t'} f(0^{+},0^{+}) h_{1}(t) h_{2}(t') + \partial_{t'} \partial_{t} f(0^{+},0^{+}) h_{2}(t) h_{2}(t').
\end{align}
Since $f_{3}(t,t')$ is extrapolated from $f_{2}(t,t')$, we have to check whether $f_{3}(t,t')$ is smoothly connected with $f_{4}(t,t')$ at $t = 0$.
To this end, let us consider obtaining $f_{3}(t,t')$ by extrapolating $f_{4}(t,t')$ along the $t$ direction.
To avoid confusion, we denote the function obtained in this way as $\tilde{f}_{3}(t,t')$.
As a result, we have
\begin{align}
	&\tilde{f}_{3}(t,t') = f(0^{+},0^{+}) h_{1}(t) h_{1}(t') + \partial_{t} f(0^{+},0^{+}) h_{2}(t) h_{1}(t') \notag \\
	&\hspace{2.5cm} +  \partial_{t'} f(0^{+},0^{+}) h_{1}(t) h_{2}(t') + \partial_{t} \partial_{t'} f(0^{+},0^{+}) h_{2}(t) h_{2}(t').
\end{align}
We see that $f_{3}(t,t')$ and $\tilde{f}_{3}(t,t')$ are identical if the mixed derivatives satisfy $\partial_{t'} \partial_{t} f(0^{+},0^{+}) = \partial_{t} \partial_{t'} f(0^{+},0^{+})$.
Since $f(t,t')$ is smooth enough on $[0, t_{\mathrm{max}}]^2$, we can safely assume that this condition is satisfied, which ensures that $f_{3}(t,t')$ is smoothly connected with $f_{4}(t,t')$ at $t = 0$.

Finally, we discuss the choice of the extrapolation parameter $\lambda$.
Although the tensor-train approximation is constructed using the extrapolated integrand, the integration in Eq.~\eqref{eq:recursive convolution} is performed only over the original simplex domains, where the integrand is unchanged. 
Consequently, any choice of $\lambda>0$ yields the same result once convergence with respect to the bond dimension and time discretization is reached. 
The choice of $\lambda$ can nevertheless affect the bond dimension required to achieve a given accuracy. 
For the calculations presented in this paper, we use $\lambda=1$, for which we typically obtain small bond dimensions at the desired accuracy.

\twocolumngrid
\bibliography{noneq_dmft}

\end{document}